\documentclass[10pt]{article}

\usepackage[a4paper]{geometry}
\usepackage[utf8]{inputenc}
\usepackage{extarrows}
\usepackage{verbatim}
\usepackage{stackengine}
\usepackage{amsfonts}
\usepackage{amsmath}
\usepackage{graphicx}
\usepackage{comment}
\usepackage[noend]{algorithmic}
\usepackage{orcidlink}
\usepackage{wasysym}
\usepackage{booktabs}

\usepackage[ruled]{algorithm}
\usepackage{algorithmic}

\usepackage{url}
\newcommand{\true}{{\sf True}}
\newcommand{\false}{{\sf False}}
\newcommand{\cons}{{\sf Cons}}
\newcommand{\inc}{{\sf Inc}}
\newcommand{\ans}{{\sf Ans}}
\newcommand{\consans}{{\sf C\_ans}}
\newcommand{\sconsans}{{\sf SC\_ans}}
\newcommand{\msconsans}{{\sf MSC\_ans}}
\newcommand{\wconsans}{{\sf WC\_ans}}

\newcommand{\srep}{{\sf S\_rep}}

\newcommand{\repcons}{{\sf R\_cons}}
\newcommand{\srepans}{{\sf SRep\_ans}}
\newcommand{\srepcons}{{\sf S\_cons}}

\newcommand{\tuples}{{\sf tuples}}

\newtheorem{theorem}{Theorem}
\newtheorem{definition}{Definition}
\newtheorem{lemma}{Lemma}
\newtheorem{corollary}{Corollary}
\newtheorem{proposition}{Proposition}
\newtheorem{example}{Example}

\usepackage{todonotes}

\author{Dominique Laurent  \orcidlink{0000-0002-7264-9576}\\
\texttt{dominique.laurent@u-cergy.fr}
\\
ETIS Laboratory - ENSEA, CY Cergy Paris University, CNRS\\F-95000 Cergy-Pontoise - FRANCE
\and
Nicolas Spyratos \orcidlink{0002-3432-8608}\\
\texttt{nicolas.spyratos@lri.fr}
\\
LISN Laboratory - University Paris-Saclay, CNRS\\F-91405 Orsay - FRANCE}

\title{Consistent Query Answering without Repairs in Tables with Nulls and Functional Dependencies}

\begin{document}

\maketitle

\begin{abstract}
In this paper, we study consistent query answering in tables with nulls and functional dependencies. Given such a table $T$, we consider the set $\mathcal{T}$ of all tuples that can be built up from constants appearing in $T$; and we use set theoretic semantics for tuples and functional dependencies to characterize the tuples of $\mathcal{T}$ in two orthogonal ways: first as true or false tuples; and then as consistent or inconsistent tuples. Queries are issued against $T$ and evaluated in $\mathcal{T}$. 

\noindent In this setting, we consider a query Q: {\em select X from T where Condition} over $T$ and define its consistent answer to be the set of tuples $x$ in $\mathcal{T}$ such that: (a) $x$ is a true and consistent tuple with schema $X$ and (b) there exists a true super-tuple $t$ of $x$ in $\mathcal{T}$ satisfying the condition. We show that, depending on the `status' that the super-tuple $t$ has in $\mathcal{T}$, there are different types of consistent answer to $Q$.  

\noindent The main contributions of the paper are: (a) a novel approach to consistent query answering {\em not} using table repairs; (b) polynomial algorithms for computing the sets of true/false tuples and the sets of consistent/inconsistent tuples of $\mathcal{T}$; (c) polynomial algorithms in the size of $T$ for computing different types of consistent answer for both conjunctive and disjunctive queries; and (d) a detailed discussion of the differences between our approach and the approaches using table repairs. 
\end{abstract}


\section{Introduction}\label{sec:intro}
In a relational database, each table is seen as a set of tuples that must satisfy a set of functional dependencies. Moreover, each table is assumed to be consistent with the dependencies and its tuples are assumed to be true when users query the table (these are basic assumptions in relational databases). Consistency is verified during updates in the following sense: if the update to be performed results in an inconsistent table then the update is rejected, otherwise it is accepted. 

However, the consistency of a table in general is not always easy to verify, for example if the table is the result of integrating data coming from independent sources (e.g., web sources). In such cases the table may contain inconsistent tuples, and the problem is: how to answer user queries so that the answers contain only tuples that are true and consistent. This kind of query answering process is usually referred to as consistent query answering \cite{ArenasBC99,Bertossi2011}.

\begin{figure}[ht]
\begin{center}
{\footnotesize
\begin{tabular}{cc}
\begin{tabular}{l|ccc}
$T$&$Emp$&$Dept$&$Addr$\\
\hline
$1$&$e~$&$d~$&$a~$\\
$2$&$e~$&$d~$&$a'$\\
$3$&$e'$&$d'$&$a~$\\
\end{tabular}
&
\qquad  \large{$\xLongrightarrow{\text{\sf repairs~}}$\qquad }
\end{tabular}
\begin{tabular}{l}
\begin{tabular}{c|ccc}
$R_1$&$Emp$&$Dept$&$Addr$\\
\hline
$1$&$e~$&$d~$&$a~$\\
$3$&$e'$&$d'$&$a~$\\
\end{tabular}
\\~\\
\begin{tabular}{c|ccc}
$R_2$&$Emp$&$Dept$&$Addr$\\
\hline
$2$&$e~$&$d~$&$a'$\\
$3$&$e'$&$d'$&$a~$\\
\end{tabular}
\end{tabular}
}
\end{center}
\caption{An inconsistent table and its two repairs
\label{fig:repairs}}
\end{figure}

To see an example, consider the table $T$ of Figure~\ref{fig:repairs}, with dependencies $Emp \to Dept$ and $Dept\to Addr$. The table $T$ is defined over attributes $Emp$, $Dept$ and $Addr$, standing for `employee identifier', `department identifier' and `department address', respectively.

This table is inconsistent as the pair of tuples 1 and 2 violates the dependency $Dept \to Addr$. However, if we ask the SQL query $Q: {\tt select}~Emp, Dept~{\tt from}~T$, it makes sense to return the set of tuples $\{ed, e'd'\}$ as the answer. Indeed, there is no reason to reject this answer as it is a consistent answer, in the sense that it satisfies the dependency $Emp \to Dept$. In other words, an inconsistent table might contain some consistent parts (i.e., some useful information) which can be extracted through queries.

The traditional approach to alleviate the impact of inconsistent data on query answers is to introduce the notion of repair: a repair is a maximal consistent sub-set of the table, and a tuple $t$ is in the consistent answer if $t$ is present  in the answer from every repair \cite{ArenasBC99,Wijsen05}. To illustrate this approach, consider again the table $T$ in Figure~\ref{fig:repairs} with its two repairs, namely $R_1=\{eda, e'd'a \}$ and $R_2=\{eda', e'd'a \}$. Both these repairs are consistent with the given dependencies, and maximal with respect to set-theoretic inclusion. The answer to $Q$ from $R_1$ is $\{ed, e'd' \}$ and so is the answer from $R_2$. Therefore the consistent answer to $Q$ is $\{ed, e'd'\}$. 
We note here that the complexity of the query evaluation algorithm in the repairs approach is APX-complete\footnote{We recall that APX is the set of NP optimization problems that allow polynomial-time approximation algorithms (source Wikipedia).} \cite{LivshitsKR20}. 

\smallskip
Now, in several applications today we have to deal with tables containing nulls (i.e., missing values) and having to satisfy a set of functional dependencies. The presence of nulls in a table is due to various reasons such as: `value does not exist' (e.g., the maiden name of a male employee); `value exists but is currently unknown' (e.g., the department of a newly hired employee currently following a training course before being assigned to a specific department); and so on. A survey of the different kinds of missing values considered in the literature can be found in \cite{LevLoizou99}. In our approach we assume that missing values are of the kind `value exists but is currently unknown'.

In the context of the previous example, let $T=\{ed, da, ea'\}$ in which the $Addr$-value in $ed$, the $Emp$-value in $da$ and the $Dept$-value in $ea'$ are nulls that is they exist but are currently unknown. Considering the query $Q: {\tt select}~Emp,Addr~{\tt from}~T$, if $T$ is seen as a regular table, the consistent answer to $Q$ is $\{ea'\}$, as the tuples in $T$ `seem' to satisfy the functional dependencies $Emp \to Dept$ and $Dept \to Addr$.
 
However, these functional dependencies allow to infer the missing values: from tuples $ed$ and $da$, it can be inferred that $eda$ is true, and from tuples $ed$, $ea'$ it can be inferred that $eda'$ is true. This shows that $Dept \to Addr$ is not satisfied, and thus $\{ea'\}$ should {\em not} be returned as a consistent answer. The process of inferring missing values in a table is known in the literature as the chase procedure \cite{FaginMU82,Spyratos87,Ullman}, and it is well-known that this procedure {\em fails} (i.e., stops and returns no table) when encountering an inconsistency. Therefore, the repair-based approaches do not work in the presence of nulls. 

\smallskip
In the light of the previous example, we propose a novel approach to consistent query answering that {\em does not rely on repairs}, but that works in the case where the given set of tuples contains missing values.

In our approach one starts again with a set of tuples $T$ that are not required to be all of them defined over the same set of attributes. Therefore $T$ is represented as a table with nulls. However, nulls simply act as place holders that may receive values implied by functional dependencies, as shown in the previous example. What is significant in our approach is the set $\mathcal{T}$, called the {\em universe of discourse} of $T$, containing all tuples that can be built up from values appearing in $T$. In our previous example, $\mathcal{T}$ consists of the tuples $eda$ and $eda'$ together with all their sub-tuples.

Queries are addressed to $T$ and consistent answers are obtained from $\mathcal{T}$. To achieve this we associate every tuple $t$ in $\mathcal{T}$ with a {\em set} of identifiers, referred to as the interpretation of $t$. In doing so, we define set-theoretic semantics for tuples and for functional dependencies which allows us to characterize the tuples of $\mathcal{T}$ along two orthogonal dimensions: a tuple of $\mathcal{T}$ can be {\em true} or {\em false} on one hand and {\em consistent} or {\em inconsistent} on the other hand. 

In this setting, we consider a query $Q:$ {\tt select $X$ from $T$ where $Condition$} over $T$ and we define its consistent answer as the set of tuples $x$ in $\mathcal{T}$ such that: (a) $x$ is a true and consistent tuple with schema $X$ and (b) there exists a true super-tuple $t$ of $x$ in $\mathcal{T}$ satisfying the condition. We show that, depending on the `status' that the super-tuple $t$ has in $\mathcal{T}$, there are different types of consistent answer to $Q$.  

The important point to emphasize again is that, in our approach, queries are addressed to $T$ and evaluated using the universe of discourse $\mathcal{T}$. As mentioned earlier, $\mathcal{T}$ is the set of all tuples that one can define using values appearing in $T$. Actually, the fundamental difference between our approach and all approaches based on repairs is that their universe of discourse is the table $T$ itself. Therefore our approach is simply not comparable to approaches based on repairs. To see this, consider again our previous example but this time with  $Emp \to Addr$ as the only functional dependency.

Now suppose that $T= \{eda, eda', e'da'\}$ and that we ask for the set of all addresses. Then the answer in the repairs approach is $\{a'\}$ because there are only two repairs, $R_1=\{eda,e'da'\}$ and $R_2=\{eda',e'da'\}$, and the answers from each repair are respectively $\{a,a'\}$ and $\{a'\}$; whereas in our approach, the answer  is empty because $a$ and $a'$ are both inconsistent tuples of $\mathcal{T}$. On the other hand,  if $T= \{eda, ed'a'\}$ and we ask for the set of all departments then the answer  in the repairs approach is empty because there are only two repairs, $R_1=\{eda\}$ and $R_2=\{ed'a'\}$ and none of $d$, $d'$ is in the answer from both repairs; whereas, in our approach, the answer is $\{d,d'\}$ because both $d$ and $d'$ are true and consistent tuples of $\mathcal{T}$.

\smallskip
Our work builds upon earlier work on partition semantics \cite{CKS86,Spyratos87} and on inconsistent tables with missing values \cite{JIIS}. The main contributions of this paper are as follows:
\begin{enumerate}
    \item We propose a novel approach to consistent query answering in tables with nulls and functional dependencies, without using table repairs. 
    \item We provide polynomial algorithms for computing (a) the sets of true/false tuples and the sets of consistent/inconsistent tuples of $\mathcal{T}$ and (b) for computing the consistent answer to any conjunctive or disjunctive query over $T$. 
    \item  We offer a detailed discussion of the differences between our approach and other approaches including the approaches by table repairs. 
\end{enumerate}
The remaining of the paper is organized as follows. In Section~\ref{sec:semantics} we first introduce the basic notions and notation of our approach, and we present our set-theoretic semantics for tuples and functional dependencies; in Section~\ref{sec:cons-query} we address the issue of consistent query answering, introducing the notion of consistency with respect to a selection condition; in Section~\ref{sec:comp-issues} we first define the m-Chase algorithm, a chase-like algorithm for computing the sets of true/false tuples and the sets of consistent/inconsistent tuples of $\mathcal{T}$, and then we propose a polynomial algorithm for computing consistent answers; in Section~\ref{sec:rel-work} we compare our approach to the repair approaches and discuss other related work; and finally, in Section~\ref{sec:conclusion} we offer concluding remarks and outline current work and future perspectives.
\section{The Semantics of our Model}\label{sec:semantics}
In this section we recall basic definitions from the relational model regarding tuples and tables, and we present the set-theoretic semantics that we use for tuples and functional dependencies. Our approach builds upon the so-called ``partition model'' introduced in \cite{Spyratos87}.

\subsection{Terminology and Notation}

Following \cite{Spyratos87}, we consider a universe $U =\{A_1, \ldots , A_n\}$ in which every attribute $A_i$ is associated with a set of atomic values called the domain of $A_i$ and denoted by $dom(A_i)$. An element of $\bigcup_{A \in U}dom(A)$ is called a  {\em constant}.

We call {\em relation schema} (or simply {\em schema}) any nonempty sub-set of $U$ and we denote it by the concatenation of its elements; for example $\{A_1, A_2\}$ is simply denoted by $A_1A_2$. Similarly, the union of schemes $S_1$ and $S_2$ is denoted as $S_1S_2$ instead of $S_1 \cup S_2$.

We define a {\em tuple} $t$ over $U$ to be a partial function from $U$ to $\bigcup_{A \in U} dom(A)$ such that, for every $A$ in $U$, if $t$ is defined over $A$ then $t(A)$ belongs to $dom(A)$. The domain of definition of $t$ is called the {\em schema} of $t$, denoted by $sch(t)$. We note that tuples in our approach satisfy the {\em First Normal Form} \cite{Ullman} in the sense that each tuple component is an atomic value from the corresponding attribute domain.

Regarding notation, we follow the usual convention that, whenever possible, lower-case characters denote domain constants and  upper-case characters denote the corresponding attributes. Following this convention, the schema  of a tuple $t=ab$ is $AB$ and more generally, we denote the schema  of a tuple $s$ as $S$.

Assuming that the schema  of a tuple $t$ is understood, $t$ is denoted by the concatenation of its values, that is: $t=a_{i_1} \ldots a_{i_k}$ means that for every $j=1, \ldots ,k$, $t(A_{i_j})=a_{i_j}$, where $a_{i_j}$ is in $dom(A_{i_j})$, and $sch(t)=A_{i_1} \ldots  A_{i_k}$.

As in \cite{Spyratos87}, we assume that for any two distinct attributes $A$ and $B$, we have $dom(A) \cap dom(B)=\emptyset$. However, it might be relevant for attribute domains to share values. For instance, in our introductory example, in the presence of attribute $Mgr$ standing for `manager', the domains of $Emp$ and $Mgr$ are employee identifiers. In such cases, in order to avoid confusion, we denote attribute values as {\em pairs} of the form $\langle attribute\_name, value\rangle$ and comparisons are assessed with respect to their $value$ component only. In order to keep the notation simple we shall omit attribute names as prefixes whenever no ambiguity is possible.

We define a {\em table} $T$ over $U$ to be any {\em finite} set of tuples over $U$ (therefore duplicates are not allowed). As tuples over $U$ are partial functions over $U$, it follows that $T$ may contain nulls.

\smallskip
Given a table $T$ over $U$, we denote by $\mathcal{T}$ the set of all tuples that can be built up from constants appearing in $T$; and we call $\mathcal{T}$ the {\em universe of discourse} of $T$ as it contains all tuples of interest in query processing. Actually, as we shall see shortly, queries are issued against $T$ and consistent answers are obtained from $\mathcal{T}$. For every relation schema $X$, we denote by $\mathcal{T}(X)$ the set of all tuples in $\mathcal{T}$ whose schema is $X$. Formally, $\mathcal{T}(X)=\{t \in \mathcal{T}~|~sch(t)=X\}$.

For every $A$ in $U$, the set of all values from $dom(A)$ occurring in $T$ is referred to as the {\em active domain of $A$}, denoted by $adom(A)$. We denote by $\mathcal{AD}$ the set of all constants appearing in $T$ that is: $\mathcal{AD}= \bigcup_{A \in U} adom(A)$. We emphasize that even when attribute domains are infinite, active domains are always finite and therefore the sets $\mathcal{AD}$ and $\mathcal{T}$ are finite as well. 

\smallskip
Given a tuple $t$, for every $A$ in $sch(t)$, $t(A)$ is also denoted by $t.A$ and more generally, for every nonempty sub-set $S$ of $sch(t)$ the restriction of $t$ to $S$, also called {\em sub-tuple} of $t$, is denoted by $t.S$. In other words, if $S \subseteq sch(t)$, $t.S$ is the tuple such that $sch(t.S)=S$, and for every $A$ in $S$ we have $(t.S).A=t.A$.

 Moreover, $\sqsubseteq$ denotes the `sub-tuple' relation, defined over $\mathcal{T}$ as follows: for any tuples $t_1$ and $t_2$, $t_1 \sqsubseteq t_2$ holds if $t_1$ is a sub-tuple of $t_2$. It is thus important to keep in mind that whenever $t_1 \sqsubseteq t_2$ holds, it is understood that $sch(t_1) \subseteq sch(t_2)$ also holds. The relation $\sqsubseteq$ is clearly a partial order over $\mathcal{T}$. Given a table $T$, the set of all sub-tuples of the tuples in $T$ is called the {\em lower closure} of $T$ and it is defined by: ${\sf LoCl}(T) = \{q \in \mathcal{T}~|~(\exists t \in T)(q \sqsubseteq t)\}$.
\subsection{Partition Semantics for Tuples}
In this work, we consider tables $T$ over a set $U$ of attributes, possibly with nulls, and we assume that every tuple $t$ in $\mathcal{T}$ is associated with a unique identifier, $id(t)$. We denote by $TID$ the set of all identifiers of tuples in $\mathcal{T}$; that is: $TID= \{id(t)~|~t \in \mathcal{T}\}$. In our examples, for simplicity, we assume that $TID$ is a set of positive integers.  Denoting by $\mathcal{P}(TID)$ the power-set of $TID$, the following definition of interpretation is in the spirit of \cite{Spyratos87}.

\begin{definition}\label{def:int-T}
Let $T$ be a table over $U$. We call {\em interpretation} of $T$ any function $I$ from $\mathcal{T}$ to $\mathcal{P}(TID)$ such that:
\begin{itemize}
    \item For every tuple $t$ of $T$, $I(t) \neq \emptyset$ 
    \item For every tuple $t= a_1a_2 \ldots a_n$ of $\mathcal{T}$, $I(t)= I(a_1)\cap I(a_2) \cap \ldots \cap I(a_n)$
\end{itemize} 
We say that a tuple $t$ in $\mathcal{T}$ is {\em true} in $I$  if $I(t) \neq \emptyset$; otherwise we say that $t$ is {\em false} in $I$. We denote by $True(I)$ the set of tuples $t$ that are true in $I$ and by $False(I)$ the set of tuples $t$ that are false in $I$.
\hfill$\Box$
\end{definition} 
\begin{example}\label{ex:nico}
\rm{
Let $T= \{ab, a'c \}$ and let $I$ be a function from $\mathcal{T}$ to $\mathcal{P}(TID)$ defined as follows:  $I(a)= \{1, 2 \}$, $I(a')= \{2 \}$, $I(b)= \{1, 2 \}$, $I(c)= \{2 \}$, $I(ab)=\{1, 2 \}$, $I(ac)=\{2 \}$, $I(a'b)= \{2 \}$, $I(a'c)= \{2 \}$, $I(abc)= \{2 \}$, $I(a'bc)= \{2 \}$.
Then $I$ is an interpretation of $T$ as we have:  
\begin{itemize}
    \item $I(ab) \neq \emptyset$ and $I(a'c) \neq \emptyset$ that is the two tuples of $T$ are true in $I$
    \item For every tuple $t$ in $\mathcal{T}$, $I(t)$ is indeed the intersection of the interpretations of its components. For example, $I(ac)= I(a) \cap I(c)= \{1, 2 \} \cap \{2 \}= \{2 \}$ and one can verify easily for the remaining tuples of $\mathcal{T}$. 
    \hfill$\Box$
\end{itemize} }
\end{example}
It is important to note that the first item in Definition~\ref{def:int-T} expresses a fundamental assumption regarding a relational table $T$, namely that every tuple $t$ of $T$ is assumed to be true. Two immediate consequences are that (a) $I(a) \neq \emptyset$, for all $a$ in $\mathcal{AD}$ (i.e., every $a$ in $\mathcal{AD}$ is true in $I$) and (b) if a tuple $t$ is true in $I$ then so is every sub-tuple of $t$; and if $t$ is false in $I$ then so is every super-tuple of $t$. As a consequence the set $\mathcal{T}$ is partitioned into true and false tuples that is $True(I) \cup False(I)= \mathcal{T}$ and $True(I) \cap False(I)= \emptyset$.

The interpretations of  $T$ can be compared according to the following definition. 
\begin{definition}
Let $T$ be a table and $I, I'$ two interpretations of $T$. Then we say that $I$ is less than or equal to $I'$, denoted  $I \preceq I'$ if for every $t$ in $\mathcal{T}$, $I(t) \subseteq I'(t)$.
\hfill$\Box$
\end{definition} 
It is easy to see that the relation $ \preceq $ is a partial ordering over the set of all interpretations of $T$. Note that if we view the constants of $\mathcal{AD}$ as unary tuples then we have that: if $I \preceq I'$ then for every $a$ in $\mathcal{AD}$, $I(a) \subseteq I'(a)$ holds; and in the opposite direction,  if $I(a) \subseteq I'(a)$ holds for every $a$ in $\mathcal{AD}$ then for every $t$ in $\mathcal{T}$, $I(t) \subseteq I'(t)$ also holds that is $I \preceq I'$ holds. Therefore we have that: 
\begin{itemize}
\item
$I \preceq I'$ holds if and only if for every $a$ in $\mathcal{AD}$, $I(a) \subseteq I'(a)$.
\end{itemize}
Based on this result, in order to verify whether $I \preceq I'$, it is sufficient to do so for every constant in $\mathcal{AD}$ rather than for every tuple in $\mathcal{T}$.

\smallskip
Now, to see the intuition behind the above definitions, think of the identifiers of $TID$ as being objects and of every $a$ in $\mathcal{AD}$ as being an atomic property that each object may have. Then $I(a)$ is the set of objects having property $a$; and similarly, the intuitive idea behind the interpretation of a tuple is that a tuple, say $ab$, is the `conjunction' of the atomic properties $a$ and $b$. Therefore $I(ab)$ is the set of objects each having both properties $a$ and $b$, hence $I(ab)= I(a) \cap I(b)$. Clearly, if $I(ab)= \emptyset$ then no object has both properties $a$ and $b$ so the tuple $ab$ is false in $I$. 

As for the ordering of interpretations what $I \preceq I'$ means is that, for every property $t$ of $\mathcal{T}$, the set of objects having property $t$ in $I$ is included in the set of objects having property $t$ in $I'$.


Another fundamental assumption made in the relational model is that the components of every tuple be atomic. Actually such a tuple is said to satisfy the First Normal Form \cite{Ullman}. To express this assumption using our definition of interpretation, suppose that there are $a$, $a'$ in $adom(A)$ such that $a \neq a'$ and $I(a) \cap I(a') \ne \emptyset$. Then every tuple whose identifier is in $I(a) \cap I(a')$ violates the First Normal Form as its $A$-component is $\{a, a'\}$, therefore not atomic. In the light of this observation we introduce the following definition of inconsistent tuple.
\begin{definition}\label{def:incons-t}
Let $T$ be a table over $U$ and $I$ an interpretation of $T$. A tuple $t$ in $\mathcal{T}$ is said to be {\em inconsistent in $I$} if there exist $A$ in $sch(t)$ and $a'$ in $adom(A)$  such that $t.A \ne a'$ and $I(t) \cap I(a') \ne \emptyset$.

The set of all inconsistent tuples in $I$ is denoted by $Inc(I)$. A tuple that is not inconsistent in $I$ is said to be {\em consistent in $I$}, and the set of all consistent tuples in $I$ is denoted by $Cons(I)$.
An interpretation $I$ such that $Inc(I)=\emptyset$ is said to be in {\em First Normal Form}.
\hfill$\Box$
\end{definition} 
For instance, in Example \ref{ex:nico}, the tuple $t=abc$ is inconsistent in $I$ as $A$ is in $sch(t)$, and $a'$ is in $adom(A)$ such that $t.A \ne a'$ and $I(t) \cap I(a')= \{2\} \ne \emptyset$. Therefore $I$ is {\em not} in First Normal Form.

In view of the above definition, the set $\mathcal{T}$ is partitioned into inconsistent and consistent tuples that is $Inc(I) \cup Cons(I)= \mathcal{T}$ and $Inc(I) \cap Cons(I)= \emptyset$. Roughly speaking, inconsistent tuples are those that violate the First Normal Form. 
This relationship between inconsistency and First Normal Form is stated formally in the following lemma.
\begin{lemma}\label{lemma:PC}
Let $I$ be an interpretation of $T$. Then $I$ is in First Normal Form if and only if $I$ satisfies the following constraint:

\begin{itemize}
\item[]
{\bf {\em Partition constraint} (PC):} For all $A$ in $U$ and for all $a, a'$ in $adom(A)$, 
if $a \neq a'$ then $I(a) \cap I(a') = \emptyset$.
\end{itemize}
\end{lemma}
{\sc Proof.}
Assume that $I$ satisfies $PC$ and that $Inc(I) \ne \emptyset$. Given $t$ in $Inc(I)$, using the notation of Definition~\ref{def:incons-t} and denoting by $a$ the $A$-value occurring in $t$, we have $I(a) \cap I(a')\ne \emptyset$, because $I(t) \subseteq I(a)\cap I(a')$. Hence, $I$ does not satisfy $PC$, which is a contradiction.

Conversely, if $I$ does not satisfy $PC$, then there exist $A$ in $U$ and $a$ and $a'$ such that $I(a) \cap I(a') \ne \emptyset$. Therefore, by Definition~\ref{def:incons-t}, $a$ and $a'$ are in $Inc(I)$, thus implying that $I$ is not in First Normal Form. The proof is therefore complete.
\hfill$\Box$

\smallskip\noindent
Note that, as mentioned in \cite{JIIS}, satisfaction of the partition constraint implies that, for all $A$ in $U$, the set $\{I(a)~|~a \in adom(A) \}$ is a partition of the set $adom(A)$ (whence the term ``partition constraint'').
%

%
An important question remains regarding the definition of an interpretation: is there a systematic way for defining an interpretation $I$ of a given table $T$ which is in First Normal Form? The answer is yes, there is such a ``canonical'' interpretation for every table $T$; it is called the {\em basic interpretation} of $T$, denoted by $I^b$ and defined as follows:

\begin{itemize}
\item[]
{\bf {\em Basic Interpretation $I^b$}:} For every $A$ in $U$ and every $a$ in $adom(A)$,
$I^b(a) = \{id(t)~ |~ t \in T ~and~ t.A=a \}$.
\end{itemize}
In the context of Example \ref{ex:nico} where $T=\{ab,a'c\}$, if $ab$ and $a'c$ are respectively associated with identifiers $1$ and $2$, the associated basic interpretation $I^b$ is defined by $I^b(a)=I^b(b)=\{1\}$ and $I^b(a')=I^b(c)=\{2\}$. Considering $T_1=\{ab,bc,ac\}$ with respective identifiers $1$, $2$ and $3$, the basic interpretation $I^b_1$ is defined by  $I^b_1(a)=\{1,3\}$, $I^b_1(b)=\{1,2\}$ and $I^b_1(c)=\{2,3\}$. 

\smallskip
It is interesting to note that the definition of basic interpretation parallels that of {\em inverted file} \cite{ZobelMR98,ZobelM06}. Indeed, we first recall that an inverted file is an index data structure that maps content to its location within a database file, in a document or in a set of documents. Hence, if we assume that each tuple of $T$ is implemented as a record (possibly with missing values) then we can view $T$ as a file that is as a function $I$ assigning to each value $x$ appearing in the file record the set $I(x)$ of all identifiers of records in which $x$ appears. 

\smallskip
The following lemma summarizes important properties of the basic interpretation.
\begin{lemma}\label{lemma:basic-int}
Let $T$ be a table and $I^b$ its  basic interpretation as defined above. Then:
\begin{enumerate}
\item  $I^b$ is an interpretation satisfying the partition constraint.
\item $True(I^b)={\sf LoCl}(T)$ and $True(I^b)$ is the set of all tuples $t$ that belong to $True(I)$ for every interpretation $I$ of $T$.
\item $I^b$ is minimal with respect to $\preceq$ among all interpretations $I$ of $T$ such that for every $t$ in $T$, $id(t)$ belongs to $I(t)$.
\end{enumerate}
\end{lemma}
{\sc Proof.}
{\bf 1.} Assume that there exist $A$ in $U$ and $a$ and $a'$ in $adom(A)$ such that $I^b(a) \cap I^b(a') \ne \emptyset$. If $i$ is in $I^b(a) \cap I^b(a') \ne \emptyset$, let $t_i$ be the tuple in $T$ such that $id(t_i)=i$. This implies that the value of $t_i(A)$ is $a$ {\em and} $a'$, which is impossible because, as $t_i$ is a function, $t_i(A)$ consists of at most one value in $adom(A)$. This part of the proof is thus complete.

\smallskip\noindent
{\bf 2.} Since for every $t$ in $T$, $I^b(t)$ contains $id(t)$ and as for every $t'$ such that $t' \sqsubseteq t$, we have $I^b(t) \subseteq I^b(t')$, it follows that ${\sf LoCl}(T) \subseteq True(I^b)$. Conversely, let $t$ not in ${\sf LoCl}(T)$ and assume that $I^b(t) \ne \emptyset$. Since $t$ is not in $T$, it is not associated with an identifier. Hence, $I^b(t)$ contains identifiers of other tuples, and if $q$ is such a tuple, then by definition of $I^b$, this implies that $id(q)$ belongs to every component of $t$, that is that $t$ is a sub-tuple of $q$. A contradiction showing that $True(I^b) \subseteq {\sf LoCl}(T)$. Hence we have $True(I^b)={\sf LoCl}(T)$. Moreover, since for every interpretation $I$ of $T$, $True(I)$ must at least contain all tuples in $T$ along with their sub-tuples (as argued just above), we have ${\sf LoCl}(T) \subseteq True(I)$. Hence, $True(I^b) \subseteq True(I)$, which completes this part of the proof.

\smallskip\noindent
{\bf 3.} Let $I$ be an interpretation such that $I \preceq I^b$ and $I \ne I^b$. Thus, there exists $a$ in $\mathcal{AD}$ such that $I(a) \subset I^b(a)$. However, since $I$ is an interpretation such that for every $t$ in $T$, $id(t)$ belongs to $I(t)$, if $a$ occurs in $t$ then  $id(t)$ belongs to $I(a)$. In other words, $I^b(a) \subseteq I(a)$ holds. A contradiction showing that $I^b$ is minimal with respect to $\preceq$ among all interpretations $I$ of $T$ such that for every $t$ in $T$, $id(t)$ belongs to $I(t)$. The proof is therefore complete. 
\hfill$\Box$

\medskip\noindent
By Definition~\ref{def:incons-t} and Lemma~\ref{lemma:PC}, it turns out that, since $I^b$ satisfies the partition constraint $PC$, $I^b$ is in First Normal Form, therefore $Inc(I^b)=\emptyset$.

From now on, in all our examples we shall use the following convention: the tuples of $T$ will receive successive integer identifiers in the order of their appearance in $T$, starting with number 1. For example, if $T= \{ab, a'c \}$ as in Example~\ref{ex:nico}, then it is understood that $id(ab)=1$ and $id(a'c)=2$. Then, the basic interpretation of $T$ is defined as earlier stated, that is:  $I^b(ab)= \{1 \}$, $I^b(a'c)= \{2 \}$, $I^b(a)= \{1 \}$, $I^b(a')= \{2 \}$, $I^b(b)= \{1 \}$, $I^b(c)= \{2 \}$. Thus, $I^b(ac)=\emptyset$, $I^b(abc)=\emptyset$, $I^b(a'bc)= \emptyset$. Therefore we have: $True(I^b)= \{a, a', b, c, ab, a'c \}$ and $False(I^b)= \{ac, abc, a'bc \}$. Note that $I^b$ is an interpretation that satisfies the partition constraint.
\subsection{Partition Semantics for Functional Dependencies}
Let $T$ be a table over a set $U$ of attributes together with a set $FD$ of functional dependencies. As usual in relational databases \cite{Ullman}, a {\em functional dependency} is an expression of the form $X \to Y$ where $X$ and $Y$ are relation schemes. The notion of functional dependency satisfaction in the context of tables with nulls is stated in the literature \cite{FaginKMP05,FaginMU82} in the following terms. A table $T$ satisfies $X \to Y$ if for all tuples $t$ and $t'$ in $T$ the following holds: if $XY$ is a sub-set of $sch(t)$ and of $sch(t')$ then $t.X= t'.X$ implies $t.Y= t'.Y$.

Moreover, it is well-known that a relation satisfies $X \to Y$ if and only if it satisfies the functional dependencies $X \to A$, for every $A$ in $Y$. This justifies that functional dependencies are usually assumed to be of the form $X \to A$, where $X$ is a relation schema and $A$ is an attribute not in $X$. In what follows, we make this assumption.  

Now, the question that we answer in this section is the following: how can an interpretation $I$ of $T$ express the satisfaction of a functional dependency by $T$? The answer is provided by the following lemma. 
\begin{lemma}\label{lemma-first}
Let $I$ be an interpretation of table $T$ in First Normal Form, and let $X \to A$ be a functional dependency of $FD$. Then $T$ satisfies $X \to A$ if $I$ satisfies the following constraint:
\begin{itemize}
\item[]{\bf {\em Inclusion constraint} (IC):} For all $t$ in $\mathcal{T}$ such that  $XA \subseteq sch(t)$,
$I(t.X)\cap I(t.A)\ne \emptyset$ implies $I(t.X)\subseteq I(t.A)$.
\end{itemize}
\end{lemma} 
{\sc Proof.} 
Suppose that $I$ satisfies the inclusion constraint and consider the relation $f: adom(X) \to adom(A)$ defined by: for every true tuple $t$ in $T$, $f(t.X)= t.A$. We show that $f$ is actually a function. Indeed, suppose there is a true tuple $t'$ in $T$ such that $t'.X= t.X$ and $t'.A \neq t.A$. As $I$ is in First Normal Form, by Lemma~\ref{lemma:PC}, $I$ satisfies the partition constraint, and thus, $I(t.A) \cap I(t'.A) =\emptyset$. On the other hand, as $I$ satisfies the inclusion constraint $IC$ for $X \to A$, we have $I(t.X) \subseteq I(t.A)$ and $I(t'.X) \subseteq I(t'.A)$. As $t.X= t'.X$, we have $I(t.X) \subseteq I(t.A) \cap I(t'.A)$, thus that $I(t.X) \subseteq \emptyset$. It follows that $I(t.X)= \emptyset$, which is a contradiction to the fact that $t.X$ is true, being a sub-tuple of $t$. The proof is therefore complete.
\hfill$\Box$

\begin{example}\label{ex:first}
{\rm
Let $T= \{ab, bc \}$ and $FD= \{A \to B \}$. Then the basic interpretation $I^b$ of $T$ is as follows: $I^b(a)= \{1\}$, $I^{b}(b)= \{1, 2 \}$, $I^b(c)= \{2\}$ and we have that $I^b(a) \cap I^b(b)= \{1\} \neq \emptyset$ and $I^b(a) \subseteq I^b(b)$. As the only tuple of $T$ to which the lemma applies is $ab$ we conclude that $T$ satisfies $A \to B$. In this case $I^b$ satisfies the partition constraint $PC$ {\em and} the inclusion constraint $IC$.

As another example, let $T= \{ab, ac \}$ and $FD= \{A \to B \}$. Then the interpretation $I^b$ of $T$ is as follows: $I^b(a)= \{1, 2\}$, $I^b(b)= \{1 \}$, $I^b(c)= \{2\}$ and we have $I^b(a) \cap I^b(b)= \{1\} \neq \emptyset$ but $I^b(a) \not\subseteq I^b(b)$, showing that $I^b$ does not satisfy the inclusion constraint. Therefore, $I^b$ satisfies the partition constraint $PC$ but {\em not} the inclusion constraint $IC$.
\hfill$\Box$}
\end{example}
In view of Lemma~\ref{lemma-first}, $I^{b}$ satisfies the partition constraint, the set of true tuples of $I^{b}$ is equal to the set of true tuples of every interpretation $I$, and $I^{b}$ satisfies a minimality property with respect to $\preceq$ among interpretations of $T$. Therefore the question that arises here is whether we can ``expand'' $I^b$ to an interpretation $I'$ that satisfies the inclusion constraint as well, so that $I'$ satisfies the same properties as $I^b$, along with the inclusion constraint $IC$.

In Example~\ref{ex:first}, the answer is yes. Indeed, if we add $I^b(a)$ to $I^b(b)$ then $I^b(b)$ becomes $I^b(a) \cup I^b(b)$ and the resulting interpretation is: $I'(a)= \{1, 2\}$, $I'(b)= \{1, 2 \}$, $I'(c)= \{2\}$; and $I'$  satisfies the inclusion constraint for $A \to B$. However, the following example shows that expanding $I^b$ may lead to non satisfaction of the partition constraint $PC$.
\begin{example}\label{ex:inc-constraint}
{\rm
Let $T = \{ab, bc, ac'\}$ with $FD=\{A \to C, B \to C\}$. Here, $I^b$ is defined by $I^b(a)=\{1,3\}$, $I^b(b)=\{1,2\}$, $I^b(c)=\{2\}$ and $I^b(c')=\{3\}$. Thus, $I^b$ does not satisfy the constraint $IC$ because $I^b(a) \cap I^b(c) =\{2\} \ne \emptyset$ but $I^b(a) \not\subseteq I^b(c)$.

Notice that this example shows that the converse of Lemma~\ref{lemma-first} does not hold. It is so because $T$ satisfies $A \to C$ and $B \to C$, although $I^b$ is an interpretation of $T$ in First Normal Form that does not satisfy the inclusion constraint $IC$.

Expanding $I^b$ as in Example~\ref{ex:first} yields the interpretation $I'$ such that $I'(c)= I'(c')=\{1,2,3\}$, $I'(a)=I^b(a)$ and $I'(b)=I^b(b)$. It should be clear that $I'$ satisfies the constraint $IC$, whereas $I'$ does {\em not} satisfy the constraint $PC$ since $I'(c) \cap I'(c') \ne \emptyset$. As a consequence, by Lemma~\ref{lemma:PC}, $Inc(I') \ne \emptyset$. More precisely, as $True(I')$ is the set containing $abc$ and $abc'$ along with their sub-tuples, we have $Inc(I')=\{abc, abc', bc,bc', ac,ac', c ,c'\}$.
\hfill$\Box$}
\end{example}
In view of our discussion above, the following definition states how we can expand an interpretation $I$ so that it satisfies a given functional dependency $X \to A$.
\begin{definition}\label{def:expand}
The {\em expansion of} $I$ with respect to a functional dependency $X \to A$ and a tuple $xa$ in $\mathcal{T}(XA)$ is the interpretation $Exp(I,xa)$ defined as follows:
\begin{itemize}
\item
If $I(x) \cap I(a) \ne \emptyset$ and $I(x) \not\subseteq I(a)$ then:
$Exp(I,xa)(a)= I(a) \cup I(x)$, and $Exp(I,xa)(\alpha)= I(\alpha)$ for $\alpha$ in $\mathcal{AD}$ different than $a$.
\item
Otherwise, $Exp(I,xa)=I$.
\hfill$\Box$
\end{itemize}
\end{definition}
Of course, the expansion processing should be iterated when several tuples satisfy the above condition and when several functional dependencies are considered. As will be seen next, starting with the basic interpretation $I^b$ and applying expansion repeatedly until a fixed point is reached, we obtain an interpretation $I^*$ such that: (a) $I^*$ satisfies the inclusion constraint $IC$ (but not necessarily the partition constraint $PC$) (b) a tuple $t$ of $\mathcal{T}$ is true in $I^*$ if and only if $t$ is true in every interpretation $I$ of $T$ satisfying $IC$ and (c) a tuple $t$ of $\mathcal{T}$ is inconsistent in $I^*$ if and only if $t$ is inconsistent in every interpretation $I$ of $T$ satisfying $IC$.
\subsection{The True and the Inconsistent Tuples}
To define the limit interpretation $I^*$ of $T$ mentioned above, for a given a table $T$ over universe $U$ with a set $FD$ of functional dependencies, we consider the sequence of interpretations $(I^j)_{j \geq 0}$ of $T$  defined by:
\begin{itemize}
\item For $j=0$, we set $I^0$ to be the basic interpretation $I^b$ as defined earlier.
\item For $j \geq 0$, let $I^{j+1}$ be defined as follows:
\begin{itemize}
\item If there exist $X \to A$, $x$ and $a$ such that $Exp(I^j, xa) \ne I^j$,
then $I^{j+1}= Exp(I^j, xa)$
\item Else $I^{j+1}=I^j$.
\end{itemize}
\end{itemize}
The theorem below states that this sequence has a limit with important properties. In this theorem, given a nonempty sub-set $S$ of  $\mathcal{AD}$  and an interpretation $I$, we denote by $I(S)$ the set $\bigcap_{a \in S}I(a)$.
\begin{theorem}\label{lemma:JIIS}
The sequence $(I^j)_{j \geq 0}$ is increasing with respect to $\preceq$ and bounded above, therefore it has a limit that we denote by $I^*$. This limit has the following properties: 
\begin{enumerate}
    \item $I^*$ is unique in the sense that it is independent of the order in which expansions are applied (i.e., the construction of $I^*$ has the Church-Rosser property).
     \item $I^*$ satisfies the inclusion constraint $IC$. 
     \item For every nonempty sub-set $S$ of $\mathcal{AD}$, $I^*(S)\ne \emptyset$ holds if and only if $I(S)\ne \emptyset$ holds for every interpretation $I$ of $T$ satisfying $IC$.
\end{enumerate}
\end{theorem}
{\sc Proof.}
First, it is clear that for $j \geq 0$, $I^j \preceq I^{j+1}$ holds by definition of the sequence. Now, as for every $a$ in $\mathcal{AD}$ we have that: $I^j(a) \subseteq TID$ for every $j \geq 0$, the sequence is bounded by the interpretation $I^{TID}$ defined by:  $I^{TID}(a)=TID$ for every $a$ in $\mathcal{AD}$. Hence, the sequence has a limit, denoted by $I^*$. The proof of the items in the theorem are as follows:  

\smallskip\noindent
{\bf 1.} Uniqueness of the limit $I^*$ is shown in Appendix~\ref{append:church-rosser}. 

\smallskip\noindent
{\bf 2.} Suppose that $I^*$ does not satisfy the inclusion constraint $IC$. It follows that there exist $X \to A$ in $FD$ and $t$ in $\mathcal{T}$ such that $XA \subseteq sch(t)$, $I^*(t.X) \cap I^*(t.A) \ne \emptyset$ and $I^*(t.X) \not\subseteq I^*(a)$. Thus, assuming $j$ is such that $I^* = I^j$, we have $I^{j+1} \ne I^j$, which implies that $I^*$ is {\em not} the limit of the sequence, a contradiction.

\smallskip\noindent
{\bf 3.} Let $S$ be a nonempty sub-set of $\mathcal{AD}$ such that $I(S) \ne \emptyset$ for every $I$ satisfying $IC$. Since $I^*$ is an interpretation that satisfies $IC$, we also have $I^*(S) \ne \emptyset$. 

Conversely, if $I^*(S)\ne \emptyset$, we prove by induction that $I(S) \ne \emptyset$ holds for every interpretation $I$ satisfying $IC$.
\\
$\bullet$ If $I^0(S) \ne \emptyset$, then it follows from Lemma~\ref{lemma:basic-int} that $S$ does not contain two elements from the same active domain (because $I^b$ satisfies the partition constraint $PC$). Thus, the elements of $S$ form a tuple $t$ that belongs to $True(I^0)$. It follows that $t$ is a sub-tuple of a tuple $q$ in $T$. Hence, $t$ belongs to $\true(I)$ for every interpretation $I$ of $T$, thus for every interpretation of $T$ satisfying the inclusion constraint $IC$.
\\
$\bullet$ Assume that for every nonempty sub-set $\Sigma$ of $\mathcal{AD}$,  if $I^j(\Sigma) \ne \emptyset$ then $I(\Sigma) \ne \emptyset$ for every $I$ satisfying $IC$, and let $S$ be such that $I^{j+1}(S) \ne \emptyset$. If $I^{j}(S) \ne \emptyset$, then $I^{j+1}(S) \ne \emptyset$ holds because $I^j \preceq I^{J+1}$. Considering the case where $I^{j}(S) = \emptyset$, as $I^{j+1}(S) \ne I^j(S)$, $S$ contains a constant $a$ such that $I^j(a) \ne I^{j+1}(a)$. Denoting $S \setminus\{a\}$ by $S_a$, we have $I^j(S_a)= I^{j+1}(S_a)$ (because expansion changes the interpretation of only one constant, namely $a$ in this case), and there exist $X \to A$ in $FD$ and $x$ in $\mathcal{T}(X)$ such that $I^j(x) \cap I^j(a) \ne \emptyset$ and $I^j(x) \not\subseteq I^j(a)$. In this case, we have $I^{j+1}(a)= I^j(a) \cup I^j(x)$, and thus

\begin{tabular}{rcl}
 $I^{j+1}(S)$&$=$&$I^{j+1}(a) \cap I^{j+1}(S_a)$\\
&$=$&$(I^j(a) \cup I^j(x))\cap I^j(S_a)$\\
&$=$&$(I^j(a) \cap I^j(S_a))\cup (I^j(x) \cap I^j(S_a))$
\end{tabular}

\noindent
Since $I^j(S)=I^j(a) \cap I^j(S_a)$ is empty, it follows that $I^j(x) \cap I^j(S_a) \ne \emptyset$ and so, for every $I$ satisfying $IC$, $I(x) \cap I(a) \ne \emptyset$ and $I(x) \cap I(S_a) \ne \emptyset$. Hence, $I(x) \subseteq I(a)$, and so  $I(x) \cap I(S_a) \subseteq I(a) \cap I(S_a)$. As this implies that $I(a) \cap I(S_a)  \ne \emptyset$, we obtain that $I(S)\ne \emptyset$, and the proof is complete.
\hfill$\Box$

\medskip\noindent
We note that similar results have been obtained with a slightly different formalism in \cite{JIIS}. Moreover, as a consequence of Theorem~\ref{lemma:JIIS}, for every tuple $t$ in $\mathcal{T}$, we have:
\begin{itemize}
\item $t$ belongs to $True(I^*)$ if and only if $t$ belong to $True(I)$ for every interpretation $I$ of $T$ satisfying $IC$. We denote the set $True(I^*)$ by $\true(\mathcal{T})$ and its tuples are said to be {\em true in $\mathcal{T}$}. The set $False(I^*)$ is denoted by $\false(\mathcal{T})$ and its tuples are said to be {\em false in $\mathcal{T}$}.
\item $t$ belongs to $Inc(I^*)$ if and only if $t$ belongs to $Inc(I)$ for every interpretation $I$ of $T$ satisfying $IC$. We denote the set $Inc(I^*)$ by $\inc(\mathcal{T})$ and its tuples are said to be {\em inconsistent in $\mathcal{T}$}.  The set $Cons(I^*)$ is denoted by $\cons(\mathcal{T})$ and its tuples are said to be {\em consistent in $\mathcal{T}$}.
\end{itemize}
The following proposition lists basic properties of true, false, consistent and inconsistent tuples in $\mathcal{T}$.
\begin{proposition}\label{prop:basic}
For every table $T$  over universe $U$, the following holds:
\begin{enumerate}
\item If $t$  is in $\true(\mathcal{T})$ then every $t'$ such that $t' \sqsubseteq t$ belongs to $\true(\mathcal{T})$.
\item $\inc(\mathcal{T}) \subseteq \true(\mathcal{T})$. However, the inclusions $\cons(\mathcal{T})\subseteq \true(\mathcal{T})$ and $\cons(\mathcal{T})\subseteq \false(\mathcal{T})$ do {\em not} hold.
\item Let $t$ be in $\inc(\mathcal{T})$ and $A$ in $sch(t)$. Let $a'$ be in $adom(A)$ such that  $a' \ne t.A$ and $I^*(t) \cap I^*(a') \ne \emptyset$. Then every tuple $t'$ such that $t' \sqsubseteq t$ and $A$ belongs to $sch(t')$ is in $\inc(\mathcal{T})$.
\item It does {\em not} hold that for $t$ in $\inc(\mathcal{T})$, every true super-tuple of $t$  is in $\inc(\mathcal{T})$.
\end{enumerate}
\end{proposition}
{\sc Proof.}
{\bf 1.} Since $t$ is in $True(I^*)$, $I^*(t) \ne \emptyset$. Thus, as $I^*(t) \subseteq I^*(t')$ (because  $t' \sqsubseteq t$), we have  $I^*(t) \ne \emptyset$, showing that $t'$ is in $\true(\mathcal{T})$.

\smallskip\noindent
{\bf 2.} By Definition~\ref{def:incons-t}, if $t$ is in $\inc(\mathcal{T})$ then there is $A$ in $sch(t)$ and $a'\ne t.A$ in $adom(A)$ such that $I^*(t) \cap I^*(a') \ne \emptyset$. Thus $I^*(t) \ne \emptyset$, which implies that $t$ belong to $\true(\mathcal{T})$.

Regarding the two other inclusions in the item, for $T=\{ab,a'b'\}$ and $FD=\emptyset$, we have $\true(\mathcal{T})={\sf LoCl}(T)$ and $\inc(\mathcal{T})=\emptyset$. Thus, $\false(\mathcal{T})=\{ab', a'b\}$ and $\cons(\mathcal{T})=\mathcal{T}$, showing that $\cons(\mathcal{T})\not\subseteq \true(\mathcal{T})$ and $\cons(\mathcal{T})\not\subseteq \false(\mathcal{T})$.

\smallskip\noindent
{\bf 3.} In this case, we have $I^*(t) \subseteq I^*(t')$ (because $t \sqsubseteq t'$) and as $I^*(t) \cap I^*(a')\ne \emptyset$, we obtain that $I^*(t') \cap I^*(a')\ne \emptyset$. Since $t'.A=t.A$ (because $A \in sch(t')$ and $t \sqsubseteq t'$), we have $a' \ne t'.A$, which implies that  $t'$ is in $\inc(\mathcal{T})$.

\smallskip\noindent
{\bf 4.} Let $T= \{ab, ab', bc\}$ with $A \to B$. In this case $I^*$ is defined by $I^*(a) = I^*(b')=\{1,2\}$, $I^*(b)=\{1,2,3\}$ and $I^*(c)=\{3\}$. Thus $b$ is inconsistent, because $I^*(b) \cap I^*(b') \ne \emptyset$, and we argue that $bc$ is {\em not} inconsistent although $bc$ is a true super-tuple of $b$. This is so because  $I^*(bc)=\{3\}$ but $I^*(b') \cap I^*(bc) = \emptyset$.
\hfill$\Box$

\medskip\noindent
We illustrate now the construction of $I^*$ through the following example.
\begin{example}\label{ex:inc-constraint-i*}
\rm{
Referring to Example~\ref{ex:inc-constraint}, we recall that $T = \{ab, bc, ac'\}$ with $FD=\{A \to C, B \to C\}$, and that $I^b$ was found to be defined by $I^b(a)=\{1,3\}$, $I^b(b)=\{1,2\}$, $I^b(c)=\{2\}$ and $I^b(c')=\{3\}$. The construction of $I^*$ starts with $I^0=I^b$, and the following steps are performed:

\smallskip\noindent
(1) Considering $A \to C$, we have $I^0(a) \cap I^0(c) \ne \emptyset$ but $I^0(a) \not\subseteq I^0(c)$. Hence $I^1(c)=I^0(c) \cup I^0(a)=\{1,2,3\}$, $I^1(a)=\{1,3\},$ $I^1(b)=\{1,2\}$ and $I^1(c') = \{3\}$.
\\
(2) Considering again $A \to C$, we have $I^1(a) \cap I^1(c') \ne \emptyset$ but $I^1(a) \not\subseteq I^1(c')$. Hence $I^2(c')$ is set to $I^1(c') \cup I^1(a)=\{1,3\}$, and  $I^2(a)=\{1,3\},$ $I^2(b)=\{1,2\}$ and $I^2(c) = \{1,2,3\}$.
\\
(3) For $B \to C$, we have $I^2(b) \cap I^2(c') \ne \emptyset$ but $I^2(b) \not\subseteq I^2(c')$. Hence $I^3(c')$ is set to $I^2(c') \cup I^2(b)=\{1,2,3\}$, and  $I^3(a)=\{1,3\},$ $I^3(b)=\{1,2\}$ and $I^3(c) = \{1,2,3\}$.

\smallskip
Since $I^3$ satisfies the inclusion constraint, we obtain $I^*=I^3$. We notice that $I^*$ is equal to the expected interpretation $I'$ defined in Example~\ref{ex:inc-constraint}. Moreover, it can be seen that in this example, $\true(\mathcal{T})$ contains the tuples $abc$ and $abc'$ along with all their sub-tuples, meaning that $\true(\mathcal{T})= \mathcal{T}$, and thus that $\false(\mathcal{T})=\emptyset$.

Regarding inconsistent tuples, as we have $I^*(c) \cap I^*(c') \ne \emptyset$, it turns out that $\inc(\mathcal{T})=\{abc,abc',ac,ac',bc,bc', c,c'\}$, and thus that $\cons(\mathcal{T})= \{ab,a,b\}$.
\hfill$\Box$}
\end{example}
To further illustrate the impact of functional dependencies over true tuples and over inconsistent tuples, we mention the following two properties: given a table $T$ over $U$, $X \to A$ in $FD$ and a true tuple $xa$  in $\mathcal{T}(XA)$, then:
\begin{enumerate}
\item
For every tuple $t$ in $\true(\mathcal{T})$ such that $X \subseteq sch(t)$ and $A \not\in sch(t)$, if $t.X=x$ then the tuple $ta$ is in $\true(\mathcal{T})$. In other words, writing $t$ as $qx$ this result expresses the well-known property of relational lossless join: if $qx$ and $xa$ are true, the functional dependency $X \to A$ implies that $qxa$ is true as well.
\\
This is so because $I^*(x) \subseteq I^*(a)$, $I^*(t) \subseteq I^*(x)$ (because $x \sqsubseteq t$) and $I^*(ta)=I^*(t) \cap I^*(a)$. Hence, $I^*(t) \subseteq I^*(a)$, implying that $I^*(ta)=I^*(t)$. Therefore, $I^*(ta) \ne \emptyset$, because $I^*(t) \ne \emptyset$.
\item
For every true super-tuple $t$ of $xa$ (i.e., $t \in \true(\mathcal{T})$, $XA \subseteq sch(t)$ and $t.XA=xa $) and for every $a'$ in $adom(A)$ different than $a$ such that $xa'$ is true (i.e., $xa' \in \true(\mathcal{T})$), if we write $t$ as $qxa$ then $t'=qxa'$ also belongs to $\true(\mathcal{T})$, and $t$ and $t'$ both belong to $\inc(\mathcal{T})$.
\\
Indeed, since $I^*(xa)$ and $I^*(xa')$ are nonempty, and $I^*$ satisfies the inclusion constraint, we have $I^*(x) \subseteq I^*(a) \cap I^*(a')$ and $I^*(x) \ne \emptyset$. It follows that $I^*(a) \cap I^*(a') \ne \emptyset$. Moreover, since $t=qxa$ and $I^*(t) \ne \emptyset$, we have $I^*(qx) \ne \emptyset$ and as $I^*(qx) \subseteq I^*(x)$, it follows that $I^*(qx) \subseteq I^*(a) \cap I^*(a')$, that is, $I^*(qxa')=I^*(qx)$. Consequently, $I^*(qxa') \ne \emptyset$, entailing that $t'=qxa'$ is in $\true(\mathcal{T})$. Since $I^*(t) \ne \emptyset$, $I^*(a') \cap I^*(xa) \ne \emptyset$ and $I^*(t) \subseteq I^*(xa)$, $t$ is in $\inc(\mathcal{T})$ and a similar reasoning shows that $t'$ is also in $\inc(\mathcal{T})$.
\end{enumerate}
\section{Consistent Query Answering}\label{sec:cons-query}
In this section, we first define the syntax of queries that we consider and  then we define and discuss four kinds of consistent answer to a query, based on the semantics presented in the previous section.
\subsection{Syntax of Queries}\label{subsec:query-syntax}
We use SQL as the query language and, as we query a single table $T$, the queries $Q$ that we consider have one of the following two forms:

\smallskip\centerline{
$Q:$ {\tt select} $X$ {\tt from} $T$\qquad or \qquad $Q:$ {\tt select} $X$  {\tt from} $T$ {\tt where $\Gamma$}}

\smallskip\noindent
In either of these forms, $X$ is an attribute list seen as a relation schema, and in the second form the {\tt where} clause specifies a selection condition $\Gamma$. As in SQL the where clause in a query is optional, the generic form of a query $Q$ is denoted by 

\smallskip\centerline {$Q:$ {\tt select}~$X$~{\tt from}~$T$~{\tt [where~}$\Gamma${\tt ]}} 

\smallskip\noindent
The set of all attributes occurring in $\Gamma$ is called the {\em schema of $\Gamma$}, denoted by $sch(\Gamma)$; and the attribute set $X \cup sch(\Gamma)$ is called the {\em schema of $Q$}, denoted by $sch(Q)$. 

A selection condition $\Gamma$ is a well-formed formula involving the usual connectors $\neg$, $\vee$ and $\wedge$ and built up from atomic Boolean comparisons of one of the following forms: $A \,\theta\, a$ or $A \,\theta\, A'$, where $\theta$ is a comparison predicate, $A$ and $A'$ are attributes in $U$ whose domain elements are comparable through $\theta$, and $a$ is in $dom(A)$. 

Given a selection condition $\Gamma$,  we denote by $Sat(\Gamma)$ the set of all tuples in $\mathcal{T}(sch(\Gamma))$ satisfying $\Gamma$, as defined below:

\begin{itemize}
\item if $\Gamma$ is of the the form $A \,\theta\, a$, $Sat(\Gamma)=\{t \in \mathcal{T}(sch(\Gamma)) ~|~t.A \,\theta\, a\}$,
\item if $\Gamma$ is of the the form $A \,\theta\, B$, $Sat(\Gamma)=\{t\in \mathcal{T}(sch(\Gamma)) ~|~t.A \,\theta\, t.B\}$,
\item if $\Gamma$ is of the form $\Gamma_1 \vee \Gamma_2$, $Sat(\Gamma)= Sat(\Gamma_1) \cup Sat(\Gamma_2)$,
\item if $\Gamma$ is of the form $\Gamma_1 \wedge \Gamma_2$, $Sat(\Gamma)= Sat(\Gamma_1) \cap Sat(\Gamma_2)$,
\item if $\Gamma$ is of the form $\neg \Gamma_1$, $Sat(\Gamma)= \mathcal{T}(sch(\Gamma)) \setminus Sat(\Gamma_1)$.
\end{itemize}
Moreover, the set of tuples that do {\em not} satisfy $\Gamma$ is defined by: 
\begin{itemize}
\item $Sat^-(\Gamma)= \mathcal{T}(sch(\Gamma)) \setminus Sat(\Gamma)$.
\end{itemize}
For example, if $Sal$ is an attribute whose active domain is $\{s,s'\}$, we have:
\begin{itemize}
\item
for $\Gamma_4=(Sal=s')$, $Sat(\Gamma_4)=\{s'\}$ and $Sat^-(\Gamma_4)=\{s\}$, 
\item
for $\Gamma_5=(Sal=s \vee Sal = s')$, $Sat(\Gamma_5)=\{s, s'\}$ and $Sat^-(\Gamma_5)=\emptyset$,
\item
for $\Gamma_6=(Sal > 10K)$, if $s=5K$ and $s'=20K$, $Sat(\Gamma_6)=\{s'\}$ and $Sat^-(\Gamma_6)=\{s\}$, and if $s=15K$ and $s'=20K$, $Sat(\Gamma_6)=\{s,s'\}$ and $Sat^-(\Gamma_6)=\emptyset$.
\end{itemize}
Next, we explain how, in our context, the above syntactic definition of satisfaction of a selection condition can be `coupled' with semantic considerations, when it comes to defining the notion of consistent answer to a query $Q$.

Intuitively, given $Q:$ {\tt select}~$X$~{\tt from}~$T$~{\tt [where~}$\Gamma${\tt ]}, the least requirements for a tuple $x$ to belong to a consistent answer to $Q$ are the following:
\begin{itemize}
\item[{\sf R1}] $x$ is in $\mathcal{T}(X)$, i.e., the schema of $x$ is $X$,
\item[{\sf R2}] $x$ is in $\true(\mathcal{T})$, i.e., $x$ is a true tuple,
\item[{\sf R3}] $x$ is in $\cons(\mathcal{T})$, i.e., $x$ is consistent, and
\item[{\sf R4}] if $Q$ involves a selection condition $\Gamma$ then there exists $t$ in $\true(\mathcal{T})$ such that $sch(Q) \subseteq sch(t)$, $x \sqsubseteq t$, and $t.sch(\Gamma)$ is in $Sat(\Gamma)$ (that is, $t$ is a true super-tuple of $x$ whose restriction to attributes in $sch(\Gamma)$  satisfies $\Gamma$).
\end{itemize}
It is important to note that when all the above requirements are satisfied, then the consistent answer coincides with the standard notion of answer to projection-selection queries against a consistent table. Indeed, if $T$ is a relational consistent table whose schema contains all attributes in $sch(Q)$, then the answer to a query $Q:$  {\tt select}~$X$~{\tt from}~$T$~{\tt [where~}$\Gamma${\tt ]} is the set of all tuples $x$ such that:
\begin{itemize}
\item the schema of $x$ is $X$ (see requirement {\sf R1}), and
\item 
$T$ contains a tuple $t$ such that $t$ is a super-tuple of $x$ and the restriction of $t$ to all attributes occurring in $\Gamma$  satisfies $\Gamma$ (see requirement {\sf R4}, knowing that $t$ is  true since all tuples in $T$ are implicitly assumed to be true).
\end{itemize}
Moreover, in this case, requirement {\sf R2} is also satisfied because all tuples in a consistent relational table are implicitly assumed to be {\em true}, and requirement {\sf R3} is trivial because in a consistent table, every tuple is consistent.

On the other hand, in the presence of inconsistencies, it should be noticed that, based on our semantics as earlier defined, the status of any tuple $x$ in the consistent answer to $Q$ is clearly defined because $x$ is then a {\em true and consistent tuple of $\mathcal{T}$ whose schema is $X$}. However, the situation is less clear regarding the status of the super-tuple $t$ in requirement {\sf R4}. More precisely, the question here is whether $t$ should be consistent or not and whether $t$ should satisfy other criteria as well (e.g., is $t$ required to be consistent or is $t$ required to be maximal with respect to $\sqsubseteq$?).  The following example illustrates this discussion, showing in particular that the status of $t$ has a significant impact on the consistent answer.
\begin{example}\label{ex:selection-condition}
\rm{
Consider the universe $U=\{Emp, Sal, Dept\}$ with the functional dependency $Emp \to Sal$ and the table $T=\{es, es'd,e's'\}$ over $U$ whose tuples state that employee $e$ has $s$ and $s'$ as salaries, employee $e$ works in department $d$, and that employee $e'$ has salary $s'$. The content of $T$ clearly does not satisfy the functional dependency $Emp \to Sal$ as employee $e$ is assigned {\em two} distinct salaries $s$ and $s'$. The limit interpretation $I^*$ of $T$ is defined by $I^*(e)=\{1,2\}$, $I^*(e')=\{3\}$, $I^*(s)=\{1,2\}$, $I^*(s')=\{1,2,3\}$ and $I^*(d)=\{2\}$. Thus $\true(\mathcal{T})$ consists of the sub-tuples of $esd$, $es'd$ and of $e's'$, and $\inc(\mathcal{T})=\{esd, es'd, es, es',sd,s'd,s,s'\}$. 
Let us now consider the following queries $Q_1$, $Q_2$, $Q_3$, $Q_4$, $Q_5$ and $Q_6$: 

\smallskip
$Q_1:$ {\tt select} $Emp, Dept$  {\tt from} $T$

$Q_2:$ {\tt select} $Emp, Sal$  {\tt from} $T$

$Q_3:$ {\tt select} $Sal$  {\tt from} $T$

$Q_4:$ {\tt select} $Emp$  {\tt from} $T$ {\tt where $Sal = s'$}

$Q_5:$ {\tt select} $Emp$  {\tt from} $T$ {\tt where $Sal = s \vee Sal = s'$}

$Q_6:$ {\tt select} $Emp$  {\tt from} $T$ {\tt where $Sal > 10K$}

\smallskip\noindent
Regarding $Q_1$, since $ed$ is the only true and consistent tuple in $\mathcal{T}(Emp,Dept)$, $ed$ is the only tuple satisfying the requirements {\sf R1}, {\sf R2} and {\sf R3} above. Requirement {\sf R4} is irrelevant because $Q_1$ involves no selection condition. Hence $ed$ is a candidate tuple to belong to the consistent answer to $Q_1$. Notice however that one could object that $ed$ is {\em not} a good candidate since all its maximal (with respect to $\sqsubseteq$) true super-tuples in $\mathcal{T}$ (namely $esd$ and $es'd$) are inconsistent.

Regarding $Q_2$, $es$, $es'$ and $e's'$ are the only tuples satisfying requirements {\sf R1} (they all belong to $\mathcal{T}(Emp,Sal)$) and {\sf R2} (they all belong to $\true(\mathcal{T})$). However, since $es$ and $es'$ are in $\inc(\mathcal{T})$, they do not satisfy requirement {\sf R3}, whereas $e's'$ does since this tuple is in $\cons(\mathcal{T})$. We moreover notice that, contrary to $Q_1$ above, $e's'$ has no true strict super-tuple. Hence the consistent answer to $Q_2$ should be $\{e's'\}$.

The case of $Q_3$ is clearer because $s$ and $s'$ are the only tuples satisfying requirements {\sf R1} and {\sf R2}, and since these two tuples are in $\inc(\mathcal{T})$, none of them can satisfy requirement {\sf R3}. Therefore, the consistent answer to $Q_3$ should be $\emptyset$.

Let us now turn to queries involving a selection condition. Regarding $Q_4$, the two tuples satisfying requirements {\sf R1} and {\sf R2} are $e$ and $e'$, and since $e$ and $e'$ are in $\cons(\mathcal{T})$, they also satisfy requirement {\sf R3}. Moreover, we notice that $es'$ is a true super-tuple of $e$ satisfying $\Gamma_4=(Sal = s')$, showing that $e$ satisfies requirement {\sf R4}. However, it should be noticed that $es'$ is inconsistent and that $es$ is another true super-tuple of $e$ {\em not} satisfying $\Gamma_4$. Hence, it can thought that $e$ should not belong to the consistent answer to $Q_4$. On the other hand, $e'$ has only one true super-tuple, namely $e's'$, and since this tuple is in $Sat(\Gamma_4)$, $e'$ satisfies requirement {\sf R4} and no further information allows us to think that the salary of $e'$ could be different than $s'$. As a consequence, the consistent answer to $Q_4$ is expected to contain  $e'$ and possibly $e$.

Regarding $Q_5$, for the same reasons as for $Q_4$, $e$ and $e'$ are the two possible tuples satisfying requirements {\sf R1}, {\sf R2} and {\sf R3}. Now, and contrary to the case of $Q_4$, the two true super-tuples of $e$, namely $es$ and $es'$, belong to $Sat(\Gamma_5)$, thus ensuring that whatever $e$'s salary it satisfies the selection condition. This remark supports the fact that $e$ is a candidate to belong to the consistent answer to $Q_5$. Of course, as for query $Q_1$, one could object that since $es$ and $es'$ are inconsistent, $e$ has {\em not} to belong to the consistent answer to $Q_5$. It should be noticed here that the case of $e'$ is treated as for $Q_4$ above, because $e's'$ is a true super-tuple of $e'$ such that $s'$ is in $Sat(\Gamma_5)$. Hence, the consistent answer to $Q_5$ is expected to be $\{e,e'\}$ or $\{e'\}$, following the choice made regarding inconsistency of the tuple $t$ in requirement {\sf R4}.

A reasoning similar to those for $Q_4$ or $Q_5$ applies to $Q_6$, depending on the actual values of $s$ and $s'$. Assuming first that $s=5K$ and $s'=20K$, as for $Q_4$, the expected consistent answer is $\{e'\}$ or possibly $\{e,e'\}$, knowing that $e$'s unique salary  may not satisfy the condition. However, if $s=15K$ and $s'=20K$, the consistent answer to $Q_6$ is expected to be as for $Q_5$, that is, $\{e,e'\}$ or $\{e'\}$, following the choice made regarding inconsistency of the tuple $t$ in requirement {\sf R4}.
\hfill$\Box$}
\end{example}
\subsection{Consistent Answers}\label{subsec:query-answer}
To account for some of the remarks regarding $Q_5$ and $Q_6$ in Example~\ref{ex:selection-condition}, we introduce the notion of {\em consistency with respect to a selection condition}.
To this end, for every relation schema $X$, we denote by $\mathcal{T}(X^\uparrow)$ the set of all tuples $t$ in $\mathcal{T}$ such that  $X \subseteq sch(t)$. In other words, $\mathcal{T}(X^\uparrow)$ is the set of all super-tuples of tuples over $X$, and it  follows that $\mathcal{T}(X)$ is a sub-set of $\mathcal{T}(X^\uparrow)$.
\begin{definition}\label{def:incons-S}
Given a table $T$  over $U$, and $\Gamma$ a selection condition:

A tuple $t$ in $\mathcal{T}(sch(\Gamma)^\uparrow)$  is said to be {\em inconsistent with respect to $Sat(\Gamma)$} if  $t.sch(\Gamma)$ is in $Sat(\Gamma)$ and there exists $s$ in $Sat^-(\Gamma)$ such that $I^*(t) \cap I^*(s) \ne \emptyset$. We denote by $\inc(\Gamma, \mathcal{T})$ the set of all tuples inconsistent with respect to $Sat(\Gamma)$.

A tuple $t$ in $\mathcal{T}(sch(\Gamma)^\uparrow)$ is said to be {\em consistent with respect to $Sat(\Gamma)$} if  $t.sch(\Gamma)$ is in $Sat(\Gamma)$ and $t$ is not in $\inc(\Gamma, \mathcal{T})$, that is if for every $s$ in $Sat^-(\Gamma)$, $I^*(t) \cap I^*(s) = \emptyset$. We denote by $\cons(\Gamma, \mathcal{T})$ the set of all tuples consistent with respect to $Sat(\Gamma)$.
\hfill$\Box$
\end{definition}
The following proposition states basic properties implied by Definition~\ref{def:incons-S}.
\begin{proposition}\label{prop:incons-S}
Given a table $T$ over $U$ and a selection condition $\Gamma$, the following holds:
\begin{enumerate}
\item
$\inc(\Gamma, \mathcal{T})$ is a sub-set of $\inc(\mathcal{T})$, and $\cons(\mathcal{T}) \cap \mathcal{T}(sch(\Gamma)^\uparrow)$ is a sub-set of $\cons(\Gamma, \mathcal{T})$. But $\cons(\Gamma, \mathcal{T}) \subseteq \cons(\mathcal{T})$ does {\em not} always hold.
\item
For every $t$ in $\cons(\Gamma, \mathcal{T})$, every super-tuple $t'$ of $t$ is also in $\cons(\Gamma, \mathcal{T})$.
\item
If $I^*$ is in First Normal Form, then $\inc(\Gamma, \mathcal{T})=\emptyset$ and $\cons(\Gamma, \mathcal{T})$ is the set of all super-tuples of tuples in $Sat(\Gamma)$.
\item
For a given tuple $t$, if $Sat(\Gamma)=\{t\}$, then $t$  is in $\inc(\mathcal{T})$ if and only if $t$ is in $\inc(\Gamma, \mathcal{T})$, and $t$  is in $\cons(\mathcal{T})$ if and only if $t$ is in $\cons(\Gamma, \mathcal{T})$.
\item
If $Sat(\Gamma) =\emptyset$ then $\cons(\Gamma, \mathcal{T}) =\inc(\Gamma, \mathcal{T}) = \emptyset$. If $Sat(\Gamma) = \mathcal{T}(X)$, then $\cons(\Gamma, \mathcal{T})$ is the set of all super-tuples of tuples in $Sat(\Gamma)$, and $\inc(\Gamma, \mathcal{T})$ is empty. 
\end{enumerate}
\end{proposition}
{\sc Proof.}
{\bf 1.~}If $t$ is in $\inc(\Gamma, \mathcal{T})$, there exists $s$ in $Sat^-(\Gamma)$ such that $I^*(t) \cap I^*(s) \ne \emptyset$. Since $t.sch(\Gamma)$ is in $Sat(\Gamma)$, $t.sch(\Gamma)$ and $s$ are distinct tuples in $\mathcal{T}(sch(\Gamma))$. Hence, there exists $A$ in $sch(\Gamma)$ such that $t.A \ne s.A$. As $I^*(s) \subseteq I^*(s.A)$, $I^*(t) \cap I^*(s) \ne \emptyset$ implies that $I^*(t) \cap I^*(s.A) \ne \emptyset$. Therefore, by Definition~\ref{def:incons-t}, $t$ is in $\inc(\mathcal{T})$.

As for the inclusion $\cons(\mathcal{T}) \cap \mathcal{T}(sch(\Gamma))\subseteq \cons(\Gamma, \mathcal{T})$, given $t$ in $\mathcal{T}(sch(\Gamma))$, if $t$ is in $\cons(\mathcal{T})$ then $t$ is not in $\inc(\mathcal{T})$. Thus, $t$ is not in $\inc(\Gamma, \mathcal{T})$, which implies by Definition~\ref{def:incons-S} that $t$ is in $\cons(\Gamma, \mathcal{T})$.

To see that $\cons(\Gamma, \mathcal{T}) \subseteq \cons(\mathcal{T})$ does not always hold, consider the earlier example where $T=\{abc,$ $ab'c,$ $a'b''c'\}$ over $U=\{A,B,C\}$ and $C \to B$. In this case, $b$ and $b'$ are clearly in $\inc(\mathcal{T})$, thus, not in $\cons(\mathcal{T})$. For $\Gamma = (B=b \vee B=b')$, we have $Sat(\Gamma)=\{b,b'\}$ and $Sat^-(\Gamma)=\{b''\}$. Since $I^*$ is defined by $I^*(a)=I^*(b)=I^*(b')=I^*(c)=\{1,2\}$, $I^*(a')=I^*(b'')=I^*(c')=\{3\}$, Definition~\ref{def:incons-S} shows that $b$ and $b'$ are in $\cons(\Gamma, \mathcal{T})$ (because $I^*(b)\cap I^*(b'')=I^*(b') \cap I^*(b'')=\emptyset$).

\smallskip\noindent
{\bf 2.~}If $t'$ is a super-tuple of $t$, $t'$ is in $\mathcal{T}(sch(\Gamma)^\uparrow)$ because so is $t$. Moreover, as $t$ is in $\cons(\Gamma, \mathcal{T})$, $I^*(t) \cap I^*(s)=\emptyset$ holds for every $s$ in $Sat^-(\Gamma)$. As $I^*(t') \subseteq I^*(t)$, it follows that $I^*(t') \cap I^*(s)=\emptyset$ holds as well, showing that $t'$ is in $\cons(\Gamma, \mathcal{T})$.

\smallskip\noindent
{\bf 3.~}Assuming that $I^*$ is in First Normal Form means that $\inc(\mathcal{T})=\emptyset$. By (1) above, this implies that $\inc(\Gamma, \mathcal{T})=\emptyset$ and thus, by Definition~\ref{def:incons-S},  $\cons(\Gamma, \mathcal{T})$ is the set of all super-tuples of tuples in $Sat(\Gamma)$. 

\smallskip\noindent
{\bf 4.~}By (1) above, we have $\inc(\Gamma, \mathcal{T})\subseteq \inc(\mathcal{T})$ always holds. Conversely, if $t$  is in $\inc(\mathcal{T})$ then, by Definition~\ref{def:incons-t}, there exist $A$ in $sch(t)$ and $a'$ in $adom(A)$ such that $a' \ne t.A$ and $I^*(t) \cap I^*(a')\ne \emptyset$. Denoting $t.A$ by $a$ and writing $t$ as $qa$, the tuple $qa'$ is such that $sch(qa') = sch(t)$ and $qa' \in Sat^-(\Gamma)$. Hence, $I^*(t) \cap I^*(qa') \ne \emptyset$, showing that $t$ is in $\inc(\Gamma,\mathcal{T})$.

If $t$ is in $\cons(\Gamma, \mathcal{T})$, then $Sat(\Gamma)= \mathcal{T}(X) \setminus \{t\}$, and for every $s$ in $Sat^-(\Gamma)$, $I^*(t) \cap I^*(s) =\emptyset$, implying that for every $A$ in $sch(\Gamma)$ and every $a'$ in $adom(A)$ such that $a' \ne t.A$, $I^*(t) \cap I^*(a') = \emptyset$. By Definition~\ref{def:incons-t}, this implies that $t$ is in $\cons(\mathcal{T})$. Conversely, if $t$ is in $\cons(\mathcal{T})$, by (1) above, as $t$ belongs to $\mathcal{T}(sch(\Gamma))$, $t$ is in $\cons(\Gamma, \mathcal{T})$.

\smallskip\noindent
{\bf 5.~}The results being immediate consequences of Definition~\ref{def:incons-S}, their proofs are omitted. The proof of the proposition is therefore complete.
\hfill$\Box$

\medskip\noindent
In what follows, in order to account for the remarks in Example~\ref{ex:selection-condition}, we propose four ways of defining the consistent answer to a query $Q$. These definitions are then illustrated by examples and compared to each other.
\begin{definition}\label{def:cons-answer}
Let $T$ be a table over universe $U$, $FD$ the set of associated functional dependencies. Given a query $Q: {\tt select}$ $X$  {\tt from} $T$ {\tt [where $\Gamma$]}, we define the following:
\begin{enumerate}
\item The {\em weakly consistent answer to} $Q$, denoted $\wconsans(Q)$, is the set of all tuples $x$ in $\mathcal{T}(X)$ such that
\begin{itemize}
\item[(a)] $x$ is in $\true(\mathcal{T}) \cap \cons(\mathcal{T})$,
\item[(b)] there exists $t$ is in $\true(\mathcal{T})$ such that $t$ is in $\mathcal{T}(sch(Q)^\uparrow)$, $t.X=x$, $t.sch(\Gamma)$ is in $Sat(\Gamma)$.
\end{itemize}
\item The {\em consistent answer to} $Q$, denoted ${\consans}(Q)$, is the set of all tuples $x$ in $\mathcal{T}(X)$ such that
\begin{itemize}
\item[(a)] $x$ is in $\true(\mathcal{T}) \cap \cons(\mathcal{T})$,
\item[(b)] there exists $t$ is in $\true(\mathcal{T})$ such that $t$ is in $\mathcal{T}(sch(Q)^\uparrow)$, $t.X=x$, $t$ is in $\cons(\Gamma, \mathcal{T})$.
\end{itemize}
\item The {\em strongly consistent answer to} $Q$, denoted $\sconsans(Q)$, is the set of all tuples $x$ in $\mathcal{T}(X)$ such that
\begin{itemize}
\item[(a)] $x$ is in $\true(\mathcal{T}) \cap \cons(\mathcal{T})$,
\item[(b)] there exists $t$ is in $\true(\mathcal{T}) \cap \cons(\mathcal{T})$ such that $t$ is in $\mathcal{T}(sch(Q)^\uparrow)$, $t.X=x$, and $t.sch(\Gamma)$ is in $Sat(\Gamma)$.
\end{itemize}
\item The {\em max-strongly consistent answer to} $Q$, denoted $\msconsans(Q)$, is the set of all tuples $x$ in $\mathcal{T}(X)$ such that
\begin{itemize}
\item[(a)] $x$ is in $\true(\mathcal{T}) \cap \cons(\mathcal{T})$,
\item[(b)] there exists $t$ is in $\true(\mathcal{T}) \cap \cons(\mathcal{T})$ such that $t$ is maximal with respect to $\sqsubseteq$ in $\true(\mathcal{T})$, $t$ is in $\mathcal{T}(sch(Q)^\uparrow)$, $t.X=x$, and $t.sch(\Gamma)$ is in $Sat(\Gamma)$.
\hfill$\Box$
\end{itemize}
\end{enumerate}
\end{definition}
The four kinds of consistent answer in Definition~\ref{def:cons-answer} relate to requirements {\sf R1}, {\sf R2}, {\sf R3} and {\sf R4} in the following respects:
\begin{itemize}
\item
Statements (1.a), (2.a), (3.a) and (4.a) clearly show that any tuple $x$ in any of the four kinds of consistent answer satisfies requirements {\sf R1} ($x \in \mathcal{T}(X)$), {\sf R2} ($x \in \true(\mathcal{T})$) and {\sf R3} ($x \in \cons(\mathcal{T})$).
\item
Statements (1.b), (2.b), (3.b) and (4.b) show that any tuple $x$ in any of the four kinds of consistent answer satisfies the requirement {\sf R4} by ensuring the existence of a true super-tuple $t$ of $x$ satisfying $\Gamma$.
\end{itemize}
Moreover, as announced in our earlier discussion, the status of the tuple $t$ is clarified in Definition~\ref{def:cons-answer} as follows:
\begin{itemize}
\item
in the case of the weakly consistent answer, $t$ is only requested to be true (as stated in requirement {\sf R4}),
\item
in the case of the consistent answer, $t$ is requested to be true and consistent with respect to $\Gamma$ (which does not disallow $t$ to be inconsistent),
\item
in the case of the strongly consistent answer, $t$ is requested to be true and consistent (which of course disallows $t$ to be inconsistent), and
\item
in the case  of the max-strongly consistent answer, $t$ is requested to be true, consistent {\em and} maximal.
\end{itemize}
We also emphasize that when the query $Q$ in Definition~\ref{def:cons-answer} involves no selection condition, then statements (1.b), (2.b) and (3.b) are reduced to the existence of $t$ such that $t$ is in $\true(\mathcal{T})$, $t$ is in $\mathcal{T}(X^\uparrow)$ and $t.X=x$. As statements (1.a), (2.a) and (3.a), ensure the existence of such a tuple  $t$, statements (1.b), (2.b) and (3.b) are redundant. However, such is not the case for statement (4.b), because the true and consistent super-tuple of $x$ needed to satisfy (4.a) might {\em not} be maximal, and thus might not satisfy (4.b). We refer to query $Q_1$ in the forthcoming Example~\ref{ex:cons-answer} for an illustration of this case.

Moreover, Definition~\ref{def:cons-answer} and the results in Proposition~\ref{prop:incons-S} lead to the following observations when the query $Q$ involves a selection condition $\Gamma$:
\begin{enumerate}
\item
For every $t$ in $\inc(\Gamma, \mathcal{T})$, the tuple $t.sch(\Gamma)$ is such that its syntax satisfies the condition $\Gamma$ whereas its semantics meets that of a tuple whose syntax does {\em not} satisfy the condition $\Gamma$. Since $\inc(\Gamma, \mathcal{T})$ is a sub-set of $\inc(\mathcal{T})$, this shows that the notion of satisfaction of a selection condition might concern inconsistent tuples. It should be noticed that any super-tuple of such tuples is discarded from max-strongly consistent answers and from strongly consistent answers, whereas they may occur in consistent answers or in weakly consistent answers.
\item
Since every super-tuple of a tuple  in $\cons(\Gamma, \mathcal{T})$ is also in $\cons(\Gamma, \mathcal{T})$, $\consans(Q)$ can be computed by scanning maximal true tuples in $\mathcal{T}(sch(Q)^\uparrow)$. Obviously, $\msconsans(Q)$ should be computed by scanning only maximal true tuples in $\mathcal{T}(sch(Q)^\uparrow)$, whereas the scan should concern true tuples in $\mathcal{T}(sch(Q))$, when computing $\sconsans(Q)$ or $\wconsans(Q)$. It will however be seen in the next section, that all consistent answers are computed based on the same scan.
\item
If $I^*$ is in First Normal Form, i.e., if $\inc(\mathcal{T})= \emptyset$, then $\cons(\Gamma, \mathcal{T})$ is the set of all super-tuples of tuples in $Sat(\Gamma)$. As moreover, $t.X$ is in $\cons(\mathcal{T})$, $t.X$ is in $\consans(Q)$. This remark fits the intuition that if $T$ is consistent then the consistent answer to $Q$ is equal to the standard answer to $Q$, that is the set of projections over $X$ of true tuples in $\mathcal{T}(sch(Q)^\uparrow)$ that satisfy $\Gamma$. This comes in line with our earlier remark stating that requirements {\sf R1}, {\sf R2}, {\sf R3} and {\sf R4} are satisfied by the tuples in the answer to $Q$ when $T$ is consistent.
\item
If $Sat(\Gamma)$ is reduced to one tuple $s$, then for every  $t$ in $\mathcal{T}(sch(Q)^\uparrow)\cap \true(\mathcal{T})$ such that $t.X$ is in $\cons(\mathcal{T}) \cap \true(\mathcal{T})$ and $t.sch(\Gamma)=s$, $t.X$ is in $\wconsans(Q)$, and moreover, $t.X$ is in $\sconsans(Q)$ and in $\consans(Q)$ if and only if $s$ is in $\cons(\mathcal{T})$. The case of $\msconsans(Q)$ is more involved because maximal true tuples must be considered. For example, it can happen that, for $Q:$ {\tt select $A$ from $T$ where $B=b$}, $a$ is in $\sconsans(Q)$ and in $\consans(Q)$, but not in $\msconsans(Q)$ because $ab$ is true and consistent, whereas $abc$ is the only true super-tuple of $ab$ and $abc$ is inconsistent. 
\item
If $Sat(\Gamma)=\emptyset$ (i.e., if $\Gamma$ is not satisfiable in $\mathcal{T}$), then it is obvious that $\msconsans(Q)$, $\sconsans(Q)$ and $\wconsans(Q)$ are empty. As in this case,  $\cons(\Gamma, \mathcal{T}) =\emptyset$, $\consans(Q)$ is empty as well.\\
On the other hand, if $Sat(\Gamma)=\mathcal{T}(sch(\Gamma))$, it is easy to see that $\sconsans(Q)$ and $\wconsans(Q)$ are equal to the set of all tuples in $\mathcal{T}(X)$ that belong to $\cons(\mathcal{T}) \cap \true(\mathcal{T})$. Moreover, since $\cons(\Gamma,\mathcal{T})=\mathcal{T}(sch(\Gamma)^\uparrow)$, $\consans(Q)$ is also the set of all tuples in $\mathcal{T}(X)$ that belong to $\cons(\mathcal{T}) \cap \true(\mathcal{T})$. However, as in the previous item, the case of $\msconsans(Q)$ is more involved because maximal true tuples must be considered. 
\end{enumerate}
The following proposition states an important relationship among the four kinds of consistent answers, namely that they form an increasing chain of inclusions.
\begin{proposition}\label{prop:comp-cons-ans}
Let $T$ be a table over universe $U$, $FD$ the set of associated functional dependencies. Given a query $Q: {\tt select}$ $X$  {\tt from} $T$ {\tt [where $\Gamma$]}, the following holds: $\msconsans(Q)\subseteq \sconsans(Q) \subseteq \consans(Q) \subseteq \wconsans(Q)$.
\end{proposition}
{\sc Proof.}
To show that  $\msconsans(Q)\subseteq \sconsans(Q)$ holds, let $x$ in $\msconsans(Q)$. Then statements (1.a) and (1.b) in Definition~\ref{def:cons-answer} hold, and thus, so does statement (2.a). The result comes form the fact that it is easily seen that statement (1.b) implies statement (2.b).

Regarding the inclusion $\sconsans(Q)\subseteq \consans(Q)$, we notice that for every $x$ in $\sconsans(Q)$, statements (2.a) and (2.b) in Definition~\ref{def:cons-answer} hold, and thus, that statement (3.a) holds as well. Moreover, since statement (2.b) holds, there exists $t$ in $\mathcal{T}(sch(Q)^\uparrow)$, such that $t.X=x$ and $t.sch(\Gamma)$ is in $Sat(\Gamma)$. Thus $t$ is in $\cons(\mathcal{T}) \cap \mathcal{T}(sch(\Gamma)^\uparrow)$, which by Proposition~\ref{prop:incons-S}(1) implies that $t$ is in $\cons(\Gamma, \mathcal{T})$. Hence statement (3.b) is satisfied, showing that $x$ is in $\consans(Q)$.

Regarding the last inclusion, for every $x$ in $\consans(Q)$, statement (4.a) clearly holds and moreover, as the tuple $t$ in $\cons(\Gamma, \mathcal{T}) \cap \true(\mathcal{T})$ is obviously in $\true(\mathcal{T})$ and in $Sat(\Gamma)$, the proof is complete.
\hfill$\Box$

\medskip\noindent
We illustrate Definition~\ref{def:cons-answer} and Proposition~\ref{prop:comp-cons-ans} in the context of Example~\ref{ex:selection-condition}. Moreover, through this example, we show that the inclusions in Proposition~\ref{prop:comp-cons-ans} might be strict.
\begin{example}\label{ex:cons-answer}
\rm{
We recall that in Example~\ref{ex:selection-condition}, $U=\{Emp,$ $Sal,$ $Dept\}$, $FD= \{Emp \to Sal\}$ and $T=\{es,$ $es'd,$ $e's'\}$, which implies that $\true(\mathcal{T})$ consists of the sub-tuples of $esd$, $es'd$ and of $e's'$ and that $\inc(\mathcal{T})=\{esd,$ $es'd,$ $es,$ $es',$ $sd,$ $s'd,$ $s,s'\}$. 
The queries we are interested in are $Q_1$, $Q_2$, $Q_3$, $Q_4$, $Q_5$ and $Q_6$ as shown below: 

\smallskip
$Q_1:$ {\tt select} $Emp, Dept$ {\tt from} $T$

$Q_2:$ {\tt select} $Emp, Sal$  {\tt from} $T$

$Q_3:$ {\tt select} $Sal$  {\tt from} $T$

$Q_4:$ {\tt select} $Emp$  {\tt from} $T$ {\tt where $Sal = s'$}

$Q_5:$ {\tt select} $Emp$  {\tt from} $T$ {\tt where $Sal = s \vee Sal = s'$}

$Q_6:$ {\tt select} $Emp$  {\tt from} $T$ {\tt where $Sal > 10K$}

\smallskip\noindent
In the case of $Q_1$, the only possible true tuple in any answer is $ed$. As all maximal true super-tuples of $ed$ are $esd$ and $es'd$, none of these tuples is consistent. Thus, we have $\msconsans(Q_1)= \emptyset$. However, $\sconsans(Q_1)= \consans(Q_1)=\wconsans(Q_1)=\{ed\}$, because this tuple is true and consistent and because no selection condition is involved. Regarding the inclusions in Proposition~\ref{prop:comp-cons-ans}, this case shows that the first inclusion might be strict. 

Considering now $Q_2$, we have $\msconsans(Q_2)= \{e's'\}$, because this tuple is true and consistent, and has no strict true super-tuple. For a similar reason, we also have $\sconsans(Q_2)= \consans(Q_2)=\wconsans(Q_2)=\{e's'\}$ because no selection condition is involved in $Q_2$. In this case the inclusions in Proposition~\ref{prop:comp-cons-ans} hold  because $\msconsans(Q_2)=\sconsans(Q_2)=\consans(Q_2)=\wconsans(Q_2)$ hold.

The case of $Q_3$ is easy because as earlier mentioned, no $Sal$-value is consistent, implying that statement (a) of the four consistent answers in Definition~\ref{def:cons-answer} can not be satisfied. As $Q_3$ involves no selection condition, we have $\msconsans(Q_3)=\sconsans(Q_3)=\consans(Q_3)=\wconsans(Q_3)=\emptyset$, showing again that the inclusions in Proposition~\ref{prop:comp-cons-ans} hold.

\smallskip
Regarding $Q_4$ involving the selection condition $\Gamma_4=(Sal=s')$, since $e$ and $e'$ are both true and consistent tuples over $Emp$, they satisfy statements (1.a), (2.a), (3.a) and (4.a) in Definition~\ref{def:cons-answer}. Since $Sat(\Gamma_3) =\{s'\}$, $es'$ and $e's'$ are the only tuples of interest regarding statements (1.b), (2.b), (3.b) and (4.b) in Definition~\ref{def:cons-answer}. Thus $\wconsans(Q_4)=\{e,e'\}$. As $es'd$ is maximal and inconsistent, statement (1b) is not satisfied and since $es'$ is also inconsistent, statement (2.b) is not satisfied either. Now, regarding statement (3.b), we have $s$ in $Sat^-(\Gamma_4)$ and $I^*(es') \cap I^*(s)=\{1,2\}$ (see Example~\ref{ex:selection-condition}). Thus, by Definition~\ref{def:incons-S}, $es'$ is in $\inc(\Gamma_4,\mathcal{T})$, and so, $es'$ does not satisfy statement (3.b). As a consequence, $e$ is not in $\msconsans(Q_4)$, $\sconsans(Q_4)$ nor in $\consans(Q_4)$.

On the other hand,  $e's'$ does satisfy statements (1.b), (2.b) and (3.b) in Definition~\ref{def:cons-answer}, because $e's'$ is a maximal true and consistent tuple such that $s'$ is in $Sat(\Gamma_4)$, and regarding (3.b), $e's'$ is in $\cons(\Gamma_4,\mathcal{T})$ (because $I^*(e's') \cap I^*(s)=\emptyset$). Therefore, $\msconsans(Q_4)=\sconsans(Q_4)=\consans(Q_4) = \{e'\}$ and $\wconsans(Q_4)=\{e,e'\}$, showing that  the inclusions in Proposition~\ref{prop:comp-cons-ans} hold, and that the last one might be strict.

\smallskip
As for $Q_5$ involving the selection condition $\Gamma_5$ defined by $(Sal = s \vee Sal = s')$, for the same reasons as for $Q_4$, the two possible tuples in the answer are $e$ and $e'$ and $\wconsans(Q_5)=\{e,e'\}$. Moreover, since $Sat(\Gamma_5)=\{s,s'\}$, $e'$ is in $\msconsans(Q_5)$, $\sconsans(Q_5)$ and $\consans(Q_5)$. Regarding now $e$, the same arguments as for $Q_4$ apply regarding the statements (1.b) and (2.b) in Definition~\ref{def:cons-answer}, showing that $e$ is neither in $\msconsans(Q_5)$ nor in $\sconsans(Q_5)$. However, contrary to the case of $Q_4$, $es$ is in $\cons(\Gamma_5,\mathcal{T})$, and thus, statement (3.b) of Definition~\ref{def:cons-answer} is satisfied. We therefore obtain that $\msconsans(Q_5)=\sconsans(Q_5)=\{e'\}$ and $\consans(Q_5)=\wconsans(Q_5)=\{e,e'\}$, which also shows that the second inclusion in Proposition~\ref{prop:comp-cons-ans} might be strict.

\smallskip
Consider now $Q_6$ involving the selection condition $\Gamma_6 = (Sal > 10K)$ and suppose that $s=5K$ and $s'=20K$. Then, we have  $Sat(\Gamma_6)=\{s'\}$ and  $Sat^-(\Gamma_6)=\{s\}$, and thus, as in the case of $Q_4$, $\msconsans(Q_4) = \sconsans(Q_4) = \consans(Q_4) = \{e'\}$, and $\wconsans(Q_6)=\{e,e'\}$. On the other hand, if $s=15K$ and $s'=20K$, then  $Sat(\Gamma_6)=\{s, s'\}$ and  $Sat^-(\Gamma_3)=\emptyset$, and thus, as for $Q_5$ above, we have $\msconsans(Q_5)=\sconsans(Q_5)=\{e'\}$ and $\consans(Q_5)=\wconsans(Q_6)=\{e,e'\}$.
\hfill$\Box$}
\end{example}
To end this section, we would like to stress that proposing distinct consistent answers as done in Definition~\ref{def:cons-answer} raises the question of {\em which one should be chosen.} Although we think that this choice should be made by the final user, the comparisons stated in Proposition~\ref{prop:comp-cons-ans} should help  in making this choice. Moreover, assuming the choice of $\consans(Q)$ is made, the knowledge of the answers $\msconsans(Q)$, $\sconsans(Q)$ and $\wconsans(Q)$ are likely to be useful to the user, regarding the {\em quality} of the provided consistent answer.

For instance, in Example~\ref{ex:cons-answer} the fact that $e$ is in $\consans(Q_5)$ but not in $\sconsans(Q_5)$ implies that the database contains an inconsistency regarding $e$'s salary. This information shows that the presence of $e$ in $\consans(Q_5)$ might be considered {\em less reliable} than that of $e'$ for which such inconsistency does not exist as $e'$ is in $\sconsans(Q_5)$ and in $\sconsans(Q_5)$. On the other hand, a tuple in $\wconsans(Q)$ but not in $\consans(Q)$ shows that although this tuple occurs with values satisfying the selection condition, it also systematically occurs in an inconsistency involving values {\em not} satisfying the selection condition. This happens for $e$ when answering $Q_4$: although $e$ occurs associated with $s'$, it also occurs associated with $s$.

This example also shows that comparing the consistent answers $\msconsans(Q)$, $\sconsans(Q)$ and $\wconsans(Q)$ with $\consans(Q)$ has an impact not only on the {\em quality} of the answers, but also allows for {\em explaining} the presence or the absence of some tuples in a consistent answer. These issues related to data quality and to answer explanation are among the subjects of our future work. 
\section{Computational Issues}\label{sec:comp-issues}
In this section, we present algorithms first for computing efficiently the sets $\cons(\mathcal{T})$ and $\inc(\mathcal{T})$  (see Section~\ref{subsec:m-Chase}) and then the four consistent answers to a given query (see Section~\ref{subsec:c-ans}).
\subsection{The Tabular Representation of $I^*$}\label{subsec:m-Chase}
In this section, we show how to compute $\true(\mathcal{T})$ and $\inc(\mathcal{T})$, using a modified version of the well-known chase algorithm \cite{FaginMU82,Ullman} that we {\em call m-Chase}. Our algorithm is based on the notion of {\em multi-valued tuple}, or m-tuple for short, which generalizes that of usual tuple in the following sense: instead of associating every attribute $A$ with {\em at most one value} in $adom(A)$, a multi-valued tuple associates every attribute $A$ with a {\em sub-set} of $adom(A)$, possibly empty. Multi-valued tuples are defined as follows.
\begin{definition}\label{def:m-tuple}
Given a universe $U=\{A_1, \ldots , A_n\}$ and active domains $adom(A_i)$, $i=1, \ldots ,n$, a {\em multi-valued tuple} $\sigma$ over $U$, or m-tuple for short, is a total function from $U$ to the cross product ${\Large{\sf X}}_{A \in U}\mathcal{P}(adom(A))$, where $\mathcal{P}(adom(A))$ denotes the power set of the active domain of attribute $A$. The `empty' m-tuple, denoted by $\omega$, is defined by $\omega (A)= \emptyset$ for every $A$ in $U$.

The set of all attributes $A$ such that $\sigma(A) \ne \emptyset$, is called the {\em schema  of $\sigma$}, denoted by $sch(\sigma)$, and the schema  of $\omega$ is $\emptyset$. Given $\sigma \ne \omega$ and a sub-set $X$  of $sch(\sigma)$, the restriction of $\sigma$ to $X$, denoted $\sigma(X)$, is the m-tuple defined by $(\sigma(X))(A)=\sigma(A)$ for every $A$ in $X$ and $(\sigma(X))(A)=\emptyset$ for any $A$ not in $X$.

Given an m-tuple $\sigma$, the set $\tuples(\sigma)$ denotes the set of all tuples $t$ such that $sch(t)=sch(\sigma)$ and for every $A$ in $sch(t)$, $t(A)$ is a value of $adom(A)$ that belongs to $\sigma(A)$. As for the empty m-tuple $\omega$, we have $\tuples(\omega)=\emptyset$.

The set of all m-tuples over $U$ is denoted by $\mathcal{MT}$, and a finite set of m-tuples is called an {\em m-table}.
\hfill$\Box$
\end{definition}
It is easy to see that for every $t$ in $\mathcal{T}$, it holds that $sch(\sigma_t)=sch(t)$ and $\tuples(\sigma_t)=\{t\}$. In what follows, we use $t$ and $\sigma_t$ interchangeably, whenever no confusion is possible. 
To simplify notation, given an m-tuple $\sigma$, the set $\sigma(A)$ is denoted by the concatenation of its elements between parentheses, and $\sigma$ itself is denoted by the concatenation of all $\sigma(A)$ such that $\sigma(A) \ne \emptyset$.

Based on Definition~\ref{def:m-tuple}, a tuple $t$ in $\mathcal{T}$ yields an m-tuple $\sigma_t$ in $\mathcal{MT}$ such that for every $A$ in $sch(t)$, $\sigma_t(A)=\{t.A\}$ (or $\sigma_t(A)= (t.A)$ according to the notation just introduced), and $\sigma_t(B) = \emptyset$ for every $B$ not in $sch(t)$. 

\smallskip
For example if $U=\{A,B,C,D\}$,  the m-tuple $\sigma$ such that $\sigma(A)=\{a,a'\}$, $\sigma(B)=\{b\}$, $\sigma(C)=\emptyset$ and $\sigma(D)=\{d_1,$ $d_2,$ $d_3\}$ is denoted by $(a,a')(b)(d_1d_2d_3)$ and we have $sch(\sigma)=\{A,B,D\}$ (or simply $sch(\sigma)=ABD$). Now, to find the set $\tuples(\sigma)$ we make all combinations of values, one from each of $\sigma(A)$, $\sigma(B)$ and $\sigma(D)$. We find $\tuples(\sigma)=\{abd_1,$ $abd_2,$ $abd_3,$ $a'bd_1,$ $a'bd_2,$ $a'bd_3\}$. Moreover, we have  $\sigma(AB)=(aa')(b)$. On the other hand, for $t =abd_1$, we have $\sigma_t=(a)(b)(d_1)$, and thus $\tuples(\sigma_t)=\{abd_1\}$.

\smallskip
We draw attention to the fact that, although m-tables can be seen as a particular case of {\em embedded relations} \cite{Ullman}, we do not intend to elaborate on such relations in our approach. We rather use m-tuples as a tool to explicitly account for situations where the First Normal Form (or equivalently the partition constraint) is not satisfied. 
We define the following operations for m-tuples:
\begin{description}
\item[{\em Comparison:}] $\sigma_1 \sqsubseteq \sigma_2$ holds if for every $A$ in $U$, $\sigma_1(A) \subseteq \sigma_2(A)$\footnote{With a slight abuse of notation, we use here the same symbol $\sqsubseteq$ introduced earlier for the sub-tuple relation between usual tuples.}.
Thus if $\sigma_1 \sqsubseteq \sigma_2$ then $sch(\sigma_1) \subseteq sch(\sigma_2)$. Moreover, if $sch(\sigma_1)= sch(\sigma_2)$ then $\tuples(\sigma_1) \subseteq \tuples(\sigma_2)$. It also holds that $\omega \sqsubseteq \sigma$, for every m-tuple $\sigma$.
\item[{\em Union:}] $\sigma_1 \sqcup \sigma_2$ is the m-tuple $\sigma$ such that for every $A$ in $U$, $\sigma(A)=\sigma_1(A) \cup \sigma_2(A)$.
Thus, $sch(\sigma_1 \sqcup \sigma_2)=sch(\sigma_1) \cup sch(\sigma_2)$.
\item[{\em Intersection:}] $\sigma_1 \sqcap \sigma_2$ is the m-tuple $\sigma$ such that for every $A$ in $U$, $\sigma(A)=\sigma_1(A) \cap \sigma_2(A)$.
Thus, $sch(\sigma_1 \sqcap \sigma_2) \subseteq sch(\sigma_i)$ for $i=1,2$.
\end{description}
We now introduce a new chase algorithm, called {\em m-Chase} whose definition is shown as Algorithm~\ref{algo:chase}. The m-Chase Algorithm works on sets of m-tuples to provide an m-table from which true and inconsistent tuples can easily be retrieved. Roughly speaking, Algorithm~\ref{algo:chase} follows the same idea as the standard chase, but when a functional dependency cannot be satisfied, our algorithm does not stop, but rather collects all values inferred from the functional dependencies. Consequently, it may happen that for a given row and a given column (i.e., for a given attribute value) we may have to store more than one constant, which is achieved by means of m-tuples. It is important to note that the output of Algorithm~\ref{algo:chase} (denoted by $\Sigma^*$) allows to explicitly `show' all violations of the partition constraint $PC$. We illustrate how Algorithm~\ref{algo:chase} works using an example.

\algsetup{indent=1.5em}
\begin{algorithm}[t]
\caption{The m-Chase Algorithm \label{algo:chase}}
{\footnotesize
\begin{algorithmic}[1]
\REQUIRE
A table $T$ over $U$ and a set $FD$ of functional dependencies over $U$.

\ENSURE An m-table denoted by $\Sigma^*$ and the truth value of $fail$.

\STATE{$\Sigma^* := \{\sigma_t~|~t \in T\}$\label{line:step1-start}}\label{line:init-D}
\STATE{$change := true$}
\STATE{$fail := false$}
\WHILE{$change = true$}\label{line:main-loop-chase}
    \STATE{$change := false$}
    \FORALL{$\sigma$ in $\Sigma^*$}
        \FORALL{$X \to A$ in $FD$ such that $XA \subseteq sch(\sigma)$}
            \FORALL{$\sigma'$ in $\Sigma^*$ such that $X \subseteq sch(\sigma')$}
               	 \IF{for every $B$ in $X$, $\sigma(B) \cap \sigma'(B) \ne \emptyset$}
                        \IF{$A \in sch(\sigma)$ and $A \in sch(\sigma')$ and $\sigma(A) \ne \sigma'(A)$ and $fail = false$}
                            \STATE{$fail := true$}\label{line:fail-true}
                        \ENDIF
               	    \STATE{$\sigma_1:= \sigma \sqcup \sigma'(A)~;~\sigma'_1:= \sigma' \sqcup \sigma(A)$}
                    \IF{$\sigma \ne \sigma_1$ or $\sigma' \ne \sigma'_1$}
                   	    \STATE{$change := true$}
               	        \IF{$\sigma_1 \sqsubseteq \sigma'_1$}
               	            \STATE{$\Sigma^*:= (\Sigma^*\setminus \{\sigma, \sigma'\}) \cup \{\sigma'_1\}$}\label{line:step1-end1}
                        \ELSIF{$\sigma'_1 \sqsubseteq \sigma_1$}
               	                \STATE{$\Sigma^*:= (\Sigma^*\setminus \{\sigma, \sigma'\}) \cup \{\sigma_1\}$}\label{line:step1-end2}
               	        \ELSE
                            \STATE{$\Sigma^*:=(\Sigma^*\setminus \{\sigma, \sigma'\}) \cup \{\sigma_1, \sigma'_1\}$}\label{line:step1-end}
               	        \ENDIF
               	    \ENDIF
               	\ENDIF
            \ENDFOR
        \ENDFOR
     \ENDFOR
\ENDWHILE
\RETURN{$\Sigma^*$ and the truth value of $fail$}
\end{algorithmic}
}
\end{algorithm}

\begin{example}\label{ex:motivate-2}
\rm{
Let $T=\{ab,ab',a'b'\}$ over universe $U=\{A,B\}$  with functional dependencies $A \to B$ and $B\to A$. The computation of $\Sigma^*$ based on Algorithm~\ref{algo:chase} works as follows:
\begin{enumerate}
\item Considering tuples in $T$ as m-tuples yields the following m-tuples: $(a)(b)$, $(a)(b')$ and $(a')(b')$. Thus, $\Sigma^*$ is set to $\{(a)(b), (a)(b'),(a')(b')\}$ on line~\ref{line:step1-start}.
\item Considering $A\to B$, with $\sigma = (a)(b)$ and $\sigma'= (a)(b')$, the m-tuples $\sigma_1 = (a)(bb')$ and $\sigma'_1= (a)(b'b)$ are generated. The variable $change$ is set to $true$ and since $\sigma_1=\sigma'_1$, $\Sigma^*$ is set to $\{(a)(bb'), (a')(b')\}$. 
Moreover, $fail$ is set to true line~\ref{line:fail-true} because we have $(b) \ne (b')$. This means intuitively that when running the standard chase algorithm, a failure would be reached at this step.\\
Considering $B \to A$, since the $B$-value $b'$ occurs in $(a)(bb')$ and $(a')(b')$, these m-tuples generate respectively $(aa')(bb')$ and $(aa')(b')$. Moreover, since $(aa')(b') \sqsubseteq (aa')(bb')$, $\Sigma^*$ is set to $\{(aa')(bb')\}$.
\item
As the second execution of the while-loop line~\ref{line:main-loop-chase} does not change $\Sigma^*$, the variable $change$ has value $false$. Hence the processing stops, returning $\Sigma^*=\{(aa')(bb')\}$ along with the value $true$ assigned to $fail$.
\hfill$\Box$
\end{enumerate}}
\end{example}
We now state a basic lemma based on which it will be shown how to compute  $\true(\mathcal{T})$ and $\inc(\mathcal{T})$. In what follows, the notation {\sf LoCl} is extended to $\Sigma^*$ in a straightforward way: ${\sf LoCl}(\Sigma^*)=\{t \in \mathcal{T}~|~(\exists \sigma \in \Sigma^*)(t \sqsubseteq \sigma)\}$. Moreover, for every interpretation $I$ and every m-tuple $\sigma$, $I(\sigma)$ is defined by $I(\sigma)= \bigcap_{a \sqsubseteq \sigma}I(a)$. It should be noticed that, as for tuples in $\mathcal{T}$, if $\sigma \sqsubseteq \sigma'$, then $I(\sigma') \subseteq I(\sigma)$ holds.
\begin{lemma}\label{lemma:chase}
Let $T$ be a table over universe $U$. Then the following hold:
\begin{enumerate}
\item
Algorithm~\ref{algo:chase} applied to $T$ always terminates.
\item
For every m-tuple $s$  in $\mathcal{MT}$, $I^*(s) \ne \emptyset$ holds if and only if there exists $\sigma$ in $\Sigma^*$ such that $s \sqsubseteq \sigma$.
\item
$\inc(\mathcal{T})= \emptyset$ if and only if $fail = false$.
\item
For every $\sigma$ in $\Sigma^*$ and all $t$ and $t'$ in $\tuples(\sigma)$, $I^*(t)=I^*(t')$.
\end{enumerate}
\end{lemma}
{\sc Proof.}
See Appendix~\ref{append:lemma-chase}.
\hfill$\Box$

\medskip\noindent
Note that Algorithm~\ref{algo:chase} is an extension of the standard chase algorithm. Indeed, the standard chase algorithm works on tuples (and not on m-tuples) which is the case for $T$. Moreover, as long as $\sigma$ and $\sigma'$ are usual tuples, our processing is similar to that of standard chase. In particular, for every $X \to A$ in $FD$ if $XA$ is a sub-set of $sch(\sigma)$ and  $sch(\sigma')$ and if $\sigma(X)$ and $\sigma'(X)$ are equal tuples, while $\sigma(A)$ and $\sigma'(A)$ are distinct values, then, in Algorithm~\ref{algo:chase}, $fail$ is set to $true$, which is precisely a case of failure for the standard chase algorithm. Thus, based on Lemma~\ref{lemma:chase}, the returned value for $fail$ is $true$ if and only if the standard chase algorithm fails. 

Moreover, the statements lines~\ref{line:step1-end1}, \ref{line:step1-end2} and \ref{line:step1-end} in Algorithm~\ref{algo:chase} imply that when $\sigma$ and $\sigma'$ are in $\Sigma^*$, for every $X \to A$ in $FD$ such that $XA$ is a sub-set of $sch(\sigma)$ and  $sch(\sigma')$, if $\tuples(\sigma(X)) \cap \tuples(\sigma'(X)) \ne \emptyset$, then $\sigma(A)=\sigma'(A)$. This remark can be seen as an extension of the well-known property of the output of the standard chase, by which all rows equal over $X$ must have the same $A$-value. 
As shown in the following proposition, the m-table $\Sigma^*$ allows to compute the sets $\true(\mathcal{T})$ and $\inc(\mathcal{T})$.
\begin{proposition}\label{prop:chase}
Let $T$ be a table over $U$, $FD$ the associated set of functional dependencies, and $\Sigma^*$ the output of Algorithm~\ref{algo:chase} run with $T$ and $FD$ as input. Then:
\begin{enumerate}
\item $\true(\mathcal{T})={\sf LoCl}(\Sigma^*)$.
\item $\inc(\mathcal{T})=\{t \in \mathcal{T}~|~(\exists \sigma \in \Sigma^*)(\exists  A\in sch(t))(\exists a,a' \in adom(A))(a \sqsubseteq t \wedge t \sqsubseteq \sigma \wedge (aa') \sqsubseteq \sigma)\}$.
\item $\cons(\mathcal{T}) \cap \true(\mathcal{T})=\{t \in \true(\mathcal{T})~|~(\forall \sigma \in \Sigma^*)(t \sqsubseteq \sigma \Rightarrow \tuples(\sigma(sch(t))) =\{t\})\}$.
\end{enumerate}
\end{proposition}
{\sc Proof.}
{\bf 1.} The result is an immediate consequence of Lemma~\ref{lemma:chase}(2).\\
{\bf 2.} The result follows from Lemma~\ref{lemma:chase}(2) and Definition~\ref{def:incons-t}.\\
{\bf 3.} The result follows from Definition~\ref{def:incons-t} and the previous two items.
\hfill$\Box$

\medskip\noindent
In order to assess the complexity of Algorithm~\ref{algo:chase}, for every $X \to A$ in $FD$ and every tuple $x$ in $\mathcal{T}(X)$, we let $N(x)$ be the number of different $A$-values $a$ such that $I^*(xa) \ne \emptyset$ and $a$ is inconsistent, and we denote by $\delta$ the maximal value of all $N(x)$ for all $X \to A$ in $FD$ and all $x$ in $\true(\mathcal{T})$; in other words $\delta= \max(\{N(x)~|~X \to A \in FD \wedge x \in \true(\mathcal{T})\})$.

Using this notation, the size of the m-tuples in $\Sigma^*$ is bounded by $|U|.\delta$, and as $\Sigma^*$ contains at most $|T|$ m-tuples, a run of the core of loop line~\ref{line:main-loop-chase} is in $\mathcal{O}(|T|^2.(|U|.\delta))$. Since the number of runs of the loop line~\ref{line:main-loop-chase} is bounded by the changes in the sets in $\Sigma^*$, this number of runs is in $\mathcal{O}(|T|.|FD|.\delta)$ (that is the maximum number of constants that can be added in a set of $\Sigma^*$ times the size of $\Sigma^*$). Therefore, the computation of $\Sigma^*$ is in 
$\mathcal{O}((|T|.|FD|.\delta).(|T|^2.(|U|.\delta))$, that is in $\mathcal{O}(|T|^3.|FD|.|U|.\delta^2)$ or even in $\mathcal{O}(|T|^3.\delta^2)$, since $|U|$ and $|FD|$ are independent from $|T|$.

\smallskip
We draw attention to the following important points regarding this result:
\begin{enumerate}
    \item When the database is consistent, $\delta$ is equal to $1$, thus yielding a complexity in $\mathcal{O}(|T|^3)$. This result can be shown independently from the above computations as follows: In the case of traditional chase the maximum of nulls in $T$ being bounded by $|U|.|T|$, the number of iterations when running the algorithm is also bounded by $|U|.|T|$. Since the run of one iteration is in $|T|^2$, the overall complexity is in $\mathcal{O}(|U|.|T|^3)$, or in $\mathcal{O}(|T|^3)$, as $|U|$ is independent from $|T|$. 
    \item The above complexity study can be seen as a {\em data} complexity study as it relies on the size of $T$. Thus, this study should be further investigated in order to provide more accurate results regarding the estimation of the number of actual tests necessary to the computation of $\Sigma^*$. The results in \cite{CKS86} are likely to be useful for such a more  thorough study of this complexity.
\end{enumerate}
\subsection{Computing Consistent Answers}\label{subsec:c-ans}
In this sub-section, we show how the consistent answers to a query can be computed based on the m-table $\Sigma^*$. We stress that to state one of the results, it is assumed that $\Sigma^*$ is {\em not redundant}, that is for all $\sigma_1$ and $\sigma_2$ in $\Sigma^*$, neither $\sigma_1 \sqsubseteq \sigma_2$ nor  $\sigma_2 \sqsubseteq \sigma_1$ holds. Notice that, {\em if $\Sigma^*$ is not redundant then for all $\sigma_1$ and $\sigma_2$ in $\Sigma^*$, $I^*(\sigma_1) \cap I^*(\sigma_2) =\emptyset$}.

Indeed, if $I^*(\sigma_1) \cap I^*(\sigma_2) \ne \emptyset$, Lemma~\ref{lemma:chase}(2) implies that $\Sigma^*$ contains $\sigma$ such that $\sigma_i \sqsubseteq \sigma$ for $i=1,2$. As this entails that $\Sigma^*$ contains redundancies, we obtain a contradiction. We also notice in this respect that redundancies are eliminated at each step of the computation of $\Sigma^*$ in Algorithm~\ref{algo:chase} (see the statements lines~\ref{line:step1-end1} and \ref{line:step1-end2}). Thus, it turns out that if the table $T$ contains no redundancy, so does $\Sigma^*$.

Moreover, to state our results, we introduce the following additional notation. Let $T$ be a non-redundant table $T$ over universe $U$, and $\Sigma^*$ its associated non-redundant m-table, as computed by Algorithm~\ref{algo:chase}. Given a query $Q: {\tt select}$ $X$  {\tt from} $T$ {\tt [where $\Gamma$]}, and a tuple $x$ in $\mathcal{T}(X)$, we respectively denote by $\Sigma(x)$ and $\Sigma_Q(x)$, the sets defined by  $\Sigma(x)=\{\sigma \in \Sigma^*~|~(x \sqsubseteq \sigma)\}$ and $\Sigma_Q(x)=\{\sigma \in \Sigma(x)~|~(\exists s \in Sat(\Gamma))(s \sqsubseteq \sigma)\}$.

Considering that $\Sigma_Q(x)=\Sigma(x)$ if $Q$ involves no selection condition, the following proposition characterizes the tuples in $\msconsans(Q)$, $\sconsans(Q)$,  $\consans(Q)$ and $\wconsans(Q)$  in terms of m-tuples as follows.

\begin{proposition}\label{prop:mscons-answer}
Let $T$ be a non-redundant table over universe $U$, $FD$ the set of associated functional dependencies and $\Sigma^*$ the non-redundant m-table as computed by Algorithm~\ref{algo:chase}. Given a query $Q: {\tt select}$ $X$  {\tt from} $T$ {\tt [where $\Gamma$]}, and a tuple $x$ in $\mathcal{T}(X)$:
\begin{enumerate}
\item $x$ is in $\msconsans(Q)$ if and only if
\begin{itemize}
\item[(a)] for every $\sigma$ in $\Sigma(x)$, $\tuples(\sigma(X))=\{x\}$,
\item[(b)] there exists $\sigma$ in $\Sigma_Q(x)$ such that $|\tuples(\sigma)| =1$.
\end{itemize}
\item $x$ is in $\sconsans(Q)$ if and only if
\begin{itemize}
\item[(a)] for every $\sigma$ in $\Sigma(x)$, $\tuples(\sigma(X))=\{x\}$,
\item[(b)] there exists $s$ in $Sat(\Gamma) \cap \true(\mathcal{T})$ such that for every $\sigma$ in $\Sigma_Q(x)$, if $s$ is  in $\sigma(sch(\Gamma))$ then $\tuples(\sigma(sch(\Gamma)))) =\{s\}$.
\end{itemize}
\item $x$ is in $\consans(Q)$ if and only if
\begin{itemize}
\item[(a)] for every $\sigma$ in $\Sigma(x)$, $\tuples(\sigma(X))=\{x\}$,
\item[(b)] there exists $\sigma$ in $\Sigma_Q(x)$ such that $\tuples(\sigma(sch(\Gamma))) \subseteq Sat(\Gamma)$.
\end{itemize}
\item $x$ is in $\wconsans(Q)$ if and only if
\begin{itemize}
\item[(a)] for every $\sigma$ in $\Sigma(x)$, $\tuples(\sigma(X))=\{x\}$,
\item[(b)] $\Sigma_Q(x)$ is not empty.
\end{itemize}
\end{enumerate}
\end{proposition}
{\sc Proof.}
See Appendix~\ref{append:mscons-answer}.
\hfill$\Box$
\begin{example}\label{ex:queries}
{\rm
We illustrate Proposition~\ref{prop:mscons-answer} in the context of Example~\ref{ex:selection-condition} where  $U=\{Emp, Sal, Dept\}$, $FD =\{Emp \to Sal\}$ and $T=\{es, es'd,e's\}$. It is easy to see that Algorithm~\ref{algo:chase} returns truth value $true$ for $fail$ and $\Sigma^*=\{\sigma_1, \sigma_2\}$ where $\sigma_1=(e)(ss')(d)$ and $\sigma_2=(e')(s')$. We apply Proposition~\ref{prop:mscons-answer} to the queries $Q_1$, $Q_2$, $Q_3$, $Q_4$, $Q_5$ and $Q_6$ shown below: 

\smallskip
$Q_1:$ {\tt select} $Emp, Dept$ {\tt from} $T$

$Q_2:$ {\tt select} $Emp, Sal$  {\tt from} $T$

$Q_3:$ {\tt select} $Sal$  {\tt from} $T$

$Q_4:$ {\tt select} $Emp$  {\tt from} $T$ {\tt where $Sal = s'$}

$Q_5:$ {\tt select} $Emp$  {\tt from} $T$ {\tt where $Sal = s \vee Sal = s'$}

$Q_6:$ {\tt select} $Emp$  {\tt from} $T$ {\tt where $Sal > 10K$}

\smallskip\noindent
Since $Q_1$ involves no selection condition, and since $\sigma_1$ is the only m-tuple in $\Sigma^*$ whose schema contains $Emp$ and $Dept$, the tests concern only this m-tuple. As $\tuples(\sigma_1(Emp,Dept)) = \{ed\}$, we have $\sconsans(Q_1)= \consans(Q_1)= \{ed\}$. Moreover, as $\Sigma_{Q_1}(ed)=\{ed\}$, $\wconsans(Q_1)=\{ed\}$ and $\msconsans(Q_1)=\emptyset$, because  $|\tuples(\sigma_1)|=2$.

For $Q_2$, as $\tuples(\sigma_1(Emp,Sal))=\{es, es'\}$ and $\tuples(\sigma_2(Emp,Sal))=\{e's'\}$, it is immediate to see that $\sconsans(Q_2)=\consans(Q_2)=\wconsans(Q_2)=\{e's'\}$.  Moreover, as $e's'$ is a maximal consistent and true tuple, we also have $\msconsans(Q_2)=\{e's'\}$. In the case of $Q_3$, it is easily seen that the four answers are empty because $\tuples(\sigma_1(Sal))=\{s,s'\}$.

Considering $Q_4$, as  $\tuples(\sigma_1(Emp))=\{e\}$ and  $\tuples(\sigma_2(Emp))=\{e'\}$, and thus, these two values satisfy statement (a) in the four items in Proposition~\ref{prop:mscons-answer}. As $Sat(\Gamma_4)=\{s'\}$, it is easy to see that $\Sigma_{Q_4}(e)=\{\sigma_1\}$ and $\Sigma_{Q_4}(e')=\{\sigma_2\}$. Thus, $\wconsans(Q_4)=\{e,e'\}$. Then, as $|\tuples(\sigma_1(Emp,Sal))|=2$, $e$ does not belong to $\msconsans(Q_4)$ nor to $\sconsans(Q_4)$. Since $\tuples(\sigma_1(Sal)) = \{s,s'\}$, $e$ is not in $\consans(Q_4)$ either, because statement (3.b) is not satisfied (as  $s$ is not in 
$Sat(\Gamma_4)$). On the other hand is is easy to see that $\sigma_2$ satisfies statements (b), entailing that we have $\msconsans(Q_4)=\sconsans(Q_4)=\consans(Q_4)=\{e'\}$.

Concerning $Q_5$, the two candidate $Emp$-values are again $e$ and $e'$. As $Sat(\Gamma_5)=\{s,s'\}$, $\Sigma_{Q_5}(e)=\{\sigma_1\}$ and $\Sigma_{Q_5}(e')=\{\sigma_2\}$, thus $\wconsans(Q_5)=\{e,e'\}$. As for $Q_4$, $e$ does not belong to $\msconsans(Q_5)$ and $\sconsans(Q_5)$ because we have $|\tuples(\sigma_1(Emp,Sal))|=2$. However, $\sigma_1$ satisfies item (3.b) in Proposition~\ref{prop:mscons-answer} because $\tuples(\sigma_1(Sal))$ is indeed a sub-set of $Sat(\Gamma_5)$. Thus, $e$ belongs to $\consans(Q_5)$. The case of $e'$ and $\sigma_2$ being similar to the previous query, we obtain $\msconsans(Q_5)=\sconsans(Q_5)=\{e'\}$ and $\consans(Q_5)=\{e,e'\}$.

Regarding $Q_6$, if $s=15K$ and $s'=20K$, $\msconsans(Q_6)=\sconsans(Q_6)=\{e'\}$ and $\consans(Q_6)=\wconsans(Q_6)=\{e,e'\}$ due to arguments similar to the previous case, and if $s=5K$ and $s'=20K$, we have $\msconsans(Q_6)=\sconsans(Q_6)=\consans(Q_6)=\{e'\}$ and $\wconsans(Q_6)=\{e,e'\}$, due to arguments similar to the case of $Q_4$ above.
\hfill$\Box$}
\end{example}

\algsetup{indent=1.5em}
\begin{algorithm}[th]
\caption{Consistent answers \label{algo:answer}}
{\footnotesize
\begin{algorithmic}[1]
\REQUIRE
A query $Q: {\tt select}$ $X$  {\tt from} $T$ {\tt [where $\Gamma$]} and $\Sigma^*$

\ENSURE The sets $\msconsans(Q)$, $\sconsans(Q)$, $\consans(Q)$ and $\wconsans(Q)$
\STATE{$\msconsans(Q):= \emptyset$ ; $\sconsans(Q):= \emptyset$ ; $\consans(Q):= \emptyset$ ; $\wconsans(Q) := \emptyset$}
\STATE{$ToRemove\_X := \emptyset$ ;  $Candidate\_SC := \emptyset$ ; $ToRemove\_SC:=\emptyset$}
\FORALL{$\sigma$ in $\Sigma^*$}\label{line:loop1}
	\IF{$X \subseteq sch(\sigma)$}\label{test:X}
		\IF{$|tuples(\sigma(X))| > 1$}\label{line:test0}
			\STATE{$ToRemove\_X :=ToRemove\_X \cup \tuples(\sigma(X))$}
		\ELSE
			\STATE{Let $x$ denote the unique tuple in $\sigma(X)$}
   			 \IF{$sch(\Gamma) \subseteq sch(\sigma)$}\label{line:testX}
			 	\IF{$\tuples(\sigma(sch(\Gamma))) \cap Sat(\Gamma) \ne \emptyset$}\label{line:test-Gamma-wcons}
					\STATE{$\wconsans(Q) := \wconsans(Q) \cup \{x\}$}\label{line:insert-wconsans}
                		    		\IF{$|\tuples(\sigma)| = 1$}\label{line:test-Gamma-mscons}
				 		\STATE{$\msconsans(Q) := \msconsans(Q) \cup \{x\}$}\label{line:insert-mscons}
						\STATE{$\sconsans(Q) := \sconsans(Q) \cup \{x\}$}\label{line:insert-scons}
						\STATE{$\consans(Q) := \consans(Q) \cup \{x\}$}\label{line:insert-cons}
					\ELSE
						\IF{$|\tuples(\sigma(sch(\Gamma)))| =1$}\label{line:test-Gamma-scons}
							\STATE{$Candidate\_SC := Candidate\_SC \cup \tuples(\sigma(sch(Q)))$}\label{line:insert-candidate}
							\STATE{$\consans(Q) := \consans(Q) \cup \{x\}$}\label{line:insert-cons-bis}
						\ELSE
							\STATE{$ToRemove\_SC := ToRemove\_SC \cup \tuples(\sigma(sch(Q)))$}\label{line:remove-sc}
							\IF{$\tuples(\sigma(sch(\Gamma))) \subseteq Sat(\Gamma)$}\label{line:test-Gamma}
	 			       				\STATE{$\consans(Q) := \consans(Q) \cup \{x\}$}\label{line:insert-cons-ter}
							\ENDIF		
						\ENDIF
					\ENDIF
				\ENDIF
			\ENDIF
		\ENDIF
	\ENDIF
\ENDFOR\label{line:end-loop1}
\STATE{$\msconsans(Q) := \msconsans(Q) \setminus ToRemove\_X$}\label{line:removal-ms}
\STATE{$\sconsans(Q) := \sconsans(Q) \cup \{x~|~(\exists q \in Candidate\_SC \setminus ToRemove\_SC)(q.X=x)\}$}\label{line:add-s}
\STATE{$\sconsans(Q) := \sconsans(Q) \setminus ToRemove\_X$}\label{line:removal-s}
\STATE{$\consans(Q) := \consans(Q) \setminus ToRemove\_X$}\label{line:removal}
\STATE{$\wconsans(Q) := \wconsans(Q) \setminus ToRemove\_X$}\label{line:removal-w}
\RETURN{$\msconsans(Q)$, $\sconsans(Q)$, $\consans(Q)$, $\wconsans(Q)$}
\end{algorithmic}
}
\end{algorithm}

\noindent
Based on Proposition~\ref{prop:mscons-answer}, it can be seen that, given a query $Q$, Algorithm~\ref{algo:answer} computes the consistent answer to $Q$. Indeed, the test line~\ref{test:X} discards all m-tuples of $\Sigma^*$ whose schema  does not contain $X$, because such m-tuples have no chance to provide an $X$-value. Then, the items in Proposition~\ref{prop:mscons-answer} are checked as follows:
\begin{itemize}
\item
The test line~\ref{line:test0} allows to store in the set $ToRemove\_X$ all $X$-values of m-tuples $\sigma$ such that $\tuples(\sigma(X))$ contains more than one tuple. All these tuples being inconsistent, are then removed from the set of retrieved $X$-values in $\msconsans(Q)$ (line~\ref{line:removal-ms}), in $\sconsans(Q)$ (line~\ref{line:removal-s}),  in $\consans(Q)$ (line~\ref{line:removal}) and in $\wconsans(Q)$ (line~\ref{line:removal-w}). Therefore, the statements (1.a), (2.a), (3.a) and (4.a) in Proposition~\ref{prop:mscons-answer} are satisfied by the returned sets. In what follows, we consider that the test line~\ref{line:test0} fails, that is that $\tuples(\sigma(X))=\{x\}$.
\item
Statement (4.b) is concerned when the tests lines~\ref{line:testX}  and \ref{line:test-Gamma-wcons} succeed. Indeed, in this case, there exists a tuple $s$ in $\tuples(\sigma(sch(\Gamma)))$ such that $s$ is in $Sat(\Gamma)$, meaning that $\sigma$ is in $\Sigma_Q(x)$. Therefore, in this case, $x$ is inserted in $\wconsans(Q)$ (line~\ref{line:insert-wconsans}).
\item
Regarding the statement (1.b) in Proposition~\ref{prop:mscons-answer}, on line~\ref{line:testX} it is checked that the schema  of $\Gamma$ is a sub-set of that of $\sigma$. If yes, when the tests lines~\ref{line:test-Gamma-wcons} and \ref{line:test-Gamma-mscons} succeed, $\sigma$ is in fact a tuple $t$ satisfying $\Gamma$. As moreover, $\Sigma^*$ is not redundant, $t$ occurs nowhere else in $\Sigma^*$, meaning by Proposition~\ref{prop:chase}(3) that $t$ is in $\cons(\mathcal{T})$. Hence, $x$ is inserted in $\msconsans(Q)$ (line~\ref{line:insert-mscons}). Notice that in this case, the inclusions stated in Proposition~\ref{prop:comp-cons-ans} justify that $x$ be also inserted in  $\sconsans(Q)$ (line~\ref{line:insert-scons}) and in $\consans(Q)$ (line~\ref{line:insert-cons}).
\item
As for statement (2.b), assuming that the test line~\ref{line:testX} succeeds, the case when the test line~\ref{line:test-Gamma-mscons} succeeds is as above. Now, if this test fails, then, if the test line~\ref{line:test-Gamma-scons} succeeds, then $\sigma(sch(Q))$ might be a consistent true tuple $t$ satisfying $\Gamma$, unless $t$ occurs in some other $\tuples(\sigma'(sch(Q)))$ such that $|\tuples(\sigma'(sch(Q)))| > 1$, meaning that $t$ is inconsistent. This is why $t$ is inserted in the set $Candidate\_SC$ when the test line~\ref{line:test-Gamma-scons} succeeds, and when this test fails, the tuples in the set $\tuples(\sigma(sch(Q)))$ are stored in $ToRemove\_SC$. Proposition~\ref{prop:chase}(3) then ensures that the tuples in  $Candidate\_SC \setminus ToRemove\_SC$ are indeed consistent true tuples satisfying $\Gamma$ (other than maximal ones earlier detected). The statement line~\ref{line:add-s} adds the corresponding $X$-values in the set $\sconsans(Q)$.
\item
Considering now statement (3.b)  when the test line~\ref{line:testX} succeeds, we first notice that, thanks to  Proposition~\ref{prop:comp-cons-ans}, the insertions in $\consans(Q)$ in statements lines~\ref{line:insert-cons} and \ref{line:insert-cons-bis} are correct. Moreover, when the test line~\ref{line:test-Gamma} succeeds, the condition in the statement (3.b) is satisfied, and so, $x$ is inserted in $\consans(Q)$ as stated line~\ref{line:insert-cons-ter}.
\end{itemize}
We illustrate Algorithm~\ref{algo:answer} using the queries $Q_3$ and $Q_4$ of Example~\ref{ex:queries}, respectively defined by $Q_3:$ {\tt select} $Sal$  {\tt from} $T$ and
$Q_5:$ {\tt select} $Emp$  {\tt from} $T$ {\tt where $Sal = s \vee Sal = s'$}.

Regarding $Q_3$, when considering $\sigma_1$, $ToRemove\_X$ is set to $\{s,s'\}$ because the test line~\ref{line:test0} succeeds. Then, with $\sigma_2$,  the test line~\ref{line:test0} fails and the test line~\ref{line:test-Gamma-wcons} succeeds, entailing that $s'$ is inserted in $\wconsans(Q_3)$. Then, as  the test line~\ref{line:test-Gamma-mscons} succeeds, $s'$ is inserted in $\msconsans(Q)$ (line~\ref{line:insert-mscons}), $\sconsans(Q)$ (line~\ref{line:insert-scons}) and in $\consans(Q)$ (line~\ref{line:insert-cons}). When processing the statements lines~\ref{line:removal-ms}, \ref{line:removal-s} and \ref{line:removal}, $s'$ is removed, thus producing empty answers, as expected. 

As for $Q_5$, when processing $\sigma_1$, the test line~\ref{line:test0} fails for $\sigma_1$ and  $\sigma_2$ because $|\tuples(\sigma_i(Emp))|=1$ ($i=1,2$). Thus $ToRemove\_X$ remains empty in this case. When processing $\sigma_1$, the tests line~\ref{line:test-Gamma-mscons} and line~\ref{line:test-Gamma-scons} fail (because $|\tuples(\sigma_1)|=2$ and $|\tuples(\sigma_1(Sal))|=2$) and the test line~\ref{line:test-Gamma} succeeds  (because $\tuples(\sigma_1(Sal))=Sat(\Gamma_5)=\{s,s'\}$). Thus, $ToRemove\_SC$ is set to $\{es,es'\}$ line~\ref{line:remove-sc} and $\consans(Q)$ is set to $\{e\}$ line~\ref{line:insert-cons-ter}. Then, when processing $\sigma_2$, all tests succeed (because $|\tuples(\sigma_2)|=|\tuples(\sigma_2(Sal))|=1$ and $\tuples(\sigma_2(Sal))=\{s'\}$). Thus, $e'$ is inserted in $\wconsans(Q_5)$, $\msconsans(Q_5)$, $\sconsans(Q_5)$ and $\consans(Q_5)$. The loop line~\ref{line:loop1} returns $ToRemove\_X=\emptyset$, $Candidate\_SC = \emptyset$, $ToRemove\_SC=\emptyset$, along with $\wconsans(Q_5)= \msconsans(Q_5)=\sconsans(Q_5)= \consans(Q_5)=\{e,e'\}$, which is not changed by the statements lines~\ref{line:removal-ms}-\ref{line:removal-w}. We thus obtain all four consistent answers equal to $\{e,e'\}$, as expected.

In the  next example, we further  illustrate the role of the sets $Candidate\_SC$ and $ToRemove\_SC$ in Algorithm~\ref{algo:answer}.
\begin{example}
{\rm
Let  $U=\{A,B,C,D\}$ over which the functional dependencies $C \to B$ and $C \to D$ are assumed. For $T=\{abcd,abcd',abc',ab'c'\}$, Algorithm~\ref{algo:chase} returns $\Sigma^*=\{(a)(b)(c)(dd'),$ $(a)(bb')(c')\}$, and the processing of $\sconsans(Q)$ for $Q : {\tt select }~A~{\tt from}~T{\tt ~where~}B = b$, is as follows:
\begin{itemize}
\item
For $\sigma=(a)(b)(c)(dd')$, the tests line~\ref{test:X} and  line~\ref{line:test-Gamma-mscons} fail (as $|\tuples(\sigma(A))|=1$ and $|\tuples(\sigma)|=2$), whereas the test line~\ref{line:test-Gamma-wcons} succeeds (because $Sat(\Gamma)=\{b\}$). Thus $ToRemove\_X$ remains empty and as the test line~\ref{line:test-Gamma-scons} succeeds (because $\tuples(\sigma(B))=\{b\}$), $ab$ is inserted in $Candidate\_SC$ on line~\ref{line:insert-candidate}, and $\sconsans(Q)$ remains empty.
\item
For $\sigma=(a)(bb')(c')$, as the test line~\ref{test:X} fails, $ToRemove\_X$ remains empty, and as the tests line~\ref{line:test-Gamma-mscons} and line~\ref{line:test-Gamma-scons} also fail, $ab$ and $ab'$ are inserted in $ToRemove\_SC$ due to the statement line~\ref{line:remove-sc}.
\end{itemize}
Thus, the loop line~\ref{line:loop1} generates $ToRemove\_X=\emptyset$, $Candidate\_SC=\{ab\}$ and $ToRemove\_SC=\{ab,ab'\}$. Therefore, when processing the statement line~\ref{line:add-s}, $\sconsans(Q)$ remains empty (because $Candidate\_SC \setminus ToRemove\_SC=\emptyset$), as  expected based on Definition~\ref{def:cons-answer}(2), because $ab$ is inconsistent.

\smallskip
Considering now $T_1=\{abcd,abcd'\}$, Algorithm~\ref{algo:chase} returns $\Sigma^*_1=\{(a)(b)(c)(dd')\}$ and the computations are as in the first item above, implying that  the loop line~\ref{line:loop1} generates $ToRemove\_X=\emptyset$, $Candidate\_SC=\{ab\}$, $ToRemove\_SC= \emptyset$ and $\sconsans(Q)=\emptyset$. When processing the statement line~\ref{line:add-s}, $\sconsans(Q)$ is set to $\{a\}$ (because $Candidate\_SC \setminus ToRemove\_SC=\{ab\}$), as expected based on Definition~\ref{def:cons-answer}(2), because $ab$ is a consistent true tuple satisfying $\Gamma$.
\hfill$\Box$}
\end{example}
To assess the complexity of Algorithm~\ref{algo:answer}, we first notice that this algorithm is clearly linear in the size of $\Sigma^*$. Moreover, using the same notation as when assessing the complexity of Algorithm~\ref{algo:chase}, for $\sigma$ in $\Sigma^*$, the size of $\tuples(\sigma)$ is in $\mathcal{O}(|FD|.\delta)$, that is in  $\mathcal{O}(\delta)$, because $|FD|$ is independent from $T$. Hence, the complexity of computing the consistent answer to a query in our approach is in $\mathcal{O}(|\Sigma^*|.\delta)$. Since it has been shown that $|\Sigma^*| \leq |T|$, we state that the sizes of $\Sigma^*$ and of $T$ are of the same order. Therefore,  the complexity of computing all four consistent answers to a query in our approach is in $\mathcal{O}(|T|.\delta)$, that is {\em linear in the size of the table $T$}.

It is however important to stress that Algorithm~\ref{algo:answer} requires significant extra storage due to the sets $ToRemove\_X$, $Candidate\_SC$ and $ToRemove\_SC$. It should be noticed that the storage of $ToRemove\_X$ and $ToRemove\_SC$ can be avoided by implementing a second scan of $\Sigma^*$ whose role would be to detect the $X$-values to be removed (when $|\tuples(\sigma(X))| >1$) and the tuples in $Candidate\_SC$ to be removed (when $|\tuples(\sigma(sch(Q)))| >1$). As usual, this option saves storage at the cost of further processing, namely an additional scan of $\Sigma^*$.

\section{Related Work}\label{sec:rel-work}
The problem of consistent query answering has attracted considerable attention in the last two decades \cite{ArenasBC99} and continues to be an active area of research today \cite{Bertossi21}. The approach by repairs is one that has attracted most attention in recent years and thus has been the subject of many papers, whose overview lies outside the scope of this paper. We refer to \cite{BertossiC03,Wijsen19} for detailed surveys covering the topic. The fact that ``deciding whether a candidate answer is a consistent answer is most commonly intractable in data complexity'' \cite{CalauttiLPS22} generated several research efforts recently focusing on approximate consistent answers \cite{CalauttiGMT22,GeertsPW15}, and on issues related to the counting of repairs \cite{CalauttiLPS22,LivshitsKW21}.

Our approach to consistent query answering is radically different as it does not rely on repairs but rather on set theoretic semantics for tuples and functional dependencies. Consequently, to compare our approach with repair-based approaches, we focus only on the basic differences between the two approaches. Our goal is to show that although the problem tackled is the same, our approach is hardly comparable with those from the literature (see Section~\ref{sec:comparison}). We then introduce in Section~\ref{sec:strong-repairs} the notion of {\em strong repairs} according to which some comparison is possible, and in Section~\ref{sec:JIIS} we compare the approach presented in this paper and the approach in our earlier work \cite{JIIS}.
\subsection{Comparison with Repair-Based Approaches}\label{sec:comparison}
We first point out that repair-based approaches assume a table $T$ with no nulls.  Therefore to compare our approach to repair-based approaches we need to restrict our attention to tables with functional dependencies but {\em without} nulls. 

In this context, we identify and discuss three fundamental differences between our approach and the approach by repairs, namely : (D1) the two approaches use different universes of discourse, (D2) the two approaches use different assumptions when evaluating queries, and (D3) the two approaches use functional dependencies for different purposes. 

\smallskip\noindent
{\bf (D1) {\em different universes of discourse.}}\\We recall from Section~\ref{sec:semantics} that the universe of discourse of a table $T$ is the space in which queries are evaluated. The universe of discourse in our approach, denoted by $\mathcal{T}$, is the set of all tuples one can define using constants appearing in $T$. On the other hand, the universe of discourse in the repairs approach, that we shall denote by $T_r$, is the set of all tuples of $T$ together with all their sub-tuples. For example, if $T= \{ab, a'b'\}$ then $\mathcal{T} = \{ab, a'b', a'b, ab', a, b, a', b'\}$ whereas $T_r= \{ab, a'b', a, b, a', b'\}$.

\smallskip\noindent
{\bf (D2) {\em different assumptions when evaluating queries.}}\\
The repairs approach uses the Closed World Assumption \cite{REI} that is: 
\begin{itemize}
\item[]
the only true tuples are those in $T_r$ (implying that tuples not in $T_r$ are false) 
\end{itemize}
whereas our approach uses the Open World Assumption that is:
\begin{itemize}
\item[]
the only true tuples are those in $T$ together with all tuples derived using the functional dependencies 
(implying that all other tuples in $\mathcal{T}$ are false). 
\end{itemize}
It follows that if a tuple $t$ of $T_r$ is true then $t$ is true in $\mathcal{T}$, that is $True(T_r) \subseteq \true(\mathcal{T})$, while the opposite does not hold.

For example, consider the table $T=\{abc, ab'c'\}$ with functional dependency $A \to B$. In this case, $ac$ is in $\mathcal{T}$ and in $\true(\mathcal{T})$, and $ac$ is in $True(T_r)$ as well. On the other hand, since $\Sigma^*=\{(a)(bb')(c),$ $(a)(bb')(c')\}$, the tuple $abc'$ is in $\mathcal{T}$ but {\em not} in $T_r$. Thus $abc'$ is not in $True(T_r)$, whereas $abc'$ is in $\true(\mathcal{T})$. 

Regarding inconsistent tuples, the comparison not as easy since, to the best of our knowledge, the notion of inconsistent tuple is not explicitly defined in repair-based approaches. However, it is clear that a table $T$ is inconsistent if and only if there exists a functional dependency $X \to A$ such that the two tuples $xa$ and $xa'$ in $\mathcal{T}(XA)$ occur in $T$. The pair $(xa,xa')$ is usually called a {\em conflicting pair}, and we define inconsistent tuples in this context as follows:

\begin{itemize}
\item[]
$t$ is inconsistent  if there exists a conflicting pair $(xa, xa')$ such that $t$ is a super-tuple of $xa$.
\end{itemize}
Denoting the set of these tuples by $Conf\_Inc(T_r)$, we argue that $Conf\_Inc(T_r) \subseteq \inc(\mathcal{T})$ whereas the inverse inclusion does not always hold. Indeed, based on Definition~\ref{def:incons-t}, it is easy to see that, for every conflicting pair $(xa, xa')$, the tuples $xa$, $xa'$ and all their true  super-tuples are in $\inc(\mathcal{T})$, thus showing the inclusion $Conf\_Inc(T_r) \subseteq \inc(\mathcal{T})$. The previous example with $T=\{abc, ab'c'\}$ and $A \to B$, shows that the inverse inclusion does not always hold, because $b$ is clearly not in $Conf\_Inc(T_r)$, whereas Definition~\ref{def:incons-t} implies that $b$ is in $\inc(\mathcal{T})$.

\smallskip\noindent
{\bf (D3) {\em different purposes in the use of functional dependencies.}}\\
The repair-based approaches use functional dependencies as a means for checking the consistency of $T$ (i.e., as a means of inference of inconsistent tuples). In our approach, we also use the functional dependencies as a means of inference of inconsistent tuples. However, additionally, we use the functional dependencies as a means of inference of {\em true} tuples. For example, if $T=\{abc, ab'c'\}$ and $FD=\{A \to B\}$, our approach allows us to infer that $abc$ is true and inconsistent, as repair-based approaches do. However, additionally, our approach allows to infer that the tuple $ab'c$ is also true and inconsistent, which  repair-based approaches fail to do since $ab'c$ is not even in $T_r$.

\smallskip
Now that we have stated the differences in the ways the two approaches operate, consider a query $Q:~{\tt select}~X~{\tt from}$ $T$ ${\tt where}~\Gamma$, and recall that in repair-based approaches, the consistent answer to $Q$ is defined as the set of true tuples $x$ in $T_r$ such that, every repair contains a tuple satisfying the condition $\Gamma$ and whose restriction over $X$ is $x$. Denoting this set by $Rep\_ans(Q)$, we recall that, by Definition~\ref{prop:mscons-answer}, all consistent answers to $Q$ in our approach are sets of true and consistent tuples in $\mathcal{T}(X)$. It is thus not surprising that the two approaches to consistent query answering are not comparable.

\smallskip
Summarising our discussion so far we cannot claim that either of the two approaches is better or preferable than the other. They are simply different as they have different universes of discourse, they operate under different assumptions and they use the functional dependencies in different ways. 
 
What we can reasonably claim in favor of our approach is that: (a) it provides a formal framework based on which different kinds of consistent answers can be defined, (b) it offers a polynomial algorithm for evaluating the consistent answer to a query, whether conjunctive or disjunctive, and (c) by allowing nulls in the table, our approach broadens the angle of its application to other important areas of research as explained in the concluding Section~\ref{sec:conclusion}.

\subsection{Strong Repairs}\label{sec:strong-repairs} 
As highlighted in the previous section, repair approaches do not compare to our approach. Nevertheless, in this section, we propose a modified version of the notion of repair according to which some comparison is possible.

First we recall that given a table $T$ without nulls and a set $FD$ of functional dependencies over $T$, a repair of $T$ is defined to be a maximal consistent sub-set of $T$, where maximality is understood with respect to set-theoretic inclusion. For example given $T=\{eda,eda', e'd'a\}$ and $FD =\{Emp \to Dept, Dept \to Addr\}$ as in our introductory example, we obtain two repairs $R_1 =\{eda,e'd'a\}$ and $R_2=\{eda',e'd'a\}$ and so the consistent answer to $Q:~{\sf select}~Emp$ ${\sf from}~T~{\sf where}~Addr=a$ is $\{e,e'\} \cap \{e'\}$, that is $\{e'\}$. We refer to \cite{Bertossi2011} for a theoretical account on how to compute  such consistent query answers when $T$ is a table {\em without} nulls. 

However, when the table $T$ contains nulls then the question is: which table should be repaired? In other words: what is the definition of a repair in this case? We illustrate this point in the following example.
\begin{example}\label{ex:repair}
{\rm
Referring to Examples~\ref{ex:inc-constraint} and \ref{ex:inc-constraint-i*}, let $T = \{ab, bc, ac'\}$ and $FD=\{A \to C, B \to C\}$. Considering that $T$ shows no explicit violation of functional dependencies, repairing $T$ is trivial, and thus, the consistent answer to $Q:~{\sf select}~A,C$ ${\sf from}~T$ would be $\{ac'\}$. However, according to approaches related to universal relation \cite{Spyratos87,Ullman}, the table $T$ has to be chased before computing the answer to $Q$. In this case, the traditional chase algorithm fails because of $A \to C$ and the presence of $abc$ and $ac'$ in the chased table. Consequently, the query cannot be answered.

We cope with this difficulty in our approach by computing $\Sigma^* =\{(a)(b)(cc')\}$, and thus, showing the inconsistency. In this context, it is possible to define $R_1 =\{abc\}$ and $R_2=\{abc'\}$ as the two repairs of $T$ and then, to state that the consistent answer to $Q$ is empty. The remainder of the section elaborates on this idea and compares it to our approach explained earlier.}
\hfill$\Box$
\end{example}
As suggested by Example~\ref{ex:repair}, in the presence of nulls, the table to be repaired is not $T$ itself, but the table denoted by $T^*$ that contains all tuples in $\tuples(\sigma)$ for every $\sigma$ in $\Sigma^*$. Formally, $T^*=\bigcup_{\sigma \in \Sigma^*}\tuples(\sigma)$.

\smallskip
Going one step further, as mentioned in the previous section, consistent answers to queries in our approach and in the repair-based approach using  $T^*$ are {\em not comparable}. To formalize this statement, we denote by $\repcons(T)$ the set of all tuples occurring in every repair of $T^*$, and we give examples showing that $\repcons(T)$ and $\cons({\cal T})\cap \true({\cal T})$ {\em are not comparable with respect to set-theoretic inclusion.}

\smallskip
First, for $T=\{ab, ab',bc\}$ over $U=\{A,B,C\}$ and $FD=\{A \to B\}$, we have $\Sigma^*=\{(a)(bb'),$ $(b)(c)\}$. We thus have two repairs $R_1$ and $R_2$ where $R_1=\{ab,bc\}$ and $R_2=\{ab',bc\}$. It follows that $b$ is in $\repcons(T)$ whereas it is obvious that $b$ is in $\inc({\cal T})$, thus not in $\cons(\mathcal{T}) \cap \true(\mathcal{T})$. Therefore, $\repcons(T)\subseteq \cons({\cal T}) \cap \true({\cal T})$ does not hold in general.

Now, let $U=\{A,B,C,D\}$, $FD =\{A \to C, B \to C\}$ and $T=\{acd,$ $ac'd,$ $bcd',$ $bc'd',$ $abc,$ $abc'\}$, in which case $\Sigma^*=\{(a)(c,c')(d),$ $(b)(c,c')(d'),$ $(a)(b)(c,c')\}$. Thus, we obtain the four repairs $R_1=\{acd,$ $bcd',$ $abc\}$, $R_2=\{ac'd,$ $bc'd',$ $abc'\}$, $R_3=\{acd, bc'd'\}$ and $R_4=\{ac'd, bcd'\}$. In this case, $ab$ is clearly {\em not} in $\repcons(T)$, whereas, by Proposition~\ref{prop:chase}(3), $ab$ belongs to $\cons({\cal T}) \cap \true({\cal T})$. It therefore turns out that $\cons({\cal T}) \cap \true({\cal T}) \subseteq \repcons(T)$ does not hold in general.

\smallskip
The last counter-example suggests that, in repair-based approaches, an m-tuple in $\Sigma^*$ might have no `representatives' in some repairs (e.g., no tuple of $\tuples((a)(b)(cc'))$ is present in $R_3$ or in $R_4$). This means that the repairs $R_3$ or $R_4$ do {\em not} convey the information that, in any repair $abc$ or $abc'$ must be true, as stated by $\Sigma^*$. In order to account for this remark, we restrict repairs so as, for every $\sigma$ in $\Sigma^*$, every repair contains at least one tuple of $\tuples(\sigma)$.
\begin{definition}\label{def:strong-repairs}
Given a table $T$ over $U$, $FD$ the associated set of functional dependencies, and $\Sigma^*$ the corresponding m-table, a table $R$ over $U$ is a {\em strong repair} of $T$ if $R$ is a repair of $T^*$, and if for every $\sigma$ in $\Sigma^*$, $R \cap \tuples(\sigma)$ is nonempty. The set of all strong repairs of $T$ is denoted by $\srep(T)$.
\hfill$\Box$
\end{definition}
In the case of the last counter-example above, $R_1$ and $R_2$ are in $\srep(T)$ whereas $R_3$ and $R_4$ are not. On the other hand, since in this case $\cons({\cal T}) \cap \true({\cal T})=\{ab,$ $ad,$ $bd',$ $a$, $b,$ $d,$ $d'\}$, it should be observed that all tuples in $\cons({\cal T}) \cap \true({\cal T})$ occur in $R_1$ and in $R_2$. To show that this property always holds, we introduce the following terminology:
\begin{itemize}
\item
The {\em strong-repair-based consistent answer} to $Q$, denoted by $\srepans(Q)$, is defined as the intersection of all answers to $Q$ in every strong repair of $T$. Formally, if  $\ans(Q_{[R]})$ denotes the answer to $Q$ in $R$, $\srepans(Q) = \bigcap_{R \in \srep(T)} \ans(Q_{[R]})$.
\item
The set of all {\em strong-repair-based consistent tuples}, denoted by $\srepcons(T)$, is defined as the set of all tuples occurring in every strong repair.\end{itemize}
Before showing the main result of this sub-section to compare $\cons({\cal T})\cap \true({\cal T})$ and $\srepcons(T)$, we show the following basic lemma.
\begin{lemma}\label{lemma:strong-repairs}
Given a table $T$ over $U$ and the associated set of functional dependencies $FD$, for every $t$ in $T^*$, there exists $R$ in $\srep(T)$ such that $t$ is in $R$.
\end{lemma}
{\sc Proof.}
The repair $R$ considered here is built up as follows, based on the m-tuples of $\Sigma^*$. First let $\sigma_t$ be an m-tuple of $\Sigma^*$ such that $t \in \tuples(\sigma_t)$. For every $A$ in $sch(t)$ and every m-tuple $\sigma$ such that $A$ is in $sch(\sigma)$ and $t.A$ in $\tuples(\sigma(A))$, mark $t.A$ in $\sigma(A)$. Notice that at this stage, all components of $\sigma_t$ have a marked value. In this case, the tuple consisting of these marked values is inserted in $R$ and the process is iterated until there remains no unmarked components in any m-tuple of $\Sigma^*$. That is, at each step, a `non-fully marked' m-tuple in $\Sigma^*$ is chosen and processed similarly to $t$. When the process terminates, every component of every $\sigma$ has one value marked, which defines a tuple of $\tuples(\sigma)$ to be inserted in $R$.

We now prove that $R$ satisfies all functional dependencies in $FD$. If not, let $t$ and $t'$ in $R$ and $X \to A$ in $FD$ such that $t.X=t'X$ and $t.A \ne t'.A$. It follows that there exist $\sigma$ and $\sigma'$ in $\Sigma^*$ such that $t\in \tuples(\sigma)$ and $t'\in \tuples(\sigma')$, $t.X \in \tuples(\sigma(X)) \cap \tuples(\sigma'(X))$. Hence, by construction of $\Sigma^*$, we have $\sigma(A)=\sigma'(A)$, and thus, by construction of $R$, it must be that $t.A=t'.A$, which is a contradiction. Thus $R$ satisfies $FD$, and to end the proof, we notice that, if $R$ is not maximal with respect to set-theoretic inclusion, tuples from $T^*$ respecting the functional dependencies are added, thus leading to a strong repair containing $t$. The proof is therefore complete.
\hfill$\Box$

\smallskip\noindent
As a case of non-maximality of $R$ in the proof above, consider $\Sigma^*=\{(aa')(bb')\}$, $FD=\{A \to B, B\to A\}$ and $t=ab$. Then $a$ and $b$ are marked and the process stops producing $\{ab\}$ which is not maximal since $\{ab, a'b'\}$ is a repair of $T^*$.

It is important to notice that, as a consequence of Lemma~\ref{lemma:strong-repairs}, the set $\srep(T)$ can not be empty, when $T$ is nonempty. Moreover, given a selection condition $\Gamma$, Lemma~\ref{lemma:strong-repairs} implies that if $\Sigma^*$ contains an m-tuple $\sigma$ such that every $t$ in $\tuples(\sigma)$ satisfies $\Gamma$, then every strong repair of $T$ contains a tuple satisfying $\Gamma$.

The following proposition shows that consistent true tuples are consistent with respect to strong repairs, that is every consistent true tuple occurs in every strong repair.
\begin{proposition}\label{prop:strong-repairs}
Let $T$ be a table over $U$ and $FD$ its associated set functional dependencies. Then $\cons({\cal T})\cap \true({\cal T}) \subseteq \srepcons(T)$ holds.

Moreover, every $t$ in $\srepcons(T)$ but not in $\cons({\cal T})\cap \true({\cal T})$ is in $\inc(\mathcal{T})$.
\end{proposition}
{\sc Proof.}
Let $t$ be in $\cons({\cal T})\cap \true({\cal T})$. By Proposition~\ref{prop:chase}(3), for every $\sigma$ in $\Sigma^*$ such that $\tuples(\sigma(sch(t)))$ contains $t$, we have $\tuples(\sigma(sch(t))) = \{t\}$. Thus, for every $q$ in $\tuples(\sigma)$, $q.sch(t)=t$, and so, as every $R$ in $\srep(T)$ contains at least one tuple of $\tuples(\sigma)$, $t$ occurs in $R$, meaning that $t$ belongs to $\srepcons(T)$.

Assume now that $t$ is in $\srepcons(T)$ but not in $\cons({\cal T})\cap \true({\cal T})$. Since $t$ is in $\srepcons(T)$, then there exists $\sigma$ in $\Sigma^*$ such that $t$ is in $\tuples(\sigma(sch(t)))$. Thus, by Proposition~\ref{prop:chase}(1), $t$ is in $\true({\cal T})$. Since on the other hand, $t$ is not in $\cons({\cal T})\cap \true({\cal T})$, $t$ is in $\inc(\mathcal{T})$, and the proof is complete.
\hfill$\Box$
\begin{corollary}\label{coro:strong-repairs}
Let $T$ be a table over $U$ and $FD$ its associated set of functional dependencies. For every query $Q$, $\consans(Q) \subseteq \srepans(Q)$ holds.
\end{corollary}
{\sc Proof.}
Given $x$ in $\consans(Q)$, by  Proposition~\ref{prop:mscons-answer}, there exists $\sigma$ in $\Sigma^*$ such that $\tuples(\sigma(X))=\{x\}$ and $\tuples(\sigma(sch(\Gamma))) \subseteq Sat(\Gamma)$. As every strong repair $R$ of $T^*$ contains a tuple in $\tuples(\sigma)$, every strong repair $R$ of $T^*$ contains a tuple $q$ such that $q.X=x$ and $q.sch(\Gamma) \in Sat(\Gamma)$. Therefore, for every strong repair $R$ of $T^*$, $x$ is in the answer to $Q$ in $R$, meaning that $x$ is in $\srepans(Q)$, and the proof is complete.
\hfill$\Box$

\smallskip\noindent
As a consequence of Corollary~\ref{coro:strong-repairs} and Proposition~\ref{prop:comp-cons-ans}, the following inclusions hold for every $Q$:

\smallskip
$\msconsans(Q) \subseteq \sconsans(Q) \subseteq \consans(Q) \subseteq \srepans(Q)$.

\subsection{Comparison with the Approach in \cite{JIIS}}\label{sec:JIIS}
The approach presented in this paper and the approach in \cite{JIIS}, both rely on partition semantics \cite{Spyratos87}. However, the two approaches have important differences. First, although the limit interpretation $I^*$ as defined in Section~\ref{sec:semantics} is the same as the interpretation $\mu^*$  in \cite{JIIS}, we have proved here the uniqueness of $I^*$ which is not done for $\mu^*$. 


Second, the notions of true/false tuples and of inconsistent/consistent tuples differ between the two approaches. More precisely, since in \cite{JIIS}, truth values are defined inspired by the Four-valued logic of \cite{Belnap}, it is not possible to have tuples that are, for instance, true {\em and} consistent, as in the present work. 

Third, the universe of discourse in \cite{JIIS}, is potentially infinite as it consists of all tuples that can be built up using constants from the entire attribute domains; and this is an issue when it comes to computing the set of all inconsistent tuples. In contrast, the universe of discourse in the present work is finite as it consists of all tuples that can be built up using constants {\em only} from the active attribute domains. 

Fourth, the notion of inconsistent tuple is not defined in the same way in the two approaches. Indeed, in \cite{JIIS}, a tuple is inconsistent if its interpretation is a sub-set of the interpretations of two constants from the same domain. For example, consider the (true) tuples $xa$ and $xa'$ in the presence of $X \to A$. These tuples are inconsistent in the present approach (see Definition~\ref{def:incons-t}), as well as in the approach of \cite{JIIS}. However, the tuple $x$ is also an inconsistent tuple according to \cite{JIIS} (because its interpretation is a sub-set of those of $a$ and $a'$), but is {\em not} inconsistent in the present approach (because $A$ is not in $sch(x)$). As a consequence, the consistent answer to a query in \cite{JIIS} differs from that in the present work.

Finally,  the issue of consistent query answering has not been addressed in full details in \cite{JIIS}: it is restricted to the case where the table $T$ contains no nulls. We also note in this respect that, based on the concept of m-table as introduced in Section~\ref{sec:comp-issues}, the present approach allows to consider also {\em disjunctive} queries, an issue not addressed in \cite{JIIS}, nor in repair-based approaches, such as \cite{BertossiC03,Wijsen05}. 
\section{Concluding Remarks and Future Work}\label{sec:conclusion}
In this paper, we have presented a novel approach to consistent query answering in tables with nulls and functional dependencies. Given such a table $T$, we defined $\mathcal{T}$, the universe of discourse of $T$, and using set-theoretic semantics for tuples and functional dependencies we partitioned $\mathcal{T}$ in two orthogonal ways: into true and false tuples, on the one hand and into consistent and inconsistent tuples on the other hand. Queries are addressed to $T$ and evaluated in $\mathcal{T}$ in order to obtain consistent answers. The consistent answer to a query $Q:$ {\tt select} $X$ {\tt from} $T$ {\tt where} $Condition$ over $T$ was defined to be the set of tuples $x$ in $\mathcal{T}$ such that: (a) $x$ is a true and consistent tuple with schema $X$ and (b) there exists a true super-tuple $t$ of $x$ in $\mathcal{T}$ satisfying the condition. We have seen that, depending on the `status' of the super-tuple $t$ in $\mathcal{T}$, there are four different types of consistent answer to $Q$ and we have presented polynomial time algorithms for computing these answers. We have also seen that our approach is not comparable to the approaches based on table repairs or the approach of \cite{JIIS}, as they have different universes of discourse and operate under different assumptions. 

By allowing the presence of nulls in the table, our approach opens the way for a number of important applications. We envisage three such applications of our approach. The first is related to universal relation interfaces \cite{FaginMU82,Spyratos87,Ullman}. Given a relational database with $n$ tables $T_1, T_2, \ldots ,T_n$, over schemes $S_1, S_2, \ldots , S_n$ and sets of functional dependencies $FD_1, FD_2, \ldots ,FD_n$, respectively, the universal relation is defined to be a table $T$ defined as follows: 
\begin{itemize}
\item
The schema  of $T$ is $S=S_1 \cup S_2 \cup \ldots \cup S_n$ and the associated set of functional dependencies is $FD= FD_1 \cup FD_2 \cup \ldots \cup FD_n$.
\item
For each tuple $t$ in $T_i$, $t$ is placed in a new line of $T$ with a new tuple identifier, for all $i= 1, 2, \ldots , n$.
\end{itemize}
For example, if $n=2$, $T_1= \{ab\}$  and $T_2= \{bc\}$ then the universal relation operates from a table $T$ with schema  $ABC$ and current instance $T= \{ab, bc\}$ with two nulls, one under $C$ and one under $A$.

A prerequisite for defining the table $T$ is ``name consolidation'' that is (a) if an attribute name appears in two different database tables then it has the same meaning in the two tables and (b) if two different attribute names in the database have the same meaning then one is renamed to the other (using the renaming operation of the relational model). 

Once the table $T$ is put in place querying the interface is straightforward: the system computes a consistent answer using our polynomial Algorithm~\ref{algo:answer} and returns the result. However, updating through such an interface poses several hard problems that we are currently investigating.

 The second application that we envisage is query result explanation. As already mentioned, the m-table $\Sigma^*$ offers a way to `display' inconsistencies. We therefore plan to investigate how m-tuples in $\Sigma^*$ can be used to `explain' to the user the presence or the absence of a given tuple in a consistent answer. This issue seems particularly relevant in the context of data integration and data exchange, where it is crucial to `explain' consistent answers in reference to the sources the data come from \cite{Bertossi21}.
 
The third application that we envisage is related to data quality. Based on the two orthogonal ways of characterizing tuples, namely consistent/inconsistent and true/false, we consider that the inconsistent tuples are as good a source of information as they are the consistent tuples; and similarly, we consider that the false tuples are as good a source of information as they are the true tuples. In doing so we can develop a meaningful theory for the {\em quality of data} in a data-set (whether the data-set is a table or a query result). To this end, we plan to consider various quality measures such as percentages of consistent/inconsistent tuples in the data-set, percentages of true or false tuples, or combinations thereof. 

\bigskip\noindent
{\bf Acknowledgment}\\
Work conducted while the second author was visiting at FORTH Institute of Computer Science, Crete, Greece (https://www.ics.forth.gr/)

\bigskip\noindent
{\bf Declarations}\\
The two authors contributed to the study, conception and design. Both read and approved the manuscript.
\\
No funds, grants, or other support was received for conducting this study.


\appendix
\section{On the Church-Rosser Property of Expansion}\label{append:church-rosser}
We show that expansion has the Church-Rosser property, that is when iterating the process, the order in which the steps are preformed is irrelevant. We first recall the definition of expansion.

\smallskip\noindent
{\bf Definition~\ref{def:expand}~}
The expansion of an interpretation $I$ with respect to a functional dependency $X \to A$ and the tuple $xa$ in $\mathcal{T}(XA)$ is the interpretation $Exp(I,xa)$ defined as follows:
\begin{itemize}
\item
If $I(x) \cap I(a) \ne \emptyset$ and $I(x) \not\subseteq I(a)$ then:\\
$Exp(I,xa)(a)= I(a) \cup I(x)$, and $Exp(I,xa)(\alpha)= I(\alpha)$ for $\alpha$ in $\mathcal{AD}$ different than $a$.
\item
Otherwise, $Exp(I,xa)=I$.
\end{itemize}

\begin{figure}[htp]
    \centering
\includegraphics[width=.75\textwidth]{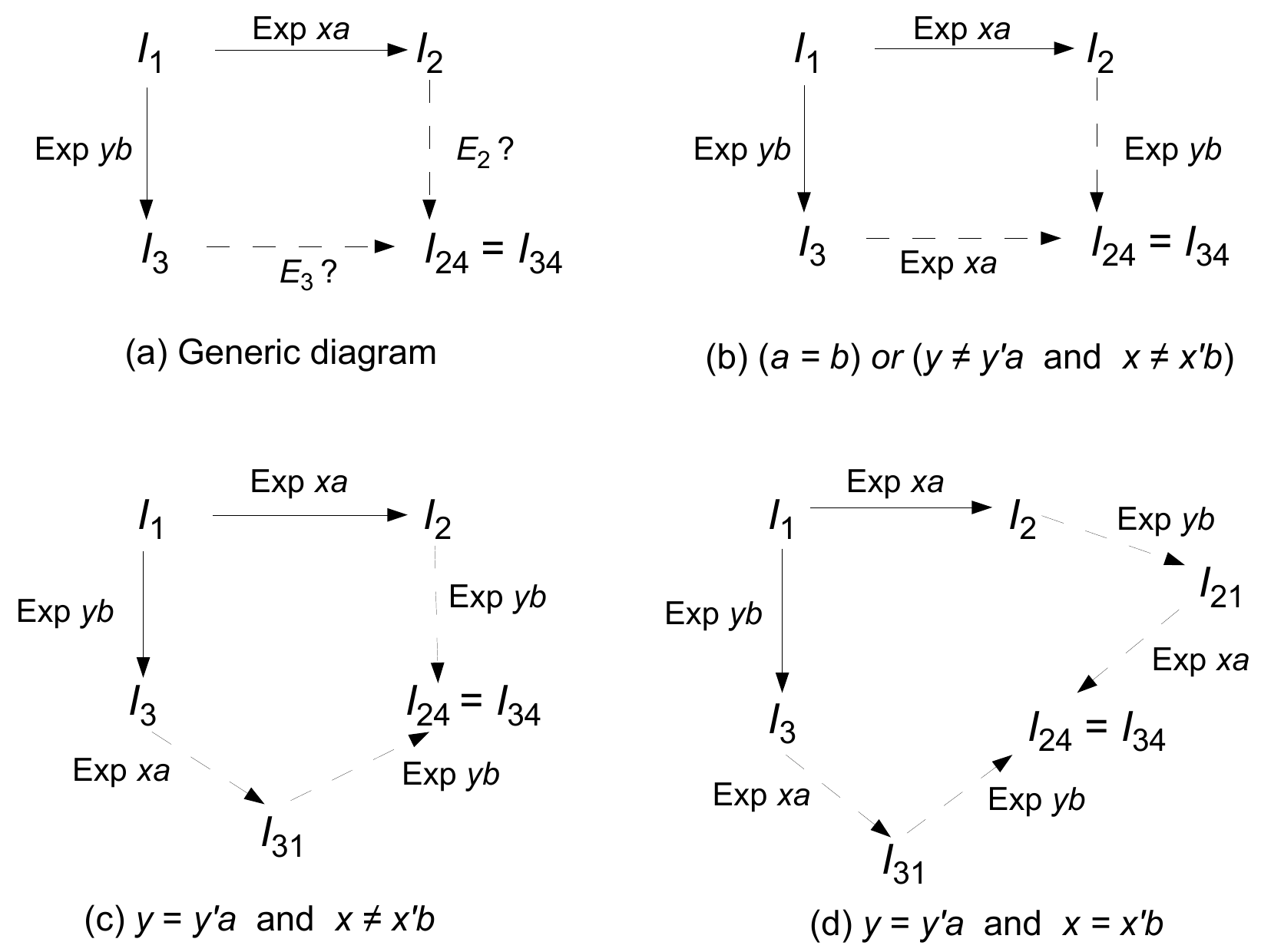}   

\caption{Expansion Diagrams}
    \label{fig-exp}
\end{figure}

\medskip\noindent
We consider two expansions one with respect to the dependency $X \to A$ and the tuple $xa$ and the other one with respect to the dependency $Y \to B$ and the tuple $yb$. Starting with an interpretation $I_1$, let $I_2= Exp(I_1, xa)$ and $I_3=Exp(I_1, yb)$. Referring to Figure~\ref{fig-exp}(a), the problem is to find sequences of expansions called $E_2$ and $E_3$ in the figure, such that $I_{24}= E_2(I_2)$ and $I_{34}=E_3(I_3)$ are equal.

It should be clear that if $I_2=I_1$ or $I_3=I_1$, then the order in which the expansions are processed is irrelevant. For example, if $I_2=I_1$ and $I_3 \ne I_1$, we obtain $I_{24}=I_{34}$ by setting $E_3= \epsilon$ (where $\epsilon$ denotes the empty sequence) and $E_2 = Exp(I_1, yb)$.

From now on, we assume that $I_2 \ne I_1$ and $I_3 \ne I_1$. Based on Definition~\ref{def:expand}, we have:
\begin{itemize}
    \item $I_2(a)=I_1(a) \cup I_1(x)$ and for every $\alpha \ne a$, $I_2(\alpha) = I_1(\alpha)$
    \item $I_3(b)=I_1(b) \cup I_1(y)$ and for every $\beta \ne b$, $I_2(\beta) = I_1(\beta)$.
\end{itemize}
As a particular case, if it is assumed that $a=b$, we have:
\begin{itemize}
    \item $I_2(a)=I_1(a) \cup I_1(x)$ and for every $\alpha \ne a$, $I_2(\alpha) = I_1(\alpha)$
    \item $I_3(a)=I_1(a) \cup I_1(y)$ and for every $\alpha \ne a$, $I_3(\alpha) = I_1(\alpha)$.
\end{itemize}
Thus, as shown in Figure~\ref{fig-exp}(b), for $E_2=Exp(I_2, ya)$ and $E_3=Exp(I_3, xa)$,  $I_{24}$ and $I_{34}$ are such that $I_{24}(a)=I_{34}(a)=I_1(a) \cup I_1(x) \cup I_1(y)$ and for every $\alpha \ne a$, $I_{24}(\alpha)=I_{34}(\alpha)=I_1(\alpha)$. We therefore have, in this case, $I_{24}=I_{34}$.

\smallskip
Now assuming that $I_2\ne I_1$ and $I_3\ne I_1$ and $a \ne b$, if $a$ does not occur in $y$ and $b$ does not occur in $x$, the two expansions are independent from each other. In this case, as illustrated in Figure~\ref{fig-exp}(b), the following lemma holds.
\begin{lemma}\label{lemma:case-1}
If $a \ne b$, if $a$ does not occur in $y$ and $b$ does not occur in $x$, for $E_2= Exp(I_2, yb)$ and $E_3= Exp(I_3, xa)$, denoting respectively by $I_{24}$ and $I_{34}$ the interpretations $Exp(I_2, yb)$ and $Exp(I_3, xa)$, we have $I_{24}=I_{34}$, that is:

\smallskip
$Exp(Exp(I_1, xa), yb)= Exp(Exp(I_1, yb), xa)$.
\end{lemma}
{\sc Proof.}
Since $a$ does not occur in $y$, we have $I_2(y)=I_1(y)$ and similarly, since $b$ does not occur in $x$, we have $I_3(x)=I_1(x)$. Therefore, for $a$ and $b$, $I_{24}=Exp(Exp(I_1,xa),yb)$ and $I_{34}=Exp(Exp(I_1,yb),xa)$ are defined by:\\
$-$ $I_{24}(a)=Exp(I_2,yb)(a)=I_2(a)=I_1(a)\cup I_1(x)$ (because $a \ne b$)\\
$-$ $I_{24}(b)=Exp(I_2,yb)(b)=I_2(b) \cup I_2(y)=I_1(b) \cup I_1(y)$\\
$-$ $I_{34}(a)=Exp(I_3,xa)(a)=I_3(a) \cup I_3(x)=I_1(a)\cup I_1(x)$\\
$-$ $I_{34}(b)=Exp(I_3,xa)(b)=I_3(b)=I_1(b) \cup I_1(y)$ (because $a \ne b$).

\smallskip
Hence, we have $I_{24}(a)=I_{34}(a)$ and $I_{24}(b)=I_{34}(b)$. Since no other symbol is changed by expansions, we have shown the commutativity property in this case, namely $I_{24}=I_{34}$.
\hfill$\Box$

\medskip\noindent
We now switch to the most general case, namely when each expansion has an impact on the other. This happens when $a$ occurs in $y$ and when $b$ occurs in $x$. Indeed, as shown in Figure~\ref{fig-exp}(d), in this case, writing $x$ and $y$ respectively as $bx'$ and $ay'$, we notice the following: $I_{2}(y) \ne I_1(y)$ because $I_2(y)=I_2(a) \cap I_2(y')$ and $I_2(a) \ne I_1(a)$, and similarly, $I_{3}(x) \ne I_1(x)$ because $I_3(x)=I_3(b) \cap I_3(x')$ and $I_3(b) \ne I_1(b)$.

An important remark regarding $x'$ and $y'$ is that neither $a$ nor $b$ occurs is these tuples, thus implying that, during the computations of the various expansions considered here, their interpretation will not change, thus remaining equal to $I_1(x')$ and $I_1(y')$, respectively. For example, $I_2(y)=I_2(a) \cap I_2(y')$ and $I_3(x)=I_3(b) \cap I_3(x')$ can be respectively written as $I_2(y)=I_2(a) \cap I_1(y')$ and $I_3(x)=I_3(b) \cap I_1(x')$. The following lemma holds in this case.
\begin{lemma}\label{lemma:case-2}
If $a \ne b$, and if $y=ay'$ and $x=bx'$, for $E_2= Exp(Exp(I_2, yb),xa)$ and $E_3= Exp(Exp(I_3, xa),yb)$, denoting respectively by $I_{24}$ and $I_{34}$ the corresponding interpretations, we have $I_{24}=I_{34}$, that is:

\smallskip
$Exp(Exp(Exp(I_1, xa), yb), xa)= Exp(Exp(Exp(I_1, yb), xa),yb)$.
\end{lemma}
{\sc Proof.}
Referring to Figure~\ref{fig-exp}(d), we first give the computations regarding $I_2$, $I_{21}$ and $I_{24}$. In each case the computations of the interpretations of $a$, $x$, $b$ and $y$ are shown.

\smallskip\noindent
$\bullet$ {\em Interpretation $I_2=Exp(I_1, xa)$.}\\
First, we have $I_2(a)=I_1(a) \cup I_1(x)$, due to expansion. Moreover, $I_2(b) = I_1(b)$ and as $I_2(y)= I_2(a) \cap I_2(y')$, we obtain $I_2(y) = (I_1(a) \cup I_1(x))\cap I_1(y')$. Since $I_1(y) = I_1(a) \cap I_1(y')$, we have $I_2(y) = I_1(y) \cup (I_1(x)\cap I_1(y'))$. Moreover, $I_2(x)=I_2(b) \cap I_2(x')$, and thus, $I_2(x)=I_1(b) \cap I_1(x')=I_1(x)$. Therefore $I_2$ is defined as shown below:

\smallskip\begin{tabular}{ll}
$I_2(a)=I_1(a) \cup I_1(x)$&\qquad$I_2(x)=I_1(x)$\\
$I_2(b) = I_1(b)$&\qquad$I_2(y) = I_1(y) \cup (I_1(x)\cap I_1(y'))$
\end{tabular}

\smallskip\noindent
$\bullet$ {\em Interpretation $I_{21}=Exp(I_2, yb)$.}\\
Now, we have $I_{21}(b)=I_2(b) \cup I_2(y)$, due to expansion. It follows that $I_{21}(b)= I_1(b) \cup I_1(y) \cup (I_1(x)\cap I_1(y'))$. Moreover, $I_{21}(a) = I_2(a)$, that is $I_{21}(a) = I_1(a) \cup I_1(x)$. As $I_{21}(x)= I_{21}(b) \cap I_{21}(x')$ and $I_{21}(x')=I_1(x')$, we obtain

\smallskip\noindent
\begin{tabular}{rcl}
$I_{21}(x)$&$=$&$(I_1(b) \cup I_1(y) \cup (I_1(x)\cap I_1(y')))\cap I_1(x')$\\
&$=$&$(I_1(b) \cap I_1(x')) \cup (I_1(y) \cap I_1(x')) \cup  (I_1(x)\cap I_1(y') \cap I_1(x'))$
\end{tabular}

\smallskip\noindent
Considering that $I_1(b) \cap I_1(x')=I_1(x)$ and that $I_1(x) \subseteq I_1(x')$ (because $x' \sqsubseteq x$), we obtain:
$I_{21}(x)=I_1(x) \cup (I_1(x') \cap I_1(y)) \cup (I_1(x) \cap I_1(y'))$. Since $(I_1(x) \cap I_1(y')) \subseteq I_1(x)$ always holds, we have: $I_{21}(x)=I_1(x) \cup (I_1(x') \cap I_1(y))$.

On the other hand, $I_{21}(y)=I_{21}(a) \cap I_{21}(y')$, and thus, $I_{21}(y)=I_2(a) \cap I_1(y')$. Therefore, $I_{21}(y)= (I_1(a) \cup I_1(x)) \cap I_1(y')$, which can be written as $I_{21}(y)= I_1(y) \cup (I_1(x) \cap I_1(y'))$. Hence $I_{21}$ is defined as shown below:

\smallskip
\begin{tabular}{l}
$I_{21}(a)= I_1(a) \cup I_1(x)$\\$I_{21}(x)=I_1(x) \cup (I_1(x') \cap I_1(y))$\\
$I_{21}(b) =  I_1(b) \cup I_1(y) \cup (I_1(x)\cap I_1(y'))$\\$I_{21}(y) = I_1(y) \cup (I_1(x) \cap I_1(y'))$
\end{tabular}

\smallskip\noindent
$\bullet$ {\em Interpretation $I_{24}=Exp(I_{21}, xa)$.}\\
Here, we have $I_{24}(a)=I_{21}(a) \cup I_{21}(x)$, due to expansion. It follows that:

\smallskip\noindent
\begin{tabular}{rcl}
$I_{24}(a)$&$=$&$(I_1(a) \cup I_1(x)) \cup (I_1(x) \cup (I_1(x') \cap I_1(y)))$\\
&$=$&$I_1(a) \cup I_1(x) \cup (I_1(x') \cap I_1(y))$\\
\end{tabular}

\smallskip\noindent
Moreover, $I_{24}(b) = I_{21}(b)$, that is $I_{24}(b) =  I_1(b) \cup I_1(y) \cup (I_1(x)\cap I_1(y'))$. As $I_{24}(x)=I_{24}(b) \cap I_{24}(x')$ and $I_{24}(x')=I_1(x')$, we have:

\smallskip\noindent
\begin{tabular}{rcl}
$I_{24}(x)$&$=$&$(I_1(b) \cup I_1(y) \cup (I_1(x)\cap I_1(y'))) \cap I_1(x')$\\
&$=$&$(I_1(b)\cap I_1(x')) \cup (I_1(y)\cap I_1(x')) \cup (I_1(x) \cap I_1(y')\cap I_1(x'))$\\
&$=$&$I_1(x) \cup (I_1(x') \cap I_1(y)) \cup (I_1(x) \cap I_1(y'))$\\
&& (because $I_1(x) \subseteq I_1(x')$ as $x' \sqsubseteq x$)\\
&$=$&$I_1(x) \cup (I_1(x') \cap I_1(y))$ (because $(I_1(x) \cap I_1(y')) \subseteq I_1(x)$)
\end{tabular}

\smallskip\noindent
As $I_{24}(y)= I_{24}(a) \cap I_{24}(y')$, we obtain

\smallskip\noindent
\begin{tabular}{rcl}
$I_{24}(y)$&$=$&$(I_1(a) \cup I_1(x) \cup (I_1(x') \cap I_1(y)))\cap I_1(y')$\\
&$=$&$(I_1(a) \cap I_1(y')) \cup (I_1(x) \cap I_1(y')) \cup  (I_1(x')\cap I_1(y) \cap I_1(y'))$\\
&$=$&$I_1(y) \cup (I_1(x) \cap I_1(y')) \cup (I_1(x') \cap I_1(y))$\\
&&(because $I_1(y) \subseteq I_1(y')$ as $y' \sqsubseteq y$)\\
&$=$&$I_1(y) \cup (I_1(x) \cap I_1(y'))$ (because $(I_1(x') \cap I_1(y)) \subseteq I_1(y)$)
\end{tabular}

\smallskip\noindent
Hence $I_{24}$ is defined as shown below:

\smallskip
\begin{tabular}{l}
$I_{24}(a)= I_1(a) \cup I_1(x) \cup (I_1(x') \cap I_1(y))$\\$I_{24}(x)=I_1(x) \cup (I_1(x') \cap I_1(y))$\\
$I_{24}(b) = I_1(b) \cup I_1(y) \cup (I_1(x)\cap I_1(y'))$\\$I_{24}(y) =I_1(y) \cup (I_1(x) \cap I_1(y'))$
\end{tabular}

\smallskip\noindent
Referring again to Figure~\ref{fig-exp}(d), we now turn to the computations regarding $I_3$, $I_{31}$ and $I_{34}$. As the computations are similar to those above, some details are skipped.

\smallskip\noindent
$\bullet$ {\em Interpretation $I_3=Exp(I_1, yb)$.}\\
Here, we have $I_3(b)=I_1(b) \cup I_1(y)$, due to expansion. Computations similar to those in the case of $I_2$ show that $I_3$ is defined as shown below:

\smallskip
\begin{tabular}{ll}
$I_3(a)=I_1(a)$&\qquad$I_3(x)=I_1(x) \cup (I_1(x')\cap I_1(y))$\\
$I_3(b) = I_1(b) \cup I_1(y)$&\qquad$I_3(y) = I_1(y)$
\end{tabular}

\smallskip\noindent
$\bullet$ {\em Interpretation $I_{31}=Exp(I_3, xa)$.}\\
Here, we have $I_{31}(a)=I_{3}(a) \cup I_{3}(x)$, due to expansion. Computations similar to those for $I_{24}$ show that $I_{31}$ is defined as shown below:
%
%
%
%
%

\smallskip
\begin{tabular}{l}
$I_{31}(a)= I_1(a) \cup I_1(x) \cup (I_1(x') \cap I_1(y))$\\$I_{31}(x)=I_1(x) \cup (I_1(x') \cap I_1(y))$\\
$I_{31}(b) = I_1(b) \cup I_1(y)$\\$I_{31}(y) =I_1(y) \cup (I_1(x) \cap I_1(y'))$
\end{tabular}

\smallskip\noindent
$\bullet$ {\em Interpretation $I_{34}=Exp(I_{31}, yb)$.}\\
Here, we have $I_{34}(b)=I_{31}(b) \cup I_{31}(y)$, due to expansion. It follows that computations similar to those for $I_{24}$ show that $I_{34}$ is defined as shown below:
%
%
%
%
%
%

\smallskip
\begin{tabular}{l}
$I_{34}(a)=I_1(a) \cup I_1(x) \cup (I_1(x')\cap I_1(y))$\\$I_{34}(x)=I_1(x) \cup (I_1(x') \cap I_1(y))$\\
$I_{34}(b) = I_1(b) \cup I_1(y) \cup (I_1(x) \cap I_1(y'))$\\$I_{34}(y) =I_1(y) \cup (I_1(x) \cap I_1(y'))$
\end{tabular}

\smallskip\noindent
It thus turns out that we indeed have $I_{24}=I_{34}$, and the proof is complete.
\hfill$\Box$

\medskip\noindent
The last case to be addressed is when $y=ay'$ and $x \ne bx'$ or when $x=bx'$ and $y \ne ay'$. We only consider the former case since the latter can be dealt with in a similar way. Moreover, as shown in Figure~\ref{fig-exp}(c), this case is in fact a simplification of the case of Lemma~\ref{lemma:case-2} where the terms involving $x'$ are omitted. We thus omit the proof and just state that is this case again $I_{24}=I_{34}$. It can be checked that in case of Figure~\ref{fig-exp}(c) (that is $y=ay'$ and $x \ne bx'$), we obtain the following interpretations:

\smallskip\noindent 
\begin{tabular}{l}
$I_{24}(a)=I_{34}(a)=I_1(a) \cup I_1(x)$\\$I_{24}(x)=I_{34}(x)=I_1(x)$\\
$I_{24}(b)=I_{34}(b) = I_1(b) \cup I_1(y) \cup (I_1(x) \cap I_1(y'))$\\$I_{24}(y)=I_{34}(y) =I_1(y) \cup  (I_1(x) \cap I_1(y'))$ 
\end{tabular}

\smallskip\noindent
We therefore state the following basic result regarding expansion:
\begin{itemize}
\item[]
{\em The process of expanding an interpretation $I$ with respect to a set $FD$ of functional dependencies satisfies the Church-Rosser property.}
\end{itemize}
Hence, given $T$ and $FD$, the process of expanding $I^b$ until a fixed point is reached does {\em not} depend on the order the expansions are processed. Thus, the limit is indeed {\em unique}.
\section{Proof of Lemma~4}\label{append:lemma-chase}
{\bf Lemma~\ref{lemma:chase}~}
{\em Let $T$ be a table over universe $U$. Then the following holds:
\begin{enumerate}
\item
Algorithm~\ref{algo:chase} applied to $T$ always terminates.
\item
For every m-tuple $s$  in $\mathcal{MT}$, $I^*(s) \ne \emptyset$ holds if and only if there exists $\sigma$ in $\Sigma^*$ such that $s \sqsubseteq \sigma$.
\item
$\inc(\mathcal{T})= \emptyset$ if and only if $fail = false$.
\item
For every $\sigma$ in $\Sigma^*$ and all $t$ and $t'$ in $\tuples(\sigma)$, $I^*(t)=I^*(t')$.
\end{enumerate}}
{\sc Proof.}
{\bf 1.~}The computation of $\Sigma^*$ terminates because $(i)$ the sets under construction for a given attribute $A$ are all bounded by the {\em finite} set $adom(A)$, and $(ii)$ the construction is monotonous, in the sense that if $\Sigma^k$ and $\Sigma^{k+1}$ are two consecutive states, for every $\sigma$ in $\Sigma^{k+1}$, there exists $\sigma'$ in $\Sigma^{k}$ such that $\sigma' \sqsubseteq \sigma$.

\smallskip\noindent
{\bf 2.~}Given  $s$ in $\mathcal{MT}$, we first show that  $I^*(s)$ is nonempty if there exists $\sigma$ in $\Sigma^*$ such that $s \sqsubseteq \sigma$. The proof is conducted by induction on the steps in the runs of the loop line~\ref{line:main-loop-chase}. We show that if $(\Sigma^k)_{k \geq 0}$ denotes the sequence of the states of $\Sigma^*$ during the computation, for every $\sigma$ in $\Sigma^k$, $I^*(\sigma) \ne \emptyset$. We first note in this respect that since $\Sigma^0 = T$, for every $\sigma$ in $\Sigma^0$,  we have $I^*(\sigma) \ne \emptyset$.

Assuming now that for $k\geq 0$, for every  $\sigma \in \Sigma^{k}$, $I^*(\sigma) \ne\emptyset$, we prove the result for every $\sigma$ in $\Sigma^{k+1}$. Let $\sigma$ in $\Sigma^{k+1}$ but not in $\Sigma^k$. Then $\sigma$ occurs in $\Sigma^{k+1}$ because there exist $X \to A$ in $FD$, $\sigma_1$ and $\sigma_2$ in $\Sigma^k$ such that for every $B$ in $X$, $\sigma_1.B \cap \sigma_2.B \ne \emptyset$. Let $\sigma'_1 = \sigma_1 \sqcup \sigma_2(A)$ and  $\sigma'_2 = \sigma_2 \sqcup \sigma_1(A)$. Then, either $\sigma = \sigma'_1$ and $\sigma'_1 \ne \sigma_1$ or $\sigma = \sigma'_2$ and $\sigma'_2 \ne \sigma_2$. The two cases being similar, we just consider the first one, that is $\sigma = \sigma'_1$ and $\sigma'_1 \ne \sigma_1$, which implies that $\sigma_2(A) \ne \emptyset$ and $\sigma_2(A) \not\sqsubseteq \sigma_1$. 
By our induction hypothesis, we have $I^*(\sigma_2) \ne \emptyset$ and thus denoting by $x$ an $X$-tuple occurring in $\tuples(\sigma_1 \sqcap \sigma_2)$, we have $I^*(x) \cap I^*(\sigma_2(A)) \ne \emptyset$. Since $I^*$ satisfies $X \to A$, $I^*(x) \subseteq I^*(\sigma_2(A))$. As $x \sqsubseteq \sigma_1$, $I^*(\sigma_1) \subseteq I^*(x)$ and thus, $I^*(\sigma_1) \subseteq I^*(\sigma_2(A))$. Therefore, $I^*(\sigma)=I^*(\sigma_1) \cap I^*(\sigma_2(A))= I^*(\sigma_1)$. As by our induction hypothesis, $I^*(\sigma_1) \ne \emptyset$, we obtain that $I^*(\sigma) \ne \emptyset$. Therefore, it holds that for every $k \geq 0$, $I^*(\sigma) \ne \emptyset$ for every $\sigma$ in $\Sigma^k$, and so, it holds that $I^*(\sigma) \ne \emptyset$ for every $\sigma$ in $\Sigma^*$. Thus, if $s$ satisfies that there exists $\sigma$ in $\Sigma^*$ such that $s \sqsubseteq \sigma$, we have $I^*(\sigma) \subseteq I^*(s)$. Hence $I^*(s) \ne \emptyset$, and this part of the proof is complete.

Conversely, we show that for every $s$, if $I^*(s)\ne \emptyset$ then there exists $\sigma$ in $\Sigma^*$ such that $s \sqsubseteq \sigma$. The proof is done by induction on the construction of $I^*$.

By definition of $I^0=I^b$, if $I^0(s) \ne \emptyset$ then $s$ is a sub-tuple of a tuple $t$ in $T$, i.e., $s \sqsubseteq t$ holds. Since $\Sigma^0 = T$, there exists $\sigma$ in $\Sigma^0$ such that $t \sqsubseteq \sigma$. By monotonicity of the construction of $\Sigma^*$ (i.e., for every $\sigma^{k+1}$ in $\Sigma^{k+1}$ there exists $\sigma^k$ in $\Sigma^{k}$ such that $\sigma^{k} \sqsubseteq \sigma^{k+1}$), there exists $\sigma^*$ in $\Sigma^*$ such that $\sigma \sqsubseteq \sigma^*$, which shows that $t \sqsubseteq \sigma^*$, thus that $s \sqsubseteq \sigma^*$.

Now, assuming that for every $k \geq 0$ and every $s$, if $I^k(s) \ne \emptyset$ then there exists $\sigma$ in $\Sigma^*$ such that $s \sqsubseteq \sigma$, we prove that this result holds for $I^{k+1}$.

Let $s$ be such that $I^k(s) =\emptyset$ and $I^{k+1}(s) \ne \emptyset$. For every $a$ in $\mathcal{AD}$ we have $I^{k+1}(a)=I^k(a) \cup I^k(x)$  if $x$ and $a$ are the tuples involved in $Exp(I^k, xa)$ and $I^{k+1}(a) = I^{k}(a)$ otherwise. Since $I^{k+1}(s) \ne I^k(s)$ and since $a$ is the only symbol for which $I^k$ has changed, it must be that $a \sqsubseteq s$. Therefore, writing $s$ as $s' \sqcup a$, we have $I^{k+1}(s)= I^{k+1}(s') \cap I^{k+1}(a)$ and $I^{k+1}(s') = I^k(s')$. It follows that $I^{k+1}(s)= I^{k}(s') \cap (I^{k}(a) \cup I^k(x))$, thus that $I^{k+1}(s)= (I^{k}(s') \cap (I^{k}(a)) \cup (I^k(s') \cap I^k(x))$, that is $I^{k+1}(s)= I^{k}(s) \cup  (I^k(s') \cap I^k(x))$.
As $I^{k+1}(s) \ne \emptyset$ and $I^k(s)=\emptyset$, we have  $I^k(s')\cap I^k(x) \ne \emptyset$, in which case our induction hypothesis entails that there exists $\sigma$ in $\Sigma^*$ such that $(s' \sqcup x) \sqsubseteq \sigma$. Since by definition of expansion, we also have $I^k(x) \cap I^k(a) \ne \emptyset$, $\Sigma^*$ contains an m-tuple $\sigma'$ such that $xa \sqsubseteq \sigma'$. In this case, Algorithm~\ref{algo:chase} shows that there exists $\sigma^*$ in $\Sigma^*$ such that $(s' \sqcup a) \sqsubseteq \sigma^*$, that is $s \sqsubseteq \sigma^*$. This part of the  proof is therefore complete.

\smallskip\noindent
{\bf 3. }If the returned value for $fail$ is false, the previous item shows that there exist $X \to A$ in $FD$, $x$ in $\mathcal{T}(X)$ and $a$ and $a'$ in $adom(A)$ such that $I^*(a) \cap I^*(a') \ne \emptyset$. Therefore in this case, $\inc(\mathcal{T})$ is not empty. Conversely, if $\inc(\mathcal{T}) \ne \emptyset$, for every $t$ in $\inc(\mathcal{T})$, there exist $A$ in $U$ and $a$ and $a'$ in $adom(A)$ such that $I^*(a) \cap I^*(a') \ne \emptyset$, and the previous item shows that there exists $\sigma$ in $\Sigma^*$ such that $(aa') \sqsubseteq \sigma$. Hence, during the execution of Algorithm~\ref{algo:chase}, the value of $fail$ must have been turned to $true$ line~\ref{line:fail-true}. 

\smallskip\noindent
{\bf 4.~}The proof is by induction on the construction of $\Sigma^*$. As all m-tuples in $\Sigma^0$ can be seen as tuples, for every $\sigma$ in $\Sigma^0$, $\tuples(\sigma)$ consists of one tuple. The result thus trivially holds in this case. Now assume that the result holds for $\Sigma^k$, and let $\sigma$ be in $\Sigma^{k+1}$ obtained by steps lines~\ref{line:step1-end1}-\ref{line:step1-end} of Algorithm~\ref{algo:chase}. Thus there exist $\sigma'$ in $\Sigma^*$, $X \to A$ in $FD$, $x_0$ in $\mathcal{T}(X)$ and $a_0$ in $adom(A)$ such that in  $\Sigma^k$, $x_0 \in \tuples(\sigma(X))\cap \tuples(\sigma'(X))$, $a_0$ is in $\tuples(\sigma'(A)) \setminus \tuples(\sigma(A))$, and in $\Sigma^{k+1}$, $\sigma(A)$ becomes $\sigma_1(A)$ such that $a_0 \in \tuples(\sigma_1(A))$.

Let $t_1$ be in $\tuples(\sigma_1)$ such that $t_1.A=a_0$, and let $x$ denote $t_1.X$. Then, $t_1$ is not in $\tuples(\sigma)$ (because $a_0 \not\in \tuples(\sigma(A))$), but $\tuples(\sigma)$ contains a tuple $t$ such that $t.X=x$ and a tuple $t'$ such that $t'.X=x_0$. Thus $\tuples(\sigma_1)$ contains $t_1$ such that $t_1.XA=xa_0$ and a tuple $t'_1$ such that $t'_1.XA=x_0a_0$.

On the other hand, the fact that $x_0a_0$ occurs in $\tuples(\sigma')$ implies that $I^*(x_0a_0) \ne \emptyset$, thus that $I^*(x_0) \subseteq I^*(a_0)$. Hence, $I^*(x_0a_0)=I^*(x_0)$. We now argue that $I^*(t'_1)=I^*(t')$. Indeed, if $\sigma(A)=\emptyset$, then $t'_1$ can be written as $t'a_0$ in which case $I^*(t'_1) = I^*(t') \cap I^*(a_0)$  holds. Since it also holds that  $I^*(t') \subseteq I^*(x_0) \subseteq I^*(a_0)$, we have $I^*(t')\cap I^*(a_0)=I^*(t')$. Otherwise, if $\sigma(A)\ne\emptyset$ then $t'$ is written as $q'a'$ and thus $t'_1$ is written as $q'a_0$ ($a'$ has been replaced by $a_0$ to obtain $t'_1$ from $t'$). In this case, $I^*(t'_1)=I^*(q')=I^*(t')$.

As $t$ and $t'$ are both in $\tuples(\sigma)$, by our induction hypothesis, in $\Sigma^{i}$ we have $I^*(t)=I^*(t')$, which implies that $I^*(t)=I^*(t'_1)=I(t')$. We therefore obtain that for every $t_1$ in $\tuples(\sigma_1)$, there exists $t$ in $\tuples(\sigma)$ such that $I^*(t_1)=I^*(t)$. As all tuples in $t$ $\tuples(\sigma)$ have the same $I^*(t)$, all tuples $t_1$ in $\tuples(\sigma_1)$ have the same $I^*(t_1)$. The proof is therefore complete.
\hfill$\Box$
\section{Proof of Proposition~5}\label{append:mscons-answer}
{\bf Proposition~\ref{prop:mscons-answer}~}
{\em 
Let $T$ be a non-redundant table over universe $U$, $FD$ the set of associated functional dependencies and $\Sigma^*$ the non-redundant m-table as computed by Algorithm~\ref{algo:chase}. Given a query $Q: {\tt select}$ $X$  {\tt from} $T$ {\tt [where $\Gamma$]}, and a tuple $x$ in $\mathcal{T}(X)$:
\begin{enumerate}
\item $x$ is in $\msconsans(Q)$ if and only if
\begin{itemize}
\item[(a)] for every $\sigma$ in $\Sigma(x)$, $\tuples(\sigma(X))=\{x\}$,
\item[(b)] there exists $\sigma$ in $\Sigma_Q(x)$ such that $|\tuples(\sigma)| =1$.
\end{itemize}
\item $x$ is in $\sconsans(Q)$ if and only if
\begin{itemize}
\item[(a)] for every $\sigma$ in $\Sigma(x)$, $\tuples(\sigma(X))=\{x\}$,
\item[(b)] there exists $s$ in $Sat(\Gamma) \cap \true(\mathcal{T})$ such that for every $\sigma$ in $\Sigma_Q(x)$, if $s$ is  in $\sigma(sch(\Gamma))$ then $\tuples(\sigma(sch(\Gamma)))) =\{s\}$.
\end{itemize}
\item $x$ is in $\consans(Q)$ if and only if
\begin{itemize}
\item[(a)] for every $\sigma$ in $\Sigma(x)$, $\tuples(\sigma(X))=\{x\}$,
\item[(b)] there exists $\sigma$ in $\Sigma_Q(x)$ such that $\tuples(\sigma(sch(\Gamma))) \subseteq Sat(\Gamma)$.
\end{itemize}
\item $x$ is in $\wconsans(Q)$ if and only if
\begin{itemize}
\item[(a)] for every $\sigma$ in $\Sigma(x)$, $\tuples(\sigma(X))=\{x\}$,
\item[(b)] $\Sigma_Q(x)$ is not empty.
\end{itemize}
\end{enumerate}
}
{\sc Proof.}
In this proof, $Q$ is assumed to involve a selection condition $\Gamma$, because the case where no selection condition is involved is a simplification, and thus, is omitted.

\smallskip\noindent
{\bf 1.} By Proposition~\ref{prop:chase}(1) and (2), statement (a) in the proposition is equivalent to statement (1.a) in Definition~\ref{def:cons-answer}. Moreover, as $\Sigma^*$ is assumed to be non-redundant, for every $\sigma$ in $\Sigma^*$ and every $t$ in $\tuples(\sigma)$, there exists no tuple $t'$  in $\true(\mathcal{T})$ such that $t \sqsubset t'$. Thus, by Proposition~\ref{prop:chase}(2), statement (b) in the proposition yields a tuple having the same properties as the tuple $t$ in statement (1.b) of Definition~\ref{def:cons-answer} (i.e., $t$ is a maximal true and consistent tuple whose restriction over $sch(\Gamma)$ satisfies $\Gamma$). Therefore this part of the proof is complete.

\smallskip\noindent
{\bf 2.} As above, by Proposition~\ref{prop:chase}(1) and (2), statement (a) in the proposition is equivalent to statement (2.a) in Definition~\ref{def:cons-answer}. Moreover, statement (b) in the proposition yields a tuple $t$ such that $t.X=x$, $t.sch(\Gamma)=s$, which is in $Sat(\Gamma)$. Moreover, Proposition~\ref{prop:chase}(3) implies that this tuple $t$ is true and consistent, thus satisfying the statement (2.b) of Definition~\ref{def:cons-answer} (i.e., $t$ is a  true and consistent tuple whose restriction over $sch(\Gamma)$ satisfies $\Gamma$). Therefore this part of the proof is complete.

\smallskip\noindent
{\bf 3.} Let us first assume that $x$ is in $\consans(Q)$. As $x$ is in $\cons(\mathcal{T}) \cap \true(\mathcal{T})$, by Proposition~\ref{prop:chase}, $\Sigma^*$ must contain at least one $\sigma$ such that $x \sqsubseteq \sigma$ and every such m-tuple must be such that $\sigma(X)=\{x\}$. Hence $\Sigma(x)$ is nonempty and the first item in the proposition holds. Moreover, by Definition~\ref{def:cons-answer}, one of these m-tuples $\sigma$ is also such that $sch(\Gamma) \subseteq sch(\sigma)$ and $\tuples(\sigma)$ contains a tuple $t$ of $\cons(\Gamma, \mathcal{T})$. By Definition~\ref{def:incons-S}, this implies that the following statement (*) holds: $t.sch(\Gamma)$ is in $Sat(\Gamma)$ and for every $s$ in $Sat^-(\Gamma)$, $I^*(t) \cap I^*(s) =\emptyset$. This in particular implies that $\sigma$ is in $\Sigma_Q(x)$, because $t.(sch(\Gamma)) \sqsubseteq t \sqsubseteq \sigma$ holds.

As $\Sigma_Q(x)\subseteq \Sigma(x)$, every $\sigma$ in $\Sigma_Q(x)$ is such that $\sigma(X) = \{x\}$. Assume now that every $\sigma$ in $\Sigma_Q(x)$ is such that $\tuples(\sigma(sch(\Gamma)))$ contains $q_0$ not in $Sat(\Gamma)$. Then $q_0$ is in $Sat^-(\Gamma)$  and $\tuples(\sigma)$ contains a tuple $t_0$ such that $t_0.sch(\Gamma) =q_0$. Thus we have two tuples $t$ and $t_0$ in $\tuples(\sigma)$ such that $t.X=t_0.X=x$, $t.sch(\Gamma) \in Sat(\Gamma)$ and $t_0.sch(\Gamma) \in Sat^-(\Gamma)$. Lemma~\ref{lemma:chase}(4) implies that $I^*(t) = I^*(t_0)$ (because $t$ and $t_0$ are in $\tuples(\sigma)$), and as $I^*(t_0) \subseteq I^*(q_0)$ (because $t_0.sch(\Gamma)=q_0$), we have $I^*(t_0) \cap I^*(q_0) = I^*(t_0)$. Thus $I^*(t_0) \cap I^*(q_0)\ne \emptyset$, implying that $I^*(t) \cap I^*(q_0) \ne \emptyset$, which is a contradiction with statement (*) above. We therefore have shown that if $x$ is in $\consans(Q)$ then the items in the proposition hold.

Conversely, if the items in the proposition are satisfied, Proposition~\ref{prop:chase} implies that $x$ is in $\cons(\mathcal{T}) \cap \true(\mathcal{T})$. Thus by Definition~\ref{def:incons-S}, we have to prove the existence of a tuple $t$ in $\cons(\Gamma, \mathcal{T}) \cap \true(\mathcal{T})$ such that $x \sqsubseteq t$. Let $\sigma$ be in $\Sigma_Q(x)$ as stated in item (3.b) of the proposition, that is such that $\tuples(\sigma(sch(\Gamma))) \subseteq Sat(\Gamma)$, and let $t_0$ be in $\tuples(\sigma)$. As $t_0$ is in $\true(\mathcal{T})$, we have to show that $t_0$ is in $\cons(\Gamma, \mathcal{T})$, that is that there exists $q$ in $Sat(\Gamma)$ such that $(i)$ $q \sqsubseteq t_0$, and $(ii)$ for every $q'$ in $Sat^-(\Gamma)$, $I^* (t_0) \cap I^*(q') = \emptyset$. Since we know that $t_0.sch(\Gamma)$ is in $Sat(\Gamma)$, $(i)$ holds for $q=t_0.sch(\Gamma)$. Regarding $(ii)$, let $q'_0$ be in $Sat^-(\Gamma)$ such that $I^*(t_0) \cap I^*(q'_0) \ne \emptyset$. As  $I^*(\sigma) = \bigcap_{u \in \tuples(\sigma)}I^*(u)$,  Lemma~\ref{lemma:chase}(4) implies that  $I^*(\sigma) =I^*(t_0)$, and thus that $I^*(\sigma) \cap I^*(q'_0) \ne \emptyset$. Applying now Lemma~\ref{lemma:chase}(2), we obtain that $\Sigma^*$ contains an m-tuple $\sigma'$ involving all attribute values in $\sigma$ or in $q'_0$. Since $q'_0$ is not in $\tuples(\sigma)$, we have $\sigma' \ne \sigma$, and thus $\sigma \sqsubset \sigma'$ holds. This is a contradiction with our hypothesis that $\Sigma^*$ contains no redundancy, showing that $t_0$ is in $\cons(\Gamma, \mathcal{T})$. This part of the proof is therefore complete.

\smallskip\noindent
{\bf 4.}  As in cases (1) and (2) above, by Proposition~\ref{prop:chase}(1) and (2), statement (4.a) in the proposition is equivalent to statement (2.a) in Definition~\ref{def:cons-answer}. Moreover, $\Sigma_Q(x) \ne \emptyset$ is equivalent to the existence of $\sigma$ in $\Sigma^*$ such that $\sigma(X)=\{x\}$ and  $\sigma(sch(\Gamma))$ contains a tuple $s$ of $Sat(\Gamma)$. Hence,  $\Sigma_Q(x) \ne \emptyset$ is equivalent to the existence of a tuple $t$ in $\true(\mathcal{T})$ (due to Proposition~\ref{prop:chase}(1)) such that $t.X=x$ and $t.sch(\Gamma)$ is in $Sat(\Gamma)$. The proof is therefore complete.
\hfill$\Box$
\end{document}